\documentclass[iop]{emulateapj}

\begin{document}

\title{Photoionization Emission Models for the Cyg X-3 X-ray Spectrum}

\author{T. Kallman\altaffilmark{1}, M. McCollough\altaffilmark{2}, L. Corrales \altaffilmark{5}, K. Koljonen\altaffilmark{3},  D. Liedahl\altaffilmark{4},  J. Miller\altaffilmark{5}, F. Paerels\altaffilmark{6},  G. Pooley\altaffilmark{7}, M. Sako\altaffilmark{8}, N. Schulz\altaffilmark{9}, S. Trushkin\altaffilmark{10,11}}

\altaffiltext{1}{NASA/GSFC, Code 662, Greenbelt MD 20771}
\altaffiltext{2}{Harvard-Smithsonian Center for Astrophysics}
\altaffiltext{3}{Finnish Centre for Astronomy with ESO (FINCA), University of Turku, V\"ais\"al\"antie 20, 21500 Piikki\"o, Finland
Aalto University Mets\"ahovi Radio Observatory, PO Box 11000, FI-00076 Aalto, Finland}
\altaffiltext{4}{Lawrence Livermore National Laboratory}
\altaffiltext{5}{Department of Astronomy, University of Michigan, Ann Arbor MI}
\altaffiltext{6}{Department of Astronomy, Columbia University}
\altaffiltext{7}{Mullard Radio Astronomy Observatory, Cambridge  UK}
\altaffiltext{8}{University of Pennsylvania}
\altaffiltext{9}{MIT Center for Space Research}
\altaffiltext{10}{Special Astrophysical Observatory RAS, Nizhnij Arkhyz 369167, Russia}
\altaffiltext{11}{Kazan Federal University, Kazan, 420008, Russia}

%\affil{NASA/GSFC, Code 662, Greenbelt MD 20771}

\begin{abstract}
We present model fits to the X-ray line spectrum of the well known High Mass X-ray binary Cyg X-3.  The 
primary observational dataset is a spectrum taken with the $Chandra$ X-ray Observatory High 
Energy Transmission Grating (HETG) in 2006, though we compare it to all the other 
observations of this source taken so far by this instrument.  
 We show that the density must be $\geq 10^{12}$ cm$^{-3}$ in the region responsible 
for most of the emission.   We discuss the influence of the dust scattering halo on the broad band spectrum 
and we argue that dust scattering and extinction is not the most likely origin for the narrow feature
seen near the Si K edge. We identify the features of a wind in the profiles of the 
strong resonance lines and we show that the wind is more apparent in the lines from the 
lighter elements.  We argue that this wind is most likely associated with the companion star.  
We show that the intensities of most lines can be fitted, crudely, by 
a single component photoionized model.  However, the iron K lines do not fit with this 
model.  We show that the iron K line variability as a function of orbital phase is 
different from the lower energy lines, which indicates that the lines arise in physically 
distinct regions.  We discuss the interpretation of these results in the context of 
what is known about the system and similar systems. 
\end{abstract}

\section{Introduction}

Cygnus X-3 (4U 2030+40, V1521 Cyg) is a high-mass X-ray binary (HMXB) which is peculiar owing to its short orbital period (P = 4.8 h; Parsignault et al. 1972) and its radio brightness (~100 mJy in a quiescent state, up to 20 Jy during outbursts; e.g. Waltman et al. 1995). The most likely distance to the source is 7.4$\pm$1.1 kpc  \citep{mcco16,mcco13}. The companion is a WN4-6 type Wolf-Rayet (WR) star \citep{vank92, vank96, kolj17}. In the X-ray band, many of the observational features of Cyg X-3 are likely due to the strong wind from the companion star \citep{paer00,szos08,zdzi10,zdzi12}. The wind absorbs and scatters X-rays from the compact object, and produces the broad sinusoidal orbital light curve \citep{will85,zdzi12}. The material responsible for this behavior also prevents unobscured observation of the compact X-ray source (the exception being jet ejection events, where the ram pressure of the jet can displace the stellar wind; \citep{kolj18}. A quantitative understanding of the wind thus can aid in disentangling its effect on the observed properties of Cyg X-3 from the intrinsic properties of the compact object.

Observations in the 1-10 keV band  reveal very strong line emission \citep{serl75}; 
in the hard state, \citet{hjal08} found  equivalent widths $\sim$0.2 keV, 
softer states have 0.2-0.4 keV \citep{kolj10, kolj18}.  The most likely scenario for this emission is
reprocessing of continuum X-rays by the wind.  If so, 
these lines provide sensitive constraints on the conditions in the 
wind and other structures in the binary \citep{paer00}.
Study of these lines is aided by the brightness of Cyg X-3, making it relatively 
well suited to study with current X-ray spectrometers.
The spectrum is highly cut off by interstellar attenuation, with an average 
equivalent hydrogen column density of $\geq 10^{22}$ cm$^{-2}$ \citep{kalb05}.
The X-ray line spectrum shows emission from ions of all abundant elements 
visible at energies above 1 keV, from Mg through Fe.  
These indicate the importance of photoionization as an 
excitation mechanism in the line emitting gas.  Observations using the Chandra
High Energy Transmission Grating (HETG) have provided insights including the 
presence of radiative recombination continua (RRCs) which constrain the 
the gas temperature \citep{paer00}, 
orbital modulation of the emission line centroids  \citep{star03}, 
and P Cygni profiles which constrain the masses of the two stars \citep{vilh09}. 

Spectra of Cyg X-3 obtained by the Chandra HETG are affected 
by the brightness of the source, leading to pileup when data is taken in the timed-event (TE) 
mode, and also by the variability of the source over timescales long compared with the 
binary period.  As we show in this paper, there is variability present in the lines also along the orbit.
  In addition to the studies of temperature and dynamics published so far, the 
X-ray line spectrum contains information on the density, elemental abundances, and ionization mechanism 
in the emitting gas.   Until now, there has not been 
an analysis of the entire spectrum obtained by the HETG which addresses these topics.  
In this paper we present model fits to the Chandra HETG spectrum.  We focus primarily on 
one observation taken during a high state, and taken in continuous clocking mode 
in order to mitigate the effects of pileup.  This spectrum was previous
reported by \cite{vilh09}. 
The strong continuum observed from Cyg X-3, together with the detection of RRCs \citep{lied96}, are indications that the ionization 
balance and excitation are dominated by photoionization and photoexcitation.  
% We focus on disentangling the effects of the wind from 
%those of static gas, which is also present in the spectrum.

\section{Data Analysis}

\subsection{Chandra Data}

The primary datasets used in this analysis were obtained with the
Chandra High Energy Transmission Grating (HETG)  (Obsids 7268 and 6601, 
PI McCollough) on 2006 January 25 – 26. The observations 
started on MJD 53760.59458 (when the source was at X-ray orbital phase 0.053) and went 
through to MJD 53761.42972 ( X-ray orbital phase 0.240). During these observations Cyg X-3 
was in a high soft state (quenched radio state) with an average RXTE/ASM 
count rate around 30 cps (corresponding to 400 mCrab; typical hard state 
count rates are less than 10 cps).  The observations in Obsid 7268 were done in CC mode 
(continuous clocking) with a window filter applied to 
the zeroth order.  The observations in Obsid 6601 were done in TE (timed event) mode 
with a  440 pixel subarray which results in a 1.4 sec frame time.  The data were reduced using standard Ciao tools, with 
appropriate modifications for CC mode data.   The average flux during the observation was 9.3 $\times 10^{-9}$ erg cm$^{-2}$ s$^{-1}$ or 
approximately 400 mcrab.    Although Cyg X-3 is bright enough to permit analysis of the third order spectra, 
in this paper we work solely with the first order HETG spectra.  We will report on analysis of third order spectra 
in a subsequent paper.

In the case of Obsid 6601 there is a difference between the HEG and MEG with the MEG falling
below the HEG spectra in the range of 2.3 to 4.4 Angstrom range
(2.8 to 5.4 keV).  This is likely evidence of  pileup as reported by \citet{corr15}.
They found the pileup was a problem with the MEG
spectra (light is dispersed over a smaller angle) but not  a
major problem with the HEG.  Also the the various orders (2 and 3rd)
are overlapping in this spectral region.  The problem with the MEG
spectra is that in addition to reducing (and distorting) the continuum
spectra the lines are also impacted even more. As a result of this, in what follows we 
ignore the MEG data in fitting to Obsid 6601.

In the case of Obsid 7268  we analyze and discuss the spectrum obtained from the +1 and -1 
orders of both the HEG and MEG arms of the HETG.  These were analyzed simultaneously with the same model
applied to both.  Analysis was done using the {\sc xspec}\citep{arna96} analysis program.
All fits were performed using the c-statistic \citep{cash79} and no rebinning was performed during the 
fitting or the plotting.  Plots of the spectrum show just the HEG +1 order spectra for clarity.
 The CC mode data has the 
advantage that it will not be affected by pileup.  In principle, it can be  affected by possible 
background contamination owing to the effectively much larger spectral extraction 
region on the ACIS chips.   However, in the case of Cyg X-3, the sky 
background is dominated by the dust scattered halo \citep{corr15}, 
and the use of CC mode data allows a more accurate treatment of this component than does TE mode.
For this reason, in what follows we illustrate many of our results using  analysis of 
the data from Obsid 7268.  Fits to Obsid 6601 are included for comparison.

In addition we have analyzed all other available HETG spectra of Cyg X-3, with significant observing time.  Exposures and observation dates are given in table 
\ref{obstable}.  
%Obsid 101 is a short $\sim$2 ksec observations in TE mode whose end time corresponds to the start time of Obsid 1456\_000.  
Here and in what follows we have experimented with joint fits which allow us to include spectra which contain less than 2 ksec of data.  We find that this results in negligible improvements or changes 
to our fits and so we have not included the short observations in the fits reported here.
%Obsid 1456 consists of two observations that were done two months apart in TE mode.  The state of Cyg X-3  had changed slightly between observations and %so we have analyzed each of the observations separately:  
Obsid 1456\_000 
was a $\sim$12 ksec observation when Cyg X-3 was transitioning from a flaring state to a 
quiescent state.  Obsid 1456\_002 was an $\sim$ 8 ksec exposure obtained two months later and Cyg X-3 was in a quiescent state. 
Obsid 425 was an observation after a major flare in graded mode.  It was done in alternating exposure mode, so 
there is an e1 (short exposure) and e2 (longer exposure).  The longer exposure is $\sim$18.5 ksec long and 
the short exposure is ~0.4 ksec long.  
Obsid 426 was an observation that was done five days after 425 in graded mode.  The longer exposure is ~15.7 ksec long 
and the short exposure is ~0.3 ksec long. 
Obsid 16622 was a calibration observation (done in CC mode) made during a quiescent state.  
%Only two arms of the HETG were on the chips. 
Results from fitting to these spectra are given in subsection \ref{otherspec}, 
and in subsection \ref{16622} since Obsid 16622 was taken in 
a very low flux state and is discussed separately.

%\begin{deluxetable}{crrrrrrrrrr}
%\tabletypesize{\scriptsize}
%\tablecaption{Table of Observations \label{obstable}}
%\tablewidth{0pt}
%\tablehead{
%\colhead{obsid}&\colhead{date (MJD)}&\colhead{tobs(s)}&\colhead{mode}&\colhead{flux (erg/s/cm$^2$)}}
%\startdata
%1456&51471.04861111&20800&TE&3.32E-09\\
%0425&51638.57531&21530&TE&3.65E-09\\
%0426&51640.83333&18320&TE&3.74E-09\\
%7268&53760.59653&70130&CC&3.60E-09\\
%6601&53761.63125&51020&TE&7.75E-09\\
%16622&56833.22431&28630&TE&1.78E-09\tablenotemark{a}\\
%\enddata
%\tablenotetext{a}{Flux is from MAXI.  All others from RXTE ASM.}
%\end{deluxetable}

\begin{deluxetable}{crrrrrrrrrr}
\tabletypesize{\scriptsize}
\tablecaption{Cygnus X-3 HETG Chandra Observations \label{obstable}}
\tablewidth{0pt}
\tablehead{
\colhead{ObsID}&\colhead{Date (MJD)}&\colhead{Instrument}&\colhead{Data Mode}&\colhead{Exp Mode}&\colhead{Exposure (ksec)}&\colhead{Count $\rm Rate^a$ (cts/sec)}&\colhead{State}}
\startdata
{$\rm 101^b$} &  51471 & ACIS-S & FAINT & TE & 1.95 & 35.8 & t/mf \\
{1456 (obi~0)$^b$}  & 51471  & ACIS-S & FAINT & TE & 12.12 & 27.7 & t/mf \\
{1456 (obi 2)} & 51531 & ACIS-S & FAINT & TE & 8.42 & 18.0 & q/qi \\
{425$^c$} & 51638 & ACIS-S & GRADED & TE & 18.54 & 88.2 & fs/mrf \\
{426$^c$} & 51640 & ACIS-S &  GRADED & TE & 15.68  & 71.0 & fi/mrf \\
{6601} & 53761 & ACIS-S & FAINT & TE & 49.56 & 106.6 & h/qu \\
{7268} & 53761 & ACIS-S & CC33-GRADED & CC & 69.86 & 121.4 & h/qu \\
{16622$^d$} & 56833 & ACIS-S & CC33-FAINT & CC & 28.51 & 25.9 & q/qi \\
\enddata
\tablecomments{The states are given as {\it k/s} where: {\it k} are those of \citet{kolj10} [q: Quiescent, t: Transition, fh: FHXR, fi: FIM, fs:
FSXR, and h: Hypersoft] and {\it s} are those of \citet{szos08a} [qi: Quiescent, mf: Minor flaring, su: Suppressed, qu:
quenched, mrf: Major flaring, and pf: Post flaring]. }
\tablenotetext{a}{This is calculated by taking the total number of 1-8 keV events in the observation and dividing it by the exposure.}
\tablenotetext{b}{The observation end time of obsid 101 is identical to the observation start time of 1456 (obi 0).  For the purpose of
the analysis in this paper these two obsids will taken to be a single observation.}
\tablenotetext{c}{Obsids 425 and 426 were done in alternating exposure mode.  The shorter frame time observation (0.3 sec) resulted in an
additional 0.4 ksec (425) and 0.3 ksec (426) exposure.  For this analysis only the observations with the longer frame time (1.8 sec) were used.}
\tablenotetext{d}{Obsid 16622 was done as a calibration observation. An offset was used such that only the HEG negative and MEG positive orders 
were on the array.}
\end{deluxetable}

%\subsection{Multiwavelength data}

\begin{figure*}[p] 
\includegraphics*[angle=90, scale=0.5]{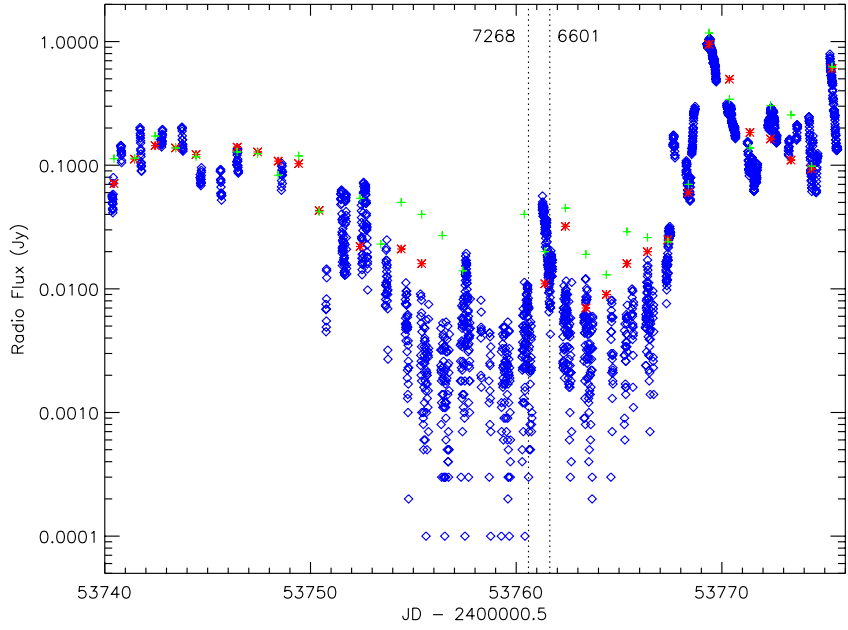}% Here is how to import EPS art 
\caption{\label{figradio} The 15 GHz (blue diamonds) radio flux densities of Cyg X-3 measured by Ryle telescope (currently the AMI-LA) with the RATAN-600 4.8 GHz (red stars) and 11.2 (green +) daily measurements overlayed. The start times of the Chandra obsids 7268 and 6601 are marked as vertical dotted lines.}
\end{figure*} 

\begin{figure*}[p] 
\includegraphics*[angle=90, scale=0.6]{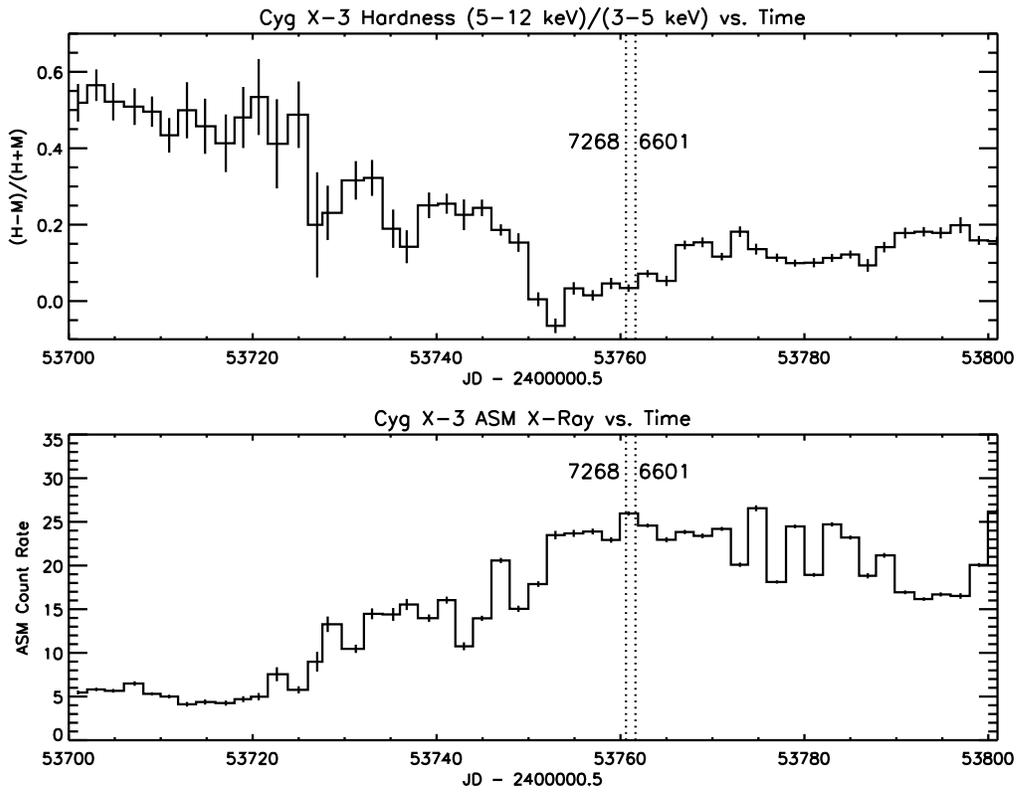}% Here is how to import EPS art 
\caption{\label{figasm} RXTE/ASM lightcurve and hardness ratio showing the long-term soft X-ray (1-10 keV) variability and spectral changes of Cyg X-3 surrounding the Chandra observations 7268 and 6601 (marked as vertical dotted lines). }
\end{figure*} 

\begin{figure*}[p] 
\includegraphics*[angle=90, scale=0.6]{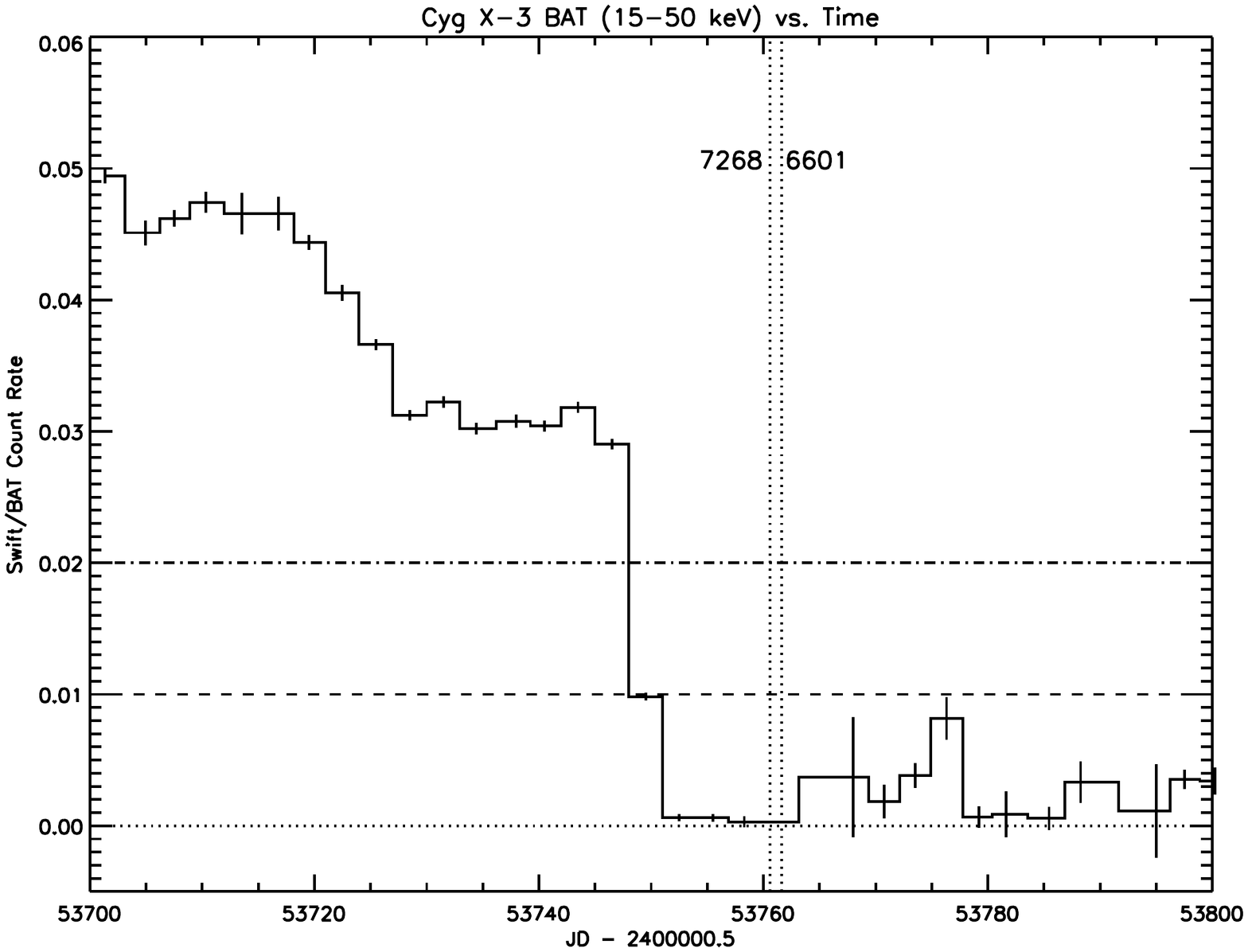}% Here is how to import EPS art 
\caption{\label{figbat} Swift/BAT lightcurve showing the long-term hard X-ray (15-50 keV) variability of Cyg X-3 surrounding the Chandra observations 7268 and 6601 (marked as vertical dotted lines).}
\end{figure*} 

For the Chandra HETG obsids 7268 and 6601 there was 
multi-wavelength coverage in the radio, and also from the RXTE and Swift/BAT 
detectors. Figure \ref{figradio}  shows the Ryle (15 GHz) and RATAN-600 (4.8, 11.2 GHz) observations vs. time.
  Cyg X-3 is clearly in a quenched radio state.  
%The expanded one
%shows that during the observations there may have been a very minor radio
%flare with no obvious HXR response.
The dotted lines are the start
times of the two Chandra grating observations.
Figure \ref{figasm} shows
the RXTE ASM count rate and hardness ratio vs. time:  This is two day averages of
the ASM count rate and the hardness ratio of the 5-12 kev / 3-5 keV bands.  
This shows Cyg X-3 going into a soft state just prior to the observations.
Figure \ref{figbat} shows the Swift/BAT vs. time.  This shows the drop of the hard X-rays (HXR) just 
prior to the Chandra observations and that Cyg X-3 was in a quenched state.
All of these observations show that during the Chandra observations Cyg X-3
was in a quenched/hypersoft state.  It is during this state that gamma-ray and major radio flares can occur.  During the time of these observations no gamma-ray observations were available and a 1 Jy radio flare occurred several days later (see Fig. \ref{figradio}).

%\subsection{Chandra HETG Light Curve}

\begin{figure*}[p] 
\includegraphics*[angle=0, scale=0.6]{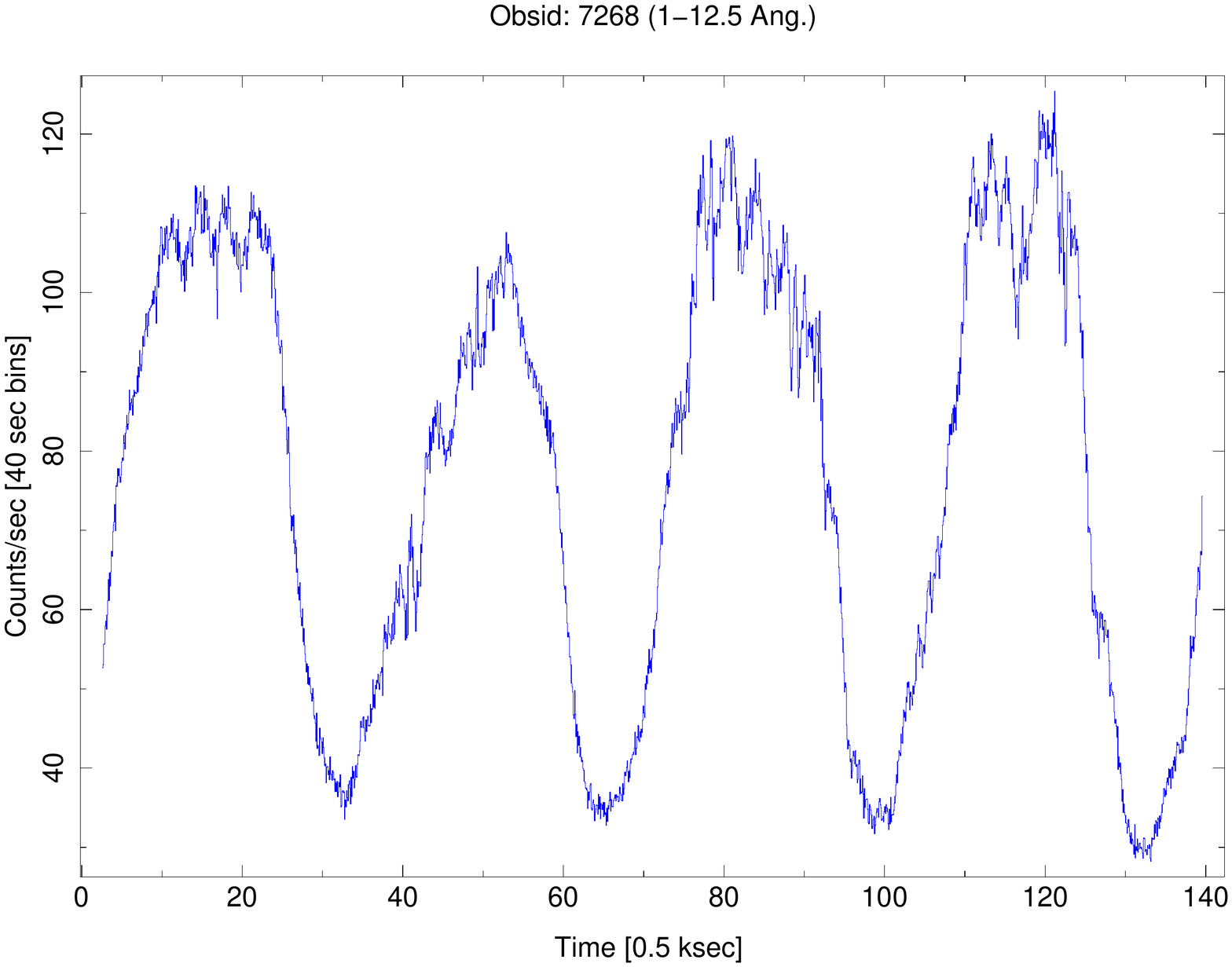}% Here is how to import EPS art 
\caption{\label{figlightcurve} Lightcurve taken from sum over orders 1 and 3, plus and minus, in the energy band 
between 1 - 12.5 \AA (12.5 -- 1 keV)  during Obsid 7268.  This is the only use of third order data in 
this paper.}  
\end{figure*}

Figure \ref{figlightcurve}  shows the lightcurve taken from the dispersed images during Obsid 7268.
This clearly shows features familiar from previous observations:  The period is  17253.3 s \cite{vilh09}; the 
minimum flux is not zero, and the lightcurve has a quasi-sinusoidal shape.  The eclipse transitions are asymmetric:
the egress from eclipse is more gradual than the ingress.  The observation spanned almost 4 orbits of the 
system, and it began just after the minimum.  The flux during the minimum is about 30$\%$ of the flux at maximum, similar 
to the modulation seen by the RXTE ASM \citep{zdzi12}.

\subsection{RXTE Data}

We searched for simultaneous pointed RXTE data during the Chandra pointings from the High Energy Astrophysics Science Archive Research Center (HEASARC). This resulted in eight pointings (91090-04-01-00, -04-02-00, -04-03-00, -05-01-00, -05-02-00, -05-03-00, -05-04-00, -05-05-00) with exposures ranging from 3 ksec to 10 ksec. The first pointing started on MJD 53760.4; about 5 hours before the start of the Chandra Obsid 7268, and the last pointing started on MJD 53762.1. We reduced each pointing using HEASOFT 6.22 and standard methods described in the RXTE cookbook. Both the Proportional Counter Array (PCA) and the High Energy X-ray Timing Experiment (HEXTE) data were reduced and spectra were obtained from PCU-2 (all layers) and cluster B, respectively. 
For spectral analysis, we group the data to a minimum of 5.5 sigma and 2 sigma significance per bin, and ignore bins below 3 keV and above 30 keV, and below 18 keV for PCA and HEXTE, respectively. In addition, 0.5$\%$ and 1$\%$ systematic error were added to all channels for PCA and HEXTE, respectively.

\begin{figure*}[p] 
\includegraphics*[angle=0, scale=0.6]{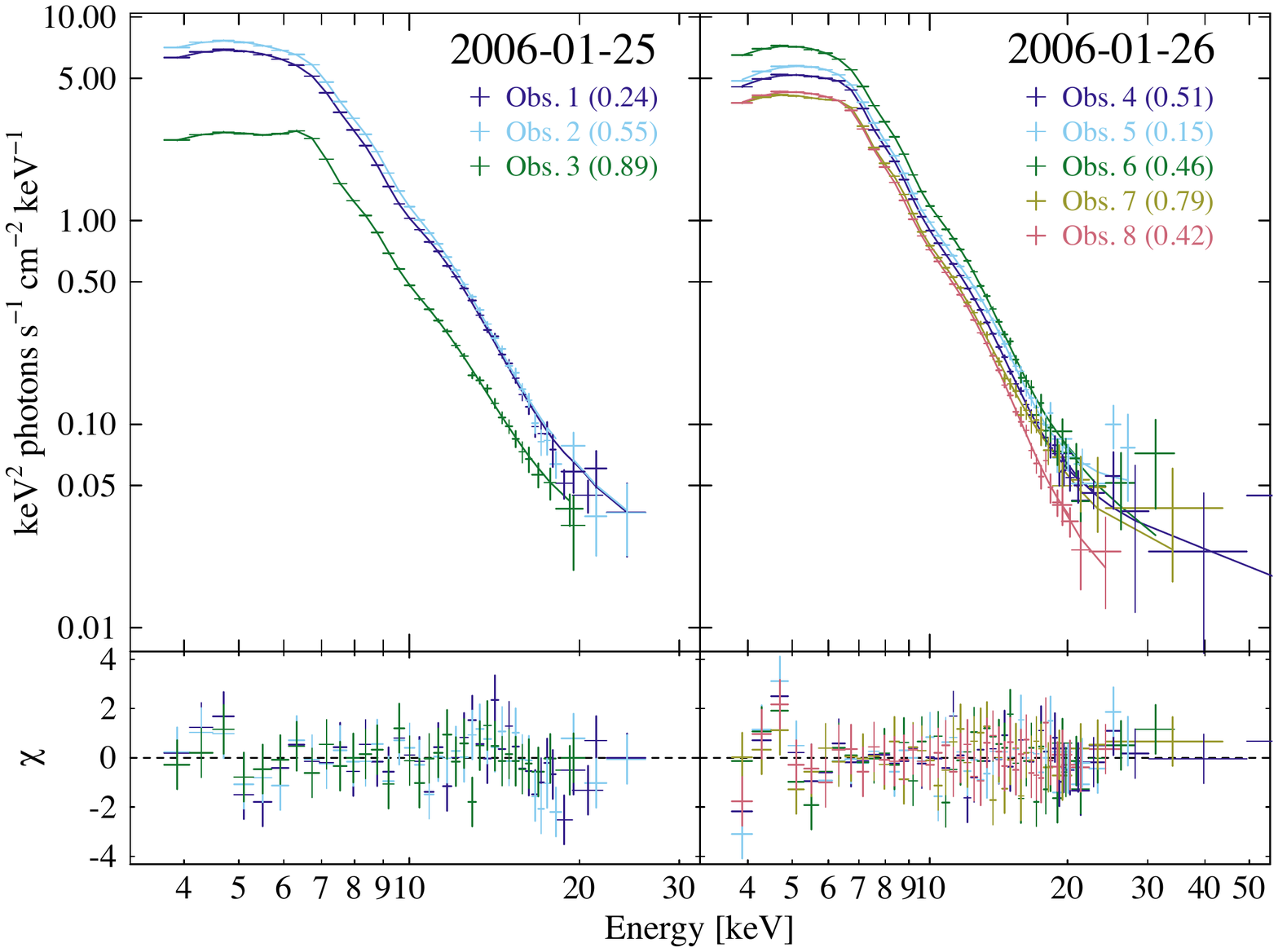}% Here is how to import EPS art 
\caption{\label{figrxtespect} Spectra taken with RXTE pointed observations.  In the figure legend, the X-ray orbital phase is shown in parentheses}
\end{figure*}

We fit the joint PCA+HEXTE spectra with a simple model including interstellar absorption (phabs), absorption edges from highly-ionized iron (edge or smedge), gaussian from iron K-line (gauss), and blackbody (bbody) convolved with a Comptonization component (simpl). The full model can be described as: phabs*edge(1)*smedge(2)*edge(3)*simpl(bbody+egauss), and it has been successfully used to fit the broad-band X-ray spectra from the hypersoft state previously \citep{kolj18}. Since the PCA is not sensitive to energies where the interstellar attenuation mostly affects the spectrum, we fix the column density to the interstellar value in the direction of Cyg X-3: 1.5 $\times 10^{22}$ cm$^{-2}$. The unabsorbed X-ray luminosity (2-100 keV) of the models range from 0.5 --- 1.2 $\times 10^{38}$ erg s$^{-1}$, the blackbody temperature is constant at 1.2 keV and the Comptonized fraction ranges from 1$\%$ to 5$\%$. Sample spectra are shown in figure \ref{figrxtespect}.

%Figure \ref{figrxtespect} shows observations including fluxes (absorbed and
%unabsorbed in units of erg/cm$^2$/s) and some model parameters (blackbody 
%luminosity
%in erg/s assuming 7.4 kpc distance, blackbody temperature in keV, and
%comptonization fraction). These are essentially hypersoft spectra. 
% The model is 
%the same as used in Koljonen et al. 2018 with very similar parameters:
%phabs*edge(1)*smedge(2)*edge(3)*simpl(bbody+egauss). For nH we use a
%fixed value of 1.5e22, which is a lower limit, thus the unabsorbed
%flux is also a lower limit.  the plot does not correspond to this description
%

\subsection{Dust Scattering Halo \label{dustsection}}

A potentially important consideration is the influence of a dust scattering halo on the spectrum.  
This is because of the relatively large optical extinction toward Cyg X-3, corresponding to $A_V\geq$10.
This results in a significant optical depth to dust scattering; $\tau_{scatt}\simeq$2 at 1 keV.
Dust scattering of X-rays is strongly peaked in the forward direction and will produce what appears to be 
diffuse emission which is extended on angular scales comparable to or greater than the $Chandra$ ACIS chip size, 
\citep{corr15, ling09, pred00}.
The total flux in this halo will be comparable to or greater than the source flux below
$\simeq$1.5 keV.  These effects have been studied in the context of Cyg X-3 by \citet{corr15}.  They showed that 
the contribution of a dust scattering halo to the flux detected in an annulus from 6 -- 100 arcsec away from the central 
image can be $\geq$ 30$\%$ below 2 keV, decreasing to $\leq$ 10$\%$ above 3 keV. 

The ACIS-S chips have a size 8.3 arcminutes perpendicular to the grating dispersion direction.  This is sufficient to 
detect a large fraction of the dust-scattered halo.  We have simulated the dust scattered halo from Cyg X-3 using the 
techniques described in \citet{corr15}.  This shows that a circular region with radius 4.1 arcminutes 
will include approximately 55$\%$ of the scattered halo at 2 keV, and 98$\%$ of the halo at 7 keV.  
Images which use CC mode result in the inclusion of all photons
hitting the chips as part of the spectrum.  Thus, such images will necessarily include the scattering halo as part of 
the signal spectrum.  As a result, the net effect of dust scattering on the spectrum will be greatly reduced; 
the photons removed from 
the direct beam by dust scattering will be emitted in the halo, and both will be included in the total spectrum when CC mode is used.  
Images taken in TE mode use a smaller extraction region, and so will include a smaller fraction of the dust-scattered photons.
The energy scale of the HETG in CC mode is based on the position of the detected photons in the direction parallel to the 
dispersion direction.  Thus, all photons which scatter from dust and are reemitted at the same position along the dispersion
as they were absorbed will be correctly assigned energies.  Dust scattering will have little effect 
on the flux or energies of these photons (though see the discussion later in this section regarding the Si K edge).  
Photons which scatter from dust and are reemitted at positions along the dispersion
which differ from where they were absorbed will be assigned energies which are incorrect; some will be eliminated 
by the ACIS energy cuts used for order separation.  Nonetheless, approximately half the dust scattered photons will 
be redistributed by this mechanism into incorrect energies.  In what follows we will not attempt to treat this process 
quantitatively.  We emphasize, however, that this affects primarily photons below 1.5 keV, where there are relatively 
few direct photons, and we do not discuss the analysis of this part of the spectrum in detail.

Photons emitted 
in the halo will have a longer path to the observatory and so will delayed compared with the direct photons.  The delay time 
is approximately $\sim D \alpha^2/2c$, for scatterers located halfway between the source and the observer,
 where $D$ is the distance and $\alpha$  is the angle of the halo photons relative to the observer's line of 
sight, giving delay time $\sim$1 day for $\alpha$=1 arcminute.  Thus, with CC mode data, timing information of the scattered 
light is smeared over several orbital periods.  This presents an ultimate limitation on the ability to study time 
variability from Cyg X-3.  But this is most important for photons below $\sim$2 keV; less than 10$\%$ of the photons emitted 
above 2 keV will be affected by this delay.

In the case of TE mode data, in which the spectrum is accumulated from a region adjacent to the dispersed spectral image, 
much of the light in the dust scattering halo will fall outside the spectral extraction region.  It will therefore not be 
included in the spectrum to be analyzed and dust scattering will result in a net loss of photons, though again primarily 
at energies less than $\sim$2 keV.  If a background spectrum is extracted from the region of the chip outside the standard
source extraction region, and if this is treated as background during analysis of the source spectrum, then there will be an 
additional loss of flux which ought to be included in the source spectrum.  In what follows in our analysis of spectra 
obtained in TE mode we do not perform any background subtraction.  However, our analysis of TE mode data is still affected 
by the loss of photons due to scattering.  Furthermore, dust scattered emission will redistribute photons in the direction 
parallel to the grating dispersion, and hence likely change their energy assignment.  The effects of this process can only be 
treated in detail with a realistic model for the combined effects of dust scattering and the spectrum extraction.  This is beyond
the scope of this paper.  The effects of this process, combined with the effects of detector pileup, can be crudely 
evaluated by comparing the spectrum taken in CC mode (obsid 7268) with the spectrum taken at a similar intensity state using 
TE mode (obsid 6601).

\subsection{Orbital Phase Resolved Spectra}

We have also extracted the spectral data from both Obsids  7268  and 6601 folded on the X-ray orbital phase.  The full observation of Obsid 7268 was
69860 seconds and the total number of counts is 1.06159 $\times 10^6$ and so covers approximately 4 binary orbits.   
We reextract the data and bin into 4 phase bins according to the orbital period of 17252.6  s \citep{vilh09}.
We choose phase bins based on the offset from the start of the obsid; this corresponds  almost to the minimum 
in the X-ray light curve, so that we can identify our phases approximately as 0.875-0.125, 0.125-0.375, 0.375-0.625, and 0.625-0.875
We denote these in what follows as phase bins 1,2,3,4.  We note that the orbital period of Cyg X-3 is increasing at a rate 
of $\dot P/P \sim 10^{-6}$ yr$^{-1}$ \citep{sing02,bhar17}, and that the minimum of the X-ray flux occurs slightly before the 
phase denoted as superior conjunction, at phase 0.97 \citep{vand89} due to asymmetry of the light curve.
%The net counts in each of the phase bins is not sufficient to tightly constrain the parameters of spectral fits
%of the type described in the previous sections.  However, the statistic are sufficient to constrain the 
%parameters of individual lines:  centroids, widths and intensities.

\begin{figure*}[p] 
\includegraphics*[angle=90, scale=0.6]{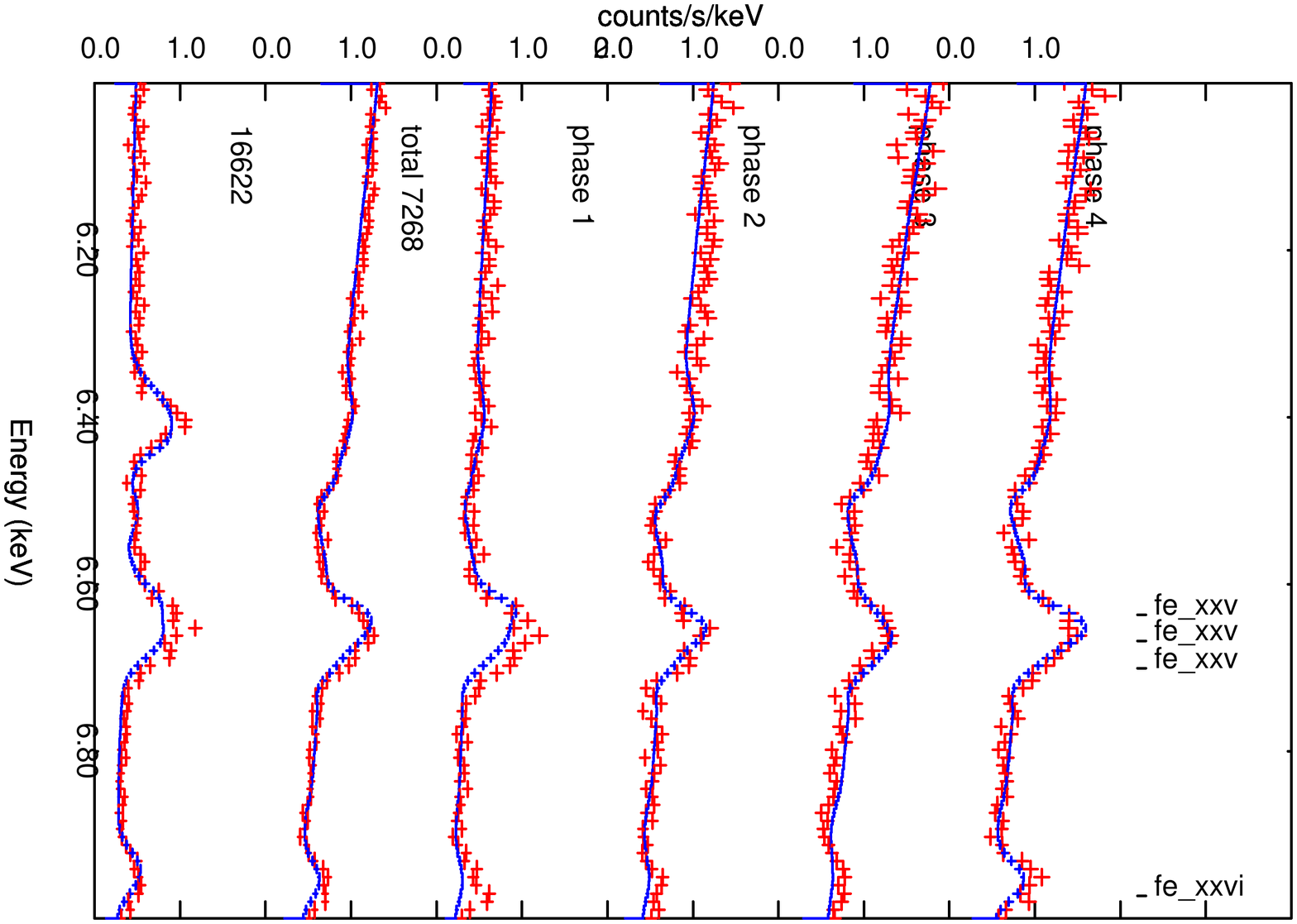}% Here is how to import EPS art 
\caption{\label{fig1} Spectrum observed by the HETG plotted vs. energy in keV
in the 6 -- 7 kev energy band.
Model shown in blue is described in the text.}
\end{figure*}

\begin{figure*}[p] 
\includegraphics*[angle=90, scale=0.6]{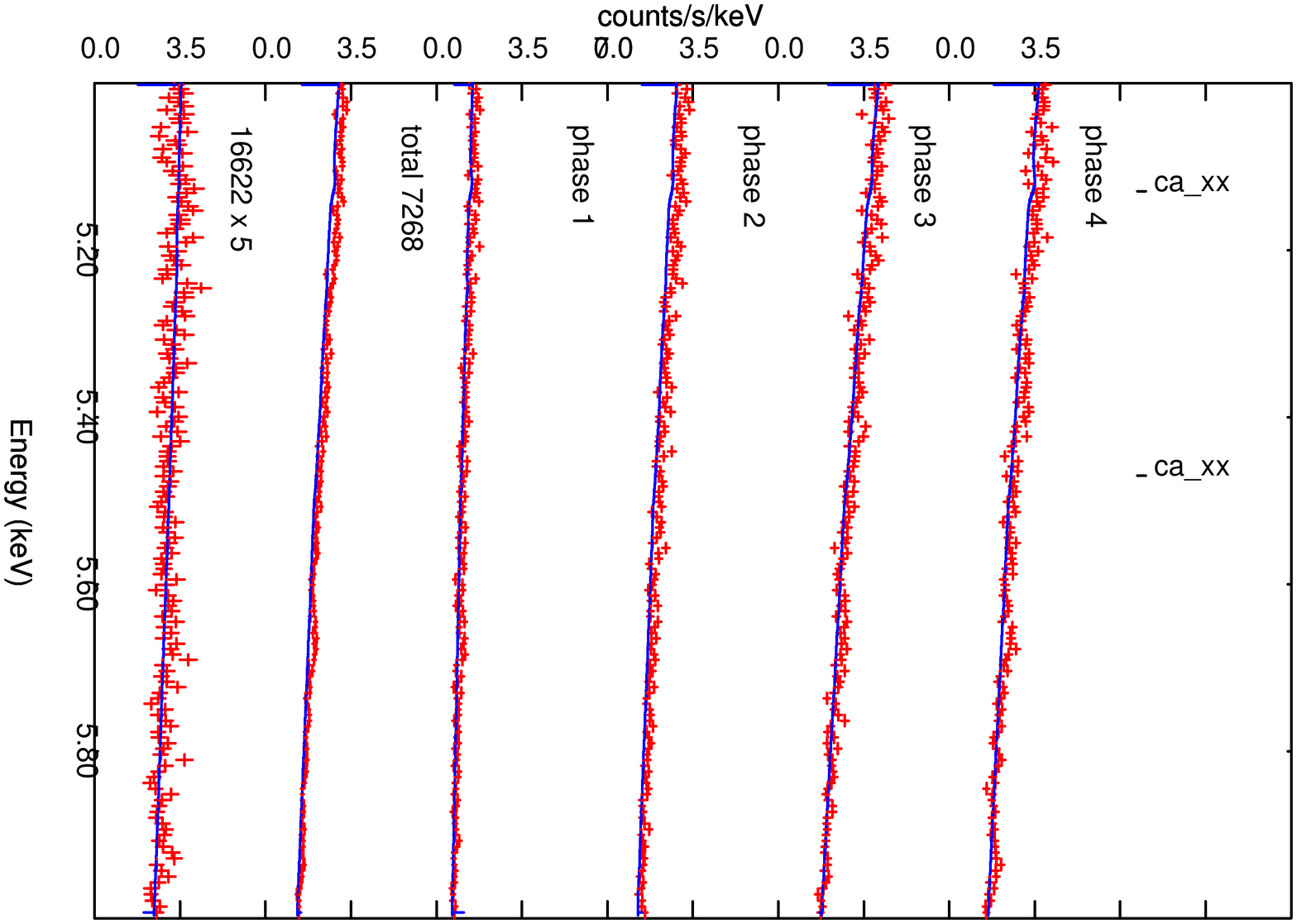}% Here is how to import EPS art 
\caption{\label{fig2} Spectrum observed by the HETG plotted vs. energy in keV
in the 5 -- 6 kev energy band.}
\end{figure*}

\begin{figure*}[p] 
\includegraphics*[angle=90, scale=0.6]{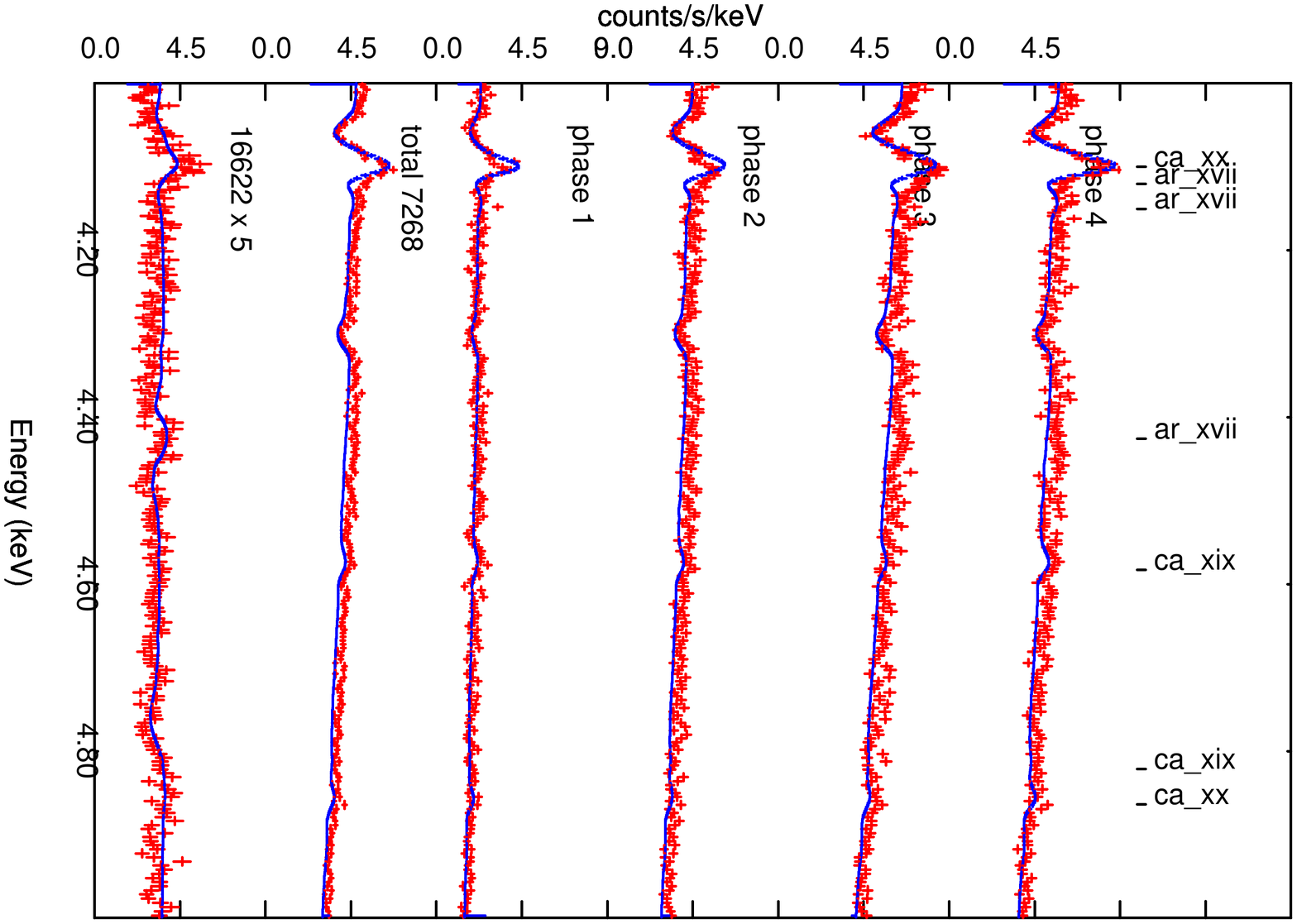}% Here is how to import EPS art 
\caption{\label{fig3} Spectrum observed by the HETG plotted vs. energy in keV.
in the 4 -- 5 kev energy band.}
\end{figure*}

\begin{figure*}[p] 
\includegraphics*[angle=90, scale=0.6]{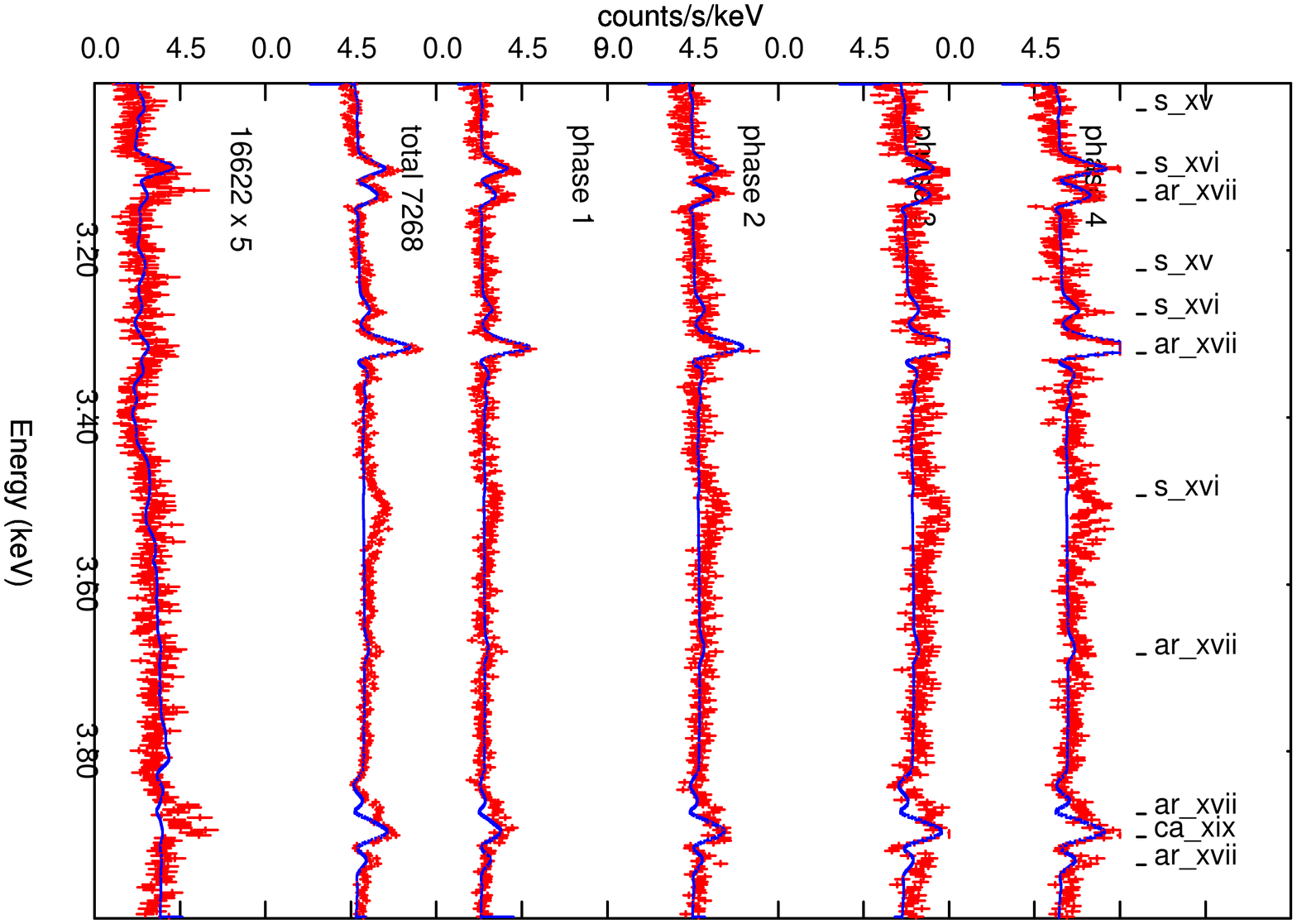}% Here is how to import EPS art 
\caption{\label{fig4} Spectrum observed by the HETG plotted vs. energy in keV 
in the 3 -- 4 keV energy band.} 
\end{figure*}

\begin{figure*}[p] 
\includegraphics*[angle=90, scale=0.6]{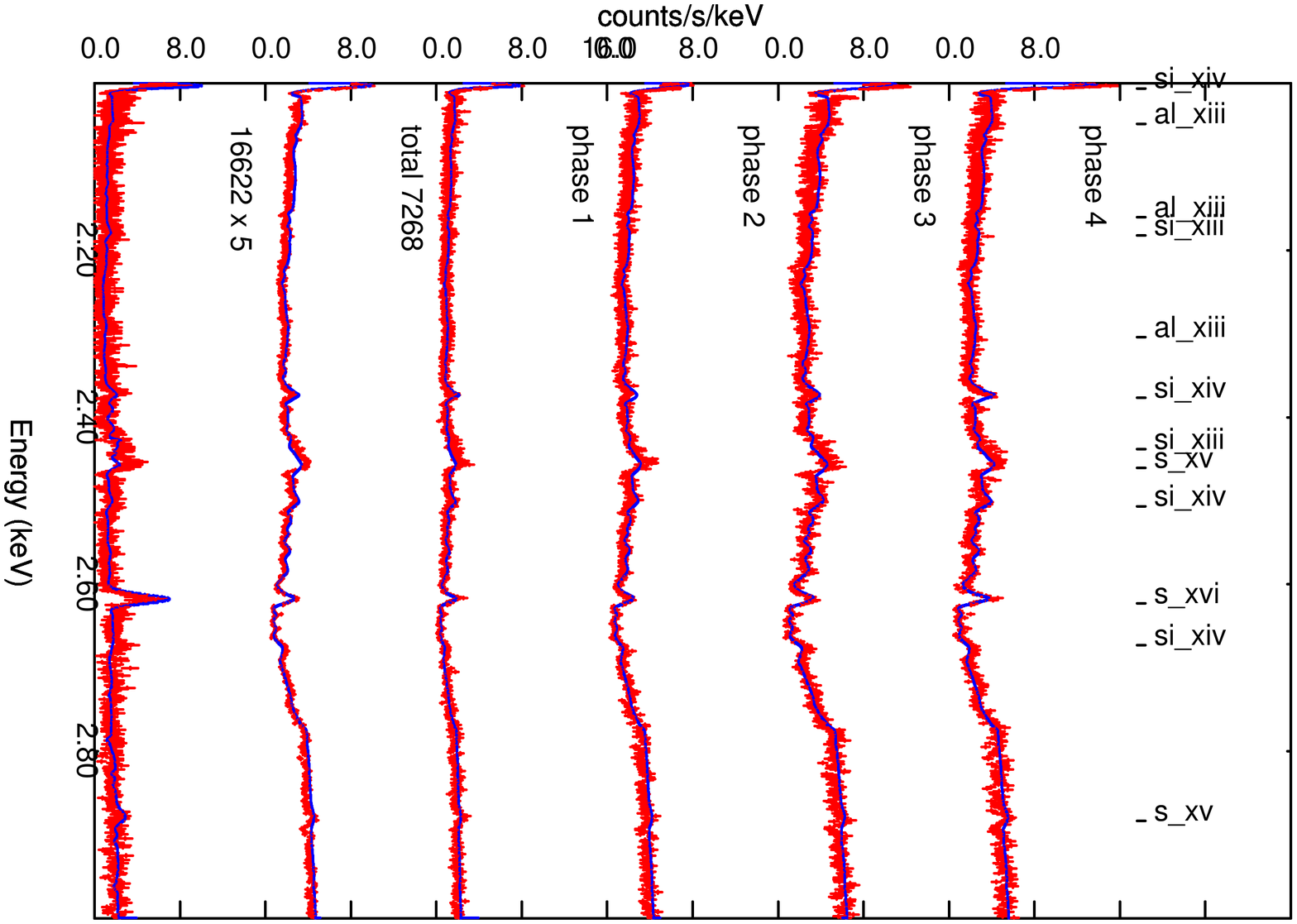}% Here is how to import EPS art 
\caption{\label{fig5} Spectrum observed by the HETG plotted vs. energy in keV
in the 2 -- 3 kev energy band.}
\end{figure*}

\begin{figure*}[p] 
\includegraphics*[angle=90, scale=0.6]{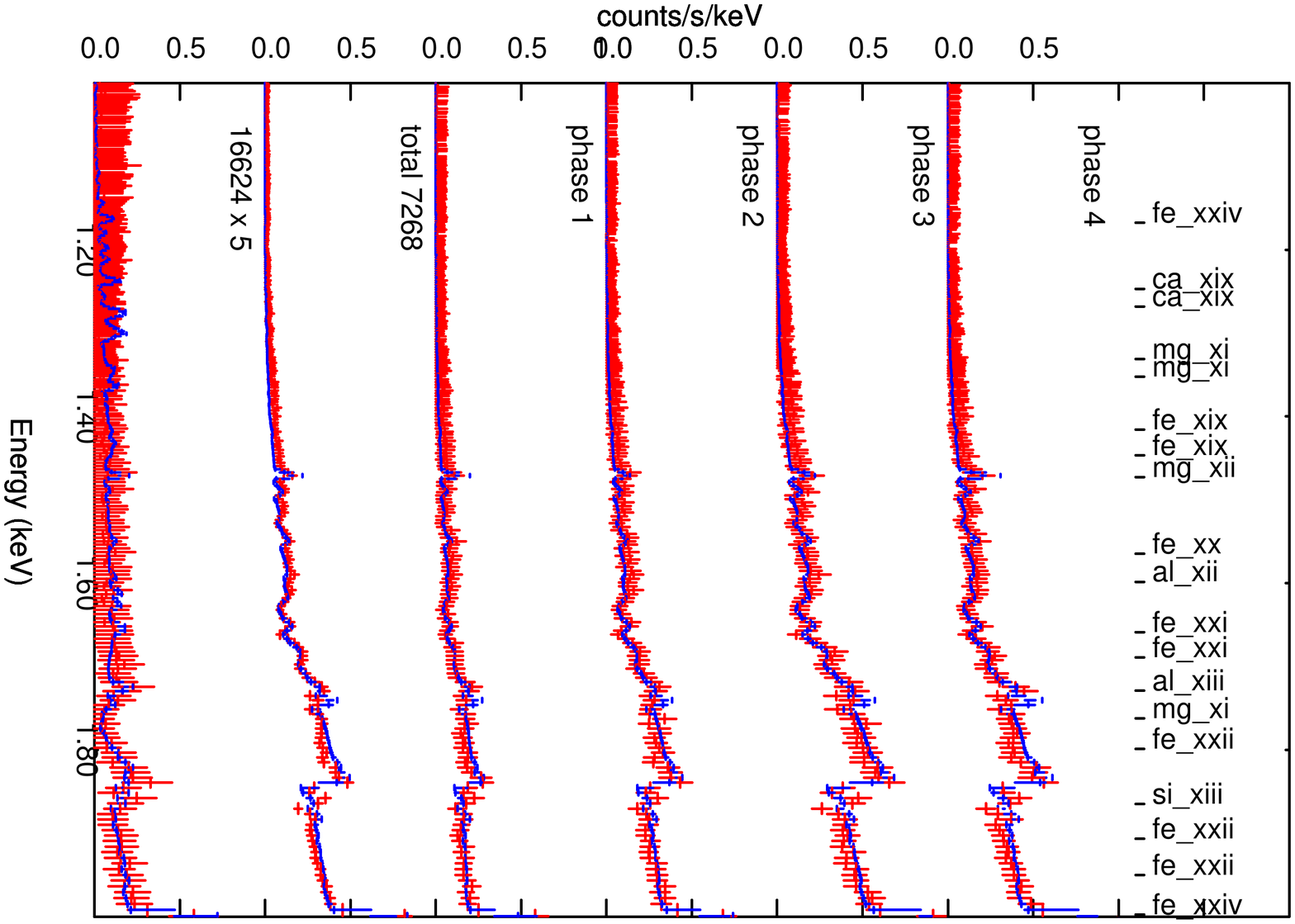}% Here is how to import EPS art 
\caption{\label{fig6} Spectrum observed by the HETG plotted vs. energy in keV
in the 1 -- 2 kev energy band.
Model is shown in blue and is described in the text.  Data has been binned by a 
factor of 8 for this figure.}
\end{figure*}

Spectra in each of the phase bins, and the total spectrum for Obsid 7268 (designated T) and the total spectrum 
for Obsid 16622 (discussed below in subsection \ref{16622}) are displayed in figures \ref{fig1} -- \ref{fig6}.
In what follows we first discuss the phase averaged spectrum, and our model fits, and then discuss the 
variation with orbital phase.

\section{Results}

\subsection{Gaussian Fitting}

As a first step toward understanding the spectrum, 
we fit Gaussian profiles for the strongest emission lines and recombination edge model for 
radiative recombination continua (RRCs) in the total spectrum from 
Obsid 7268.  The list of lines to use as trials for the Gaussians is obtained as follows:  
we make a model fit to the spectrum using the  {\sc xstar warmabs/photemis} \citep{kall01}
analytic model in {\sc xspec}.  The best fit spectrum yields a list of 194 strongest features.  
These come from the {\sc xstar} database, and are level-resolved.  
We then use the energy/wavelengths of these 
as trials for Gaussians, and test each one in turn proceeding in order of decreasing energy or increasing wavelength.
These lines then also serve as the list of lines from which to derive the 
line identifications after the fits are performed.  RRCs are fitted to a simple 
exponential shape similar to the `redge' model in {\sc xspec}.
The centroid energies, widths, fluxes, and IDs of the features are listed in table \ref{table1} for 
lines and table \ref{table2} for RRCs.   The widths of the features, both lines
and RRCs, are given in units of km s$^{-1}$.  

The fitting procedure is designed to allow for the fact that many of the features in the spectrum are broad and
represent blends of separate line components.  In addition, as we will demonstrate below, there are net offsets in 
the line peaks or centroids relative to the laboratory wavelength, and these are likely related to Doppler shifts 
associated with gas motions.  Therefore, our procedure is, first, to use the laboratory wavelengths and assume the lines 
are narrow ($\sigma$=200 km s$^{-1}$) in order to test for the presence of a given line.  We do this for all the 
lines in our trial list, and test whether each line improves the fit parameter.
Then we allow the centroid, width and flux to vary to find the errors on each quantity.
We adopt the $\chi^2$ statistic for this 
automated Gaussian fitting procedure (all other spectral fits in this section use the C-statistic); 
typical channels in the HETG have $\geq$ 100 counts.  We do not allow the line centroid to differ by more than 
1000 km s$^{-1}$ from the lab (cf. NIST) energy/wavelength, and do not include features for which the best fit 
width is greater than 10$^4$ km s$^{-1}$.  We consider a feature detected only if it improves the fit by 
$\Delta\chi^2\geq$10 \citep{avni76}.  In this way, we try to account for the fact that a given feature may 
be a blend of separate lines.  

Our Gaussian fitting procedure is imperfect, as will be shown below, owing to the fact that the lines 
are broadened and shifted, and so a given feature in the spectrum may have contributions from various lines.  Our 
level-resolved line list therefore can provide more than one identification for a given spectral feature, using the procedure
defined above.  And our procedure does not attempt to fit multiple lines to a given feature; if an identification 
for a feature is found, its region of the spectrum is excluded from further fitting attempts.  Thus, there is a bias
toward accepting the first plausible identification for a given feature.  

Here and in what follows we adopt a simple choice for the continuum in the HETG band:  a disk blackbody spectrum as 
implemented in the {\sc diskbb} model in {\sc xspec} \citep{mits84}, with a characteristic temperature $kT=1.72$ keV
and a normalization which is allowed to vary in order to achieve a good fit.
In addition, we find that an additional harder component is needed to fit to the spectrum in the low state.  For this 
we adopt a 2 keV blackbody.  We have also experimented with other shapes for these continuum components:  for the 
soft state, a 1.2 keV blackbody plus a power law with index $\gamma$=2.5, \citep{kolj18} 
and for the hard state a comptonization model \citep{hjal08,szos08,hjal09,zdzi10}.  These models differ from our choice primarily in the flux in the 
extremes of the energy band away from the peak of the thermal component, i.e. below 2 keV and above $\sim$5 keV.  The
1.2 keV blackbody plus power law, compared with the disk blackbody, 
when fitted to the soft state spectrum Obsid 7268 results in a lower cold column NH and 
also lower normalization for the emission measure of the high ionization component responsible for the iron K lines.
These continuum choices do not result in significantly different values for the goodness of fit criteria 
C-statistic or $\chi^2$, which we attribute to the relatively narrow bandpass of the $Chandra$ HETG plus the predominance 
of the thermal component within this spectral range.  The 2-10 keV flux also is affected by the continuum choice by 
$\simeq$ 5$\%$.   Such different continuum components do correspond to significantly 
different total fluxes or luminosities for Cyg X-3 when extrapolated to energies outside the HETG bandpass, such as for the 
RXTE PCA.  

\begin{figure*}[p] 
\includegraphics*[angle=0, scale=0.6]{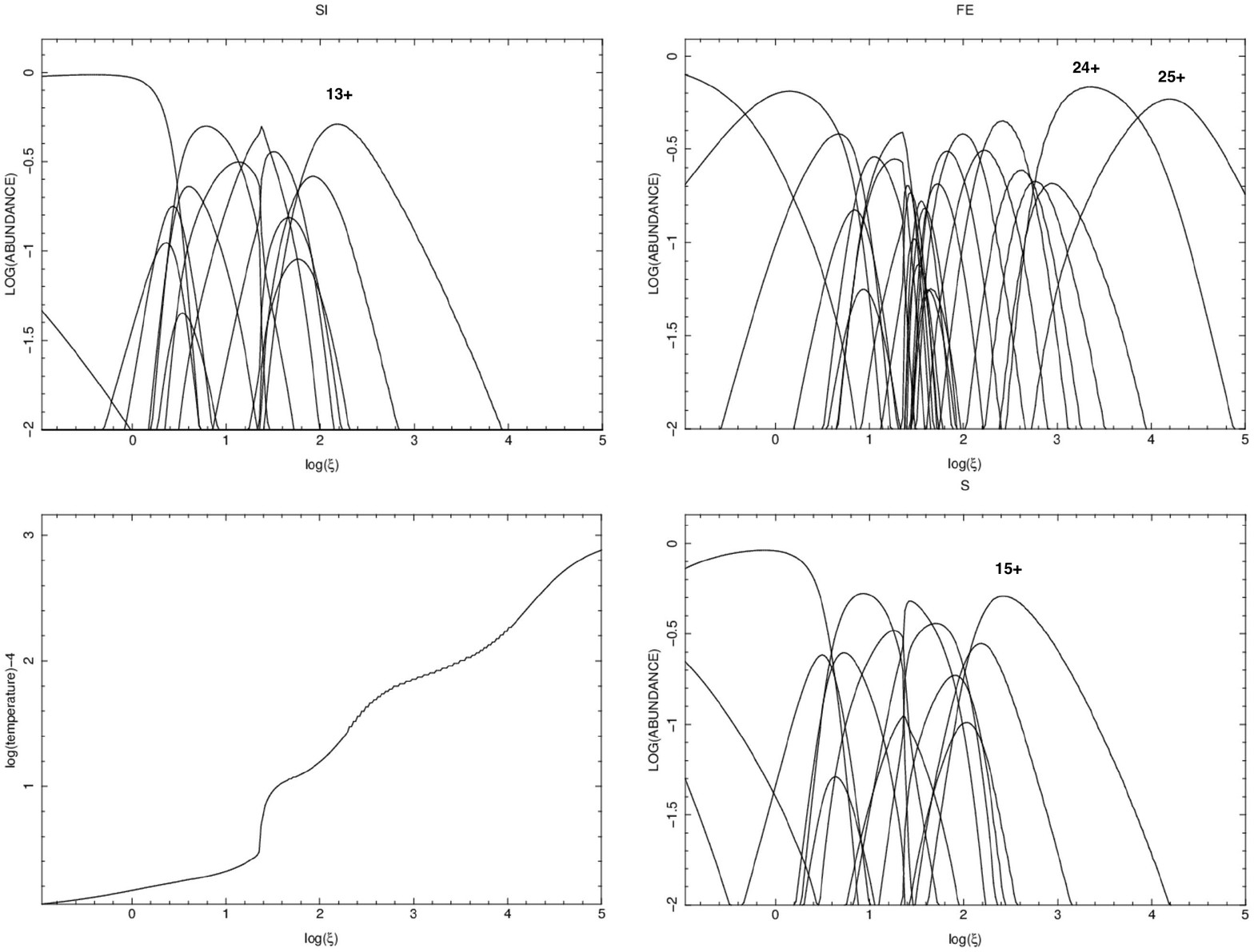}% Here is how to import EPS art 
\caption{\label{figtxi} Ion fractions and temperature for an {\sc xstar}  photoionization equilibrium 
model using a 1.72 keV diskbb illuminating spectrum. H-like curves are labeled for Si and S, and H- and He-like 
curves are labeled for Fe.  This shows that the temperature in the region where 
Si and S are hydrogenic or helium-like is $\geq 10^5$K.}
\end{figure*}

RRCs are indicative of radiative recombination.  The width of the RRC, as measured by the exponential shape above the 
threshold, is a measure of the gas temperature \citep{lied96}.  
We have fitted the strongest and least blended RRCs, i.e. those from hydrogen-like Si and S,
 to an analytic 'redge' model from {\sc xspec}.  This yields, for the 
Si$^{13+}$ RRC at 2.675 keV, a width parameter $T=8.88^{+1.36}_{-2.53} \times 10^4$K, and for the 
S$^{15+}$ RRC at 3.499 keV the width parameter is $T=3.57^{+3.71}_{-1.77} \times 10^4$K.  These are both consistent 
with a temperature $T\simeq 7 - 8 \times 10^4$K.  This can be compared with the results of 
photoionization calculations, shown in figure \ref{figtxi}, which shows that the model temperature in the region where 
Si and S are hydrogenic or helium-like is $\geq 10^5$K.  This suggests additional cooling, such as adiabatic expansion,
 is affecting the gas in Cyg X-3.  The importance of adiabatic expansion cooling depends on the timescale for 
the gas to flow across a region where the pressure changes significantly when compared with the radiative cooling timescale.  As 
shown by \citet{krol81,stev92}  this can be described by the parameter $\theta=t_{cool}/t_{flow}$; the equilibrium temperature is then 
reduced by a factor $1/(1+\theta)$.  Low temperatures inferred from RRC widths have been observed in other sources: $\gamma^2$ Velorum,
an WC + O-star binary \citep{schi04}, and the planetary nebula BD+30 and its WC9 wind \citep{nord09}.  In the latter 
it was speculated that ions are created by shocks and then cross the contact discontinuity and recombine with 
unshocked cold electrons. Such a scenario cannot be ruled out for Cyg X-3, though it is notable that the X-ray continuum is 
adequate to provide the ionization we observe.   Comparing the flux in the L$\alpha$ line to the corresponding RRC shows a ratio 
of $\simeq$13 for Si and $\simeq$11 for S.  This can be compared with a predicted ratio $\sim$1 for pure case A recombination
\citep{oste74}.  Thus it is clear that the L$\alpha$ line emission is affected by other processes.  As described 
below, we consider the most likely such mechanism to be resonance scattering.  

% The spectrum features strong H- and He-like lines from 
%Mg, Al,  Si, S, Ar, Ca, and Fe. 
% For this fit $\chi^2/\nu$=13915/7991=1.74.

\subsection{He-like Lines}

\begin{figure*}[p] 
\includegraphics*[angle=0, scale=0.6]{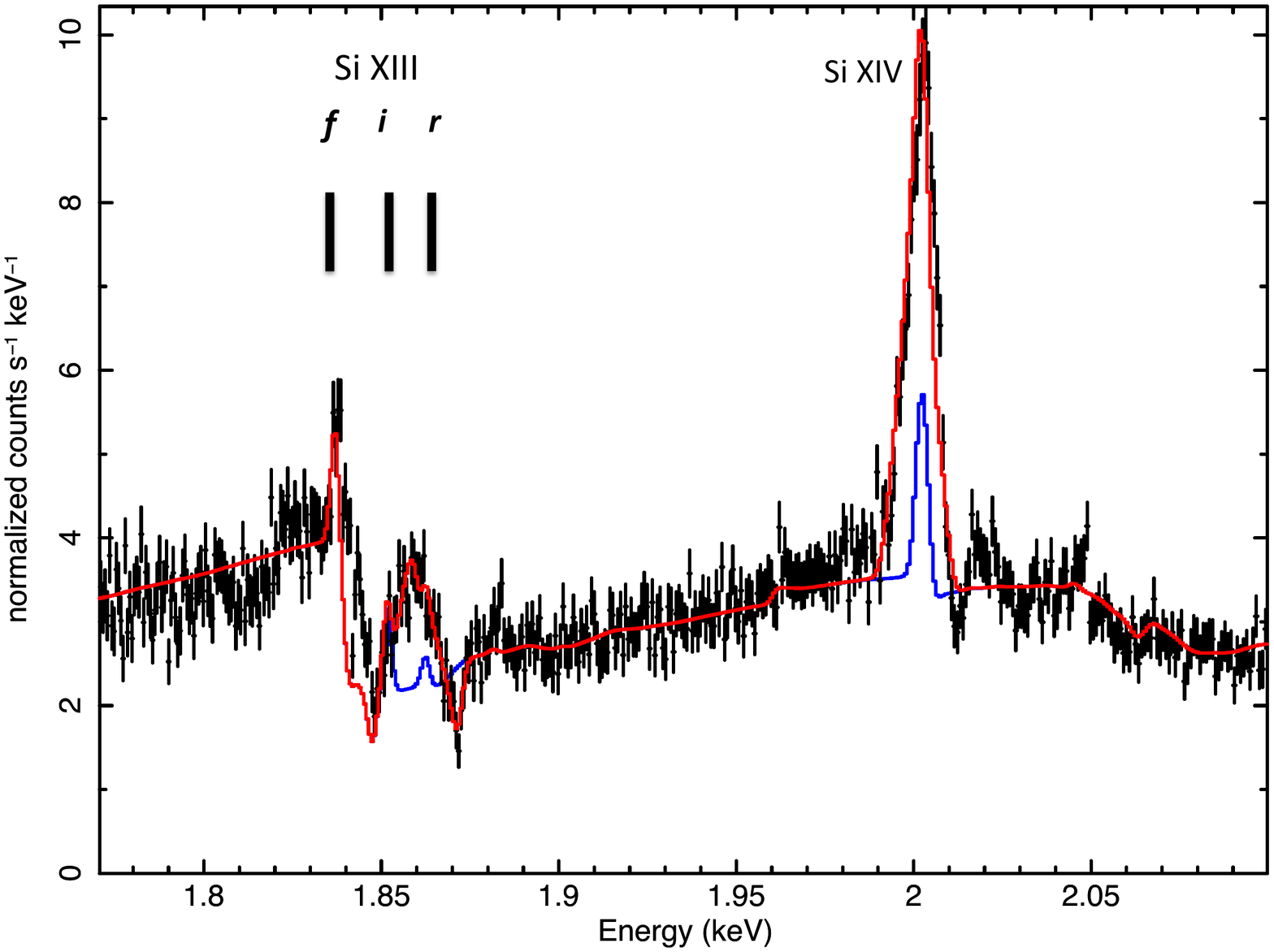}% Here is how to import EPS art 
\caption{\label{figsinarrow} Spectral region containing the He-like Si lines.  The blue curve shows 
a model consisting of a photoionized emission model, recombination dominated.
Positions of the narrow components at the rest wavelengths of the resonance ($r$), 
intercombination ($i$) and forbidden ($f$) lines are indicated schematically 
by black lines.  The blue and red curves coincide in the region of the $f$ line.  
This illustrates the fact that the region between 
the $r$ and $i$ lines is filled in by a redshifted and broadened resonance line component.
The red curve shows the total fit including the wind lines.}
\end{figure*}

Among the strongest lines in the spectrum are those from the He-like ions.
These lines are familiar from the study of other sources and can be 
crudely described as three components: the
resonance ($r$), intercombination ($i$) and forbidden ($f$) lines.  
The He-like lines are detected in the Chandra HETG spectrum of Cyg X-3 from the elements
Mg, Al, Si, S, Ar, Ca and Fe.   Table \ref{table1} gives values for the difference between the measured line peak and the lab value from NIST  \footnote{https://physics.nist.gov/PhysRefData/ASD/lines\_form.html} in units of km s$^{-1}$.  These values are constrained to have values in the range -200 km s$^{-1}$ to  200 km s$^{-1}$ and are given to within 10 km s$^{-1}$.  Examination of table \ref{table1} shows that the 
%energies of these lines agree with the lab energies to within $\simeq$100 km s$^{-1}$ 
%for the $f$ and $i$ lines.  The 
energies of the $r$ lines are systematically redshifted 
relative to the $f$ and $i$ lines by  $\sim$200 km s$^{-1}$. 
The Gaussian fits do not clearly detect distinct $r$, $i$, and $f$ components for any of the 
he-like species. 
Figure \ref{figsinarrow} shows the spectrum in the vicinity of the He-like 
Si lines near 1.8 keV.    In this figure, for the 
purposes of illustration, we force the model lines to be narrow (i.e. width $<$ 10 km s$^{-1}$) 
and at the lab wavelengths for the $r$, $i$ and $f$ components 
(cf. NIST  \footnote{https://physics.nist.gov/PhysRefData/ASD/lines\_form.html}).
The lab energies for these components are 1.8650, 1.8538 and  1.8395 keV, respectively
This clearly shows the three components are present, 
most notably the $f$ line near 1.84 keV which is not blended with any other component.
On the other hand, the feature near 1.86 keV is not centered on the rest wavelength
of any tabulated line.  It is broad, and therefore spans the region containing the $r$ and $i$ line
rest wavelengths.  This, together with the deficit of photons at energies  near 1.87 keV, 
above the $r$ line rest energy, is suggestive of a P-Cygni profile.  Such profiles 
were previously identified by \cite{vilh09}, and are indicative of outflow in a spherical 
wind.  If so, the 1.86 keV line is formed by resonance scattering of the $r$ line; the 
absorption corresponds to material seen in projection in front of the continuum source, 
and the emission corresponds to material seen in reflection.  The net blue shift of the 
absorption and redshift of the emission suggests that the outflow originates at or near the 
continuum source.

Motivated by this, we fit the lines of Si$^{12+}$ with a two component model.  One component 
consists of narrow Gaussians at the rest energies of the $r$, $i$ and $f$ line components.
The intercombination  line is further separated into three lines corresponding 
to the three fine structure levels of the 1s2p($^3$P) term.  Here and in what follows we refer to this 
as the nebular component.  We employ a recombination model for this component, the 
{\sc xstar photemis} model, which is discussed in section \ref{ionizationbalance}.   The second component is a 
P-Cygni profile which we model using the Sobolev Exact Integration (SEI) method 
of \cite{lame87} as implemented in the {\sc windabs} analytic model in {\sc xspec}.
We employ the {\sc photemis} and {\sc windabs} analytic models within {\sc xspec}, both of which are based on the {\sc xstar}
photoionization models.  {\sc photemis} calculates the emission from a gas in photoionization equilibrium with 
specified ionization parameter, element abundances, normalization  and turbulent velocity. {\sc windabs} 
uses the same calculation of ionization balance and temperature, with the important difference  
that it treats lines excited by photoabsorption from the continuum, and also adds the 
effects of the gas motion calculated using the Sobolev Exact Integration method \citep{lame87}.  
It is important to note that {\sc photemis} is an additive model; that is, it adds to the model spectrum linearly
with a given normalization.  {\sc windabs}  is a multiplicative model; it imprints features on another model, in our case
the continuum.  The strength of a multiplicative model is described by a multiplicative factor, in this case 
the column density of the model wind.  

Ion fractions and atomic level populations for both models are calculated using {\sc xstar}
using a specified photoionizing continuum shape, and assuming the gas is optically thin to the ionizing 
radiation.  Details are presented in \cite{kall01} and at the {\sc xstar} webpage 
\footnote{https://heasarc.nasa.gov/lheasoft/xstar/xstar.html}.  
{\sc windabs} is available as part of the {\sc warmabs/photemis} package
\footnote{ftp://legacy.gsfc.nasa.gov/software/plasma\_codes/xstar/warmabs.tar.gz}
and is described in the {\sc xstar} manual 
\footnote{https://heasarc.nasa.gov/lheasoft/xstar/xstar.html}.  
The SEI model is described by five parameters: the wind terminal 
velocity $v_{W}$, 
the line optical depth parameter $\tau_{line}$, the wind velocity law 
parameter $\gamma$, the optical depth velocity dependence parameter $\alpha_1$ and 
the line redshift $z$.  
{\sc windabs} fixes the values of the last two, $\gamma$=1 and  $\alpha_1$=1 and 
calculates $\tau_{line}$ from the specified column density, the elemental abundances
and the ion fraction taken from stored {\sc xstar} results.  
{\sc windabs} also allows for departures from spherical symmetry in an ad hoc way by defining 
a covering fraction, $C$, such that, when 0$\leq C \leq$1, the emission component 
is reduced by a factor of $C$.  When covering fraction $C\geq$1, the absorption 
component is reduced by a factor $2-C$. At $C=2$, the model produces pure 
scattered emission.  This allows for situations in which the wind is viewed in reflection
more than in transmission.  It also can crudely simulate situations in which the 
line thermalization parameter in the wind is not small, though in this case the 
physical meaning of the covering fraction is lost.
The fit to the model consisting of narrow emission, calculated by {\sc photemis} plus wind 
emission calculated by {\sc windabs}, is also shown as the red curve in figure 
\ref{figsinarrow}.  The blue curve in figure \ref{figsinarrow} shows just the {\sc photemis} component.
This illustrates the fact that, for a recombination-dominated gas, the $r$ line is much weaker than the 
$f$ or $i$ lines.

P-Cygni profiles were previously identified and fitted by \cite{vilh09}.  Those authors focused on 
the lines from the H-like ions, notably Si$^{13+}$.   We fit the profile of the Si$^{12+}$ $r$ line 
simultaneously with the Si$^{13+}$ L$\alpha$ line profile.  
{\sc windabs} forces the dynamical quantities 
($v_{wind}$, $z$ and $C$) to be the same for the two lines. 
%Independent values for the $\tau_{line}$ and  $\alpha_1$ parameters
%allows for the effects of differing ion fractions and differing distributions of ion fractions 
%across the wind.  
%Table \ref{si13table} shows the best-fit values and errors 
%(2 $\sigma$) for the parameters used for fitting the lines.

Similar results apply to the other He-like lines detected in the Cyg X-3 spectrum.
These come from elements S, Ar, Ca and Fe (the lines from Mg and Al are too weak, owing to the low energy cutoff,
 to permit such a decomposition).  That is, all require the presence of both a broad wind line 
feature in the $r$ line and contributions to the $r$, $i$ and $f$ components which are unshifted 
and narrow.   
%Lines from Mg are too highly absorbed to be measured accurately.  
Figure \ref{fighewidth} shows the widths of the wind and narrow components from fits to the various 
He-like lines.   There is a weak indication that the wind component becomes narrower for higher Z elements, i.e. 
the favored values for wind speed decrease for Ar and Ca, although the errors are large.
It also becomes weaker, such that it cannot be detected for Fe. 
 In the case of S the errors on the widths of the higher and lower speed 
components overlap.   The narrow 
component width is smaller for Ar; the other elements show values for this quantity which are similar to each other.

\eject 

\begin{figure*}[p] 
\includegraphics*[angle=0, scale=0.6]{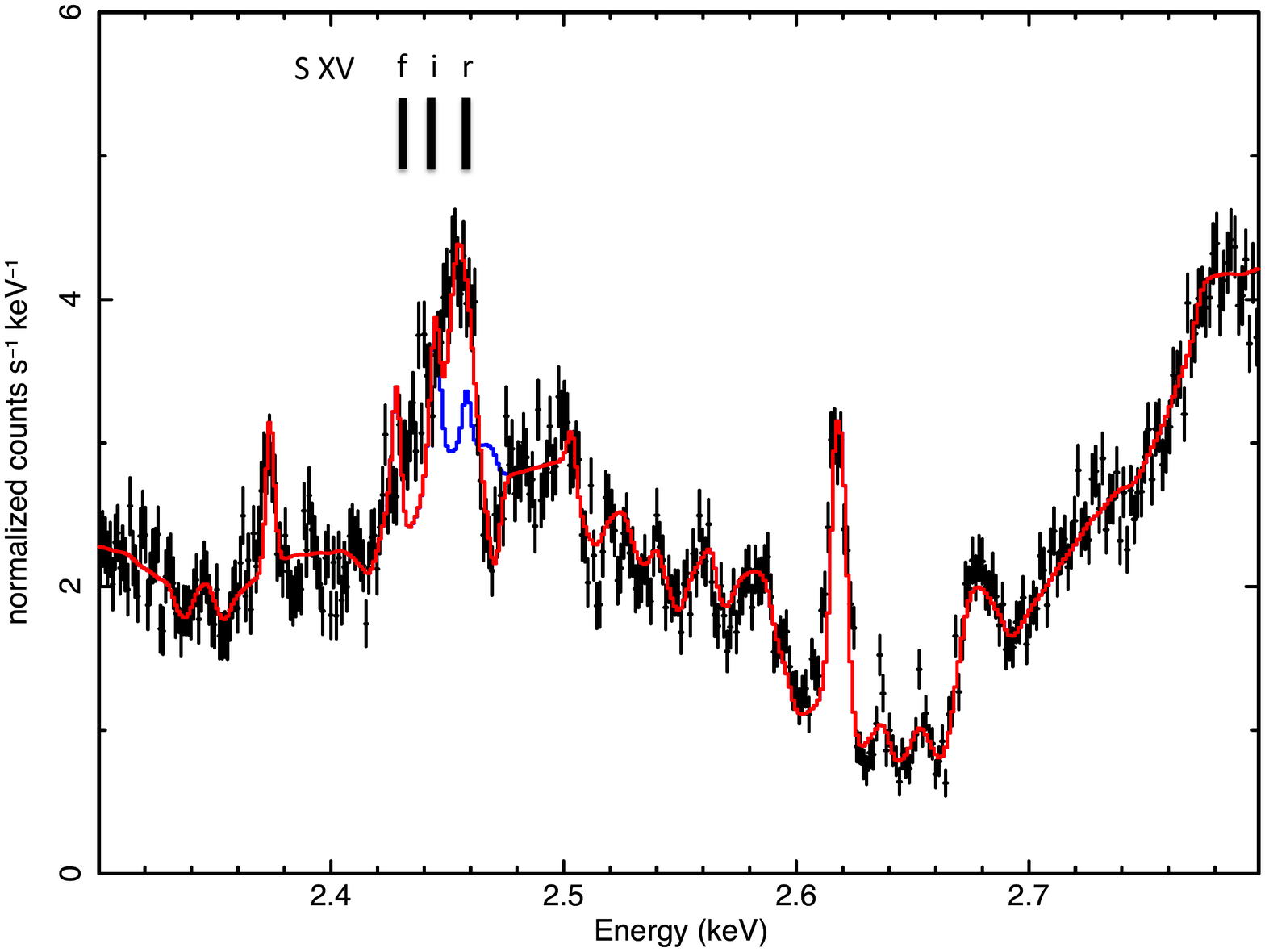}% Here is how to import EPS art 
\caption{\label{figsnarrow} Spectral region containing the He-like S lines.  The blue curve shows 
a model consisting of a photoionized emission model, recombination dominated.
Positions of the narrow components at the rest wavelengths of the resonance ($r$), 
intercombination ($i$) and forbidden ($f$) lines are indicated schematically 
by black lines.  The blue and red curves coincide in the region of the $f$ line.  
This illustrates the fact that the region between 
the $r$ and $i$ lines is filled in by a redshifted and broadened resonance line component.
The red curve shows the total fit including the wind lines.}
\end{figure*}

\begin{figure*}[p] 
\includegraphics*[angle=0, scale=0.6]{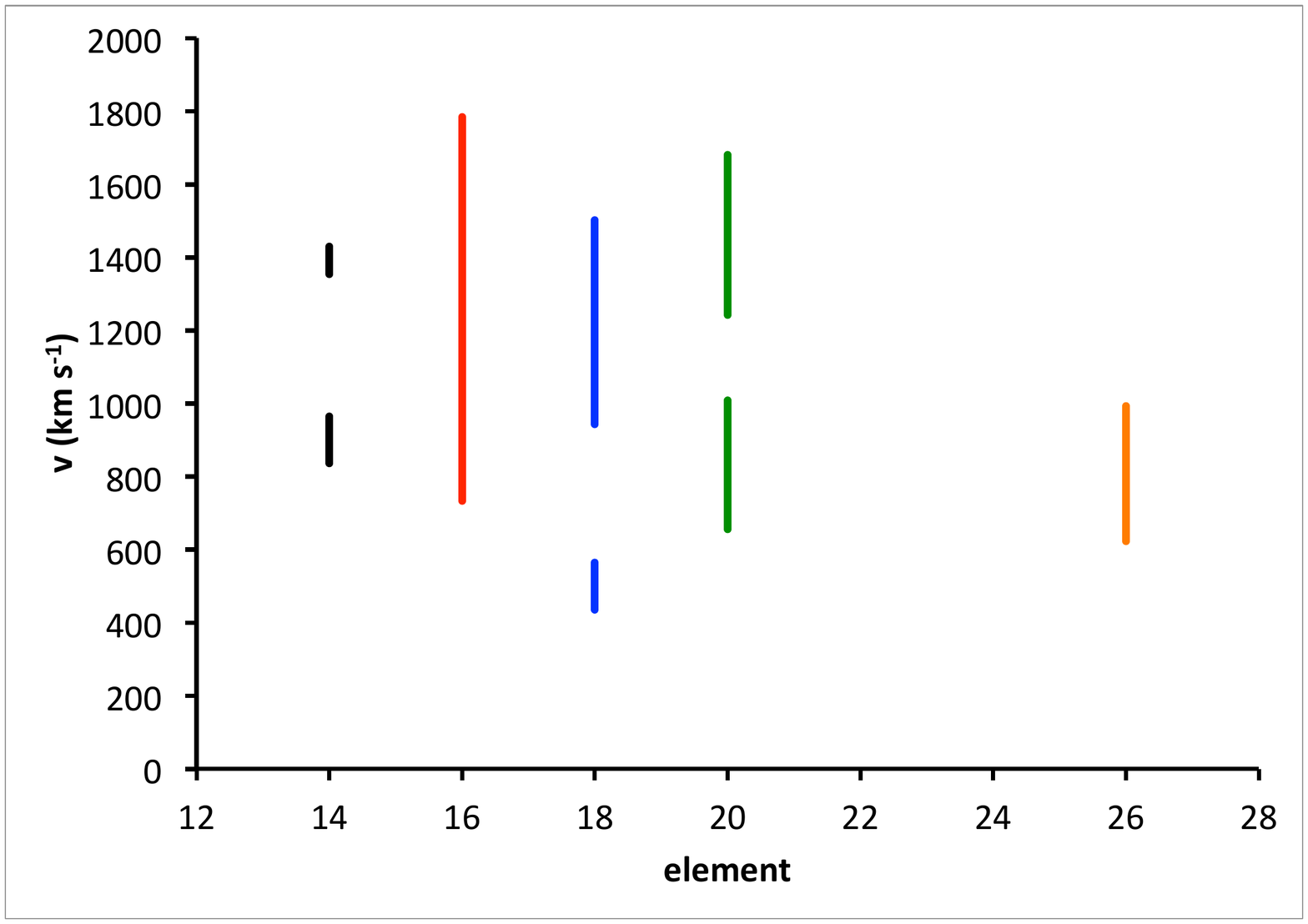}% Here is how to import EPS art 
\caption{\label{fighewidth} Widths of the wind and narrow components in km s$^{-1}$ vs the nuclear 
charge of the parent element, i.e. Si=14, S=16, Ar=18, Ca=20, Fe=26. 
Vertical bars show the allowed regions of width for the wind and nebular components.
The upper bar segment corresponding to higher speed and a broader line, eg. for Si, 
corresponds to the wind, and the lower bar segment with a lower speed and a narrower line corresponds to the 
nebular model component. }
\end{figure*} 

\begin{figure*}[p] 
\includegraphics*[angle=90, scale=0.5]{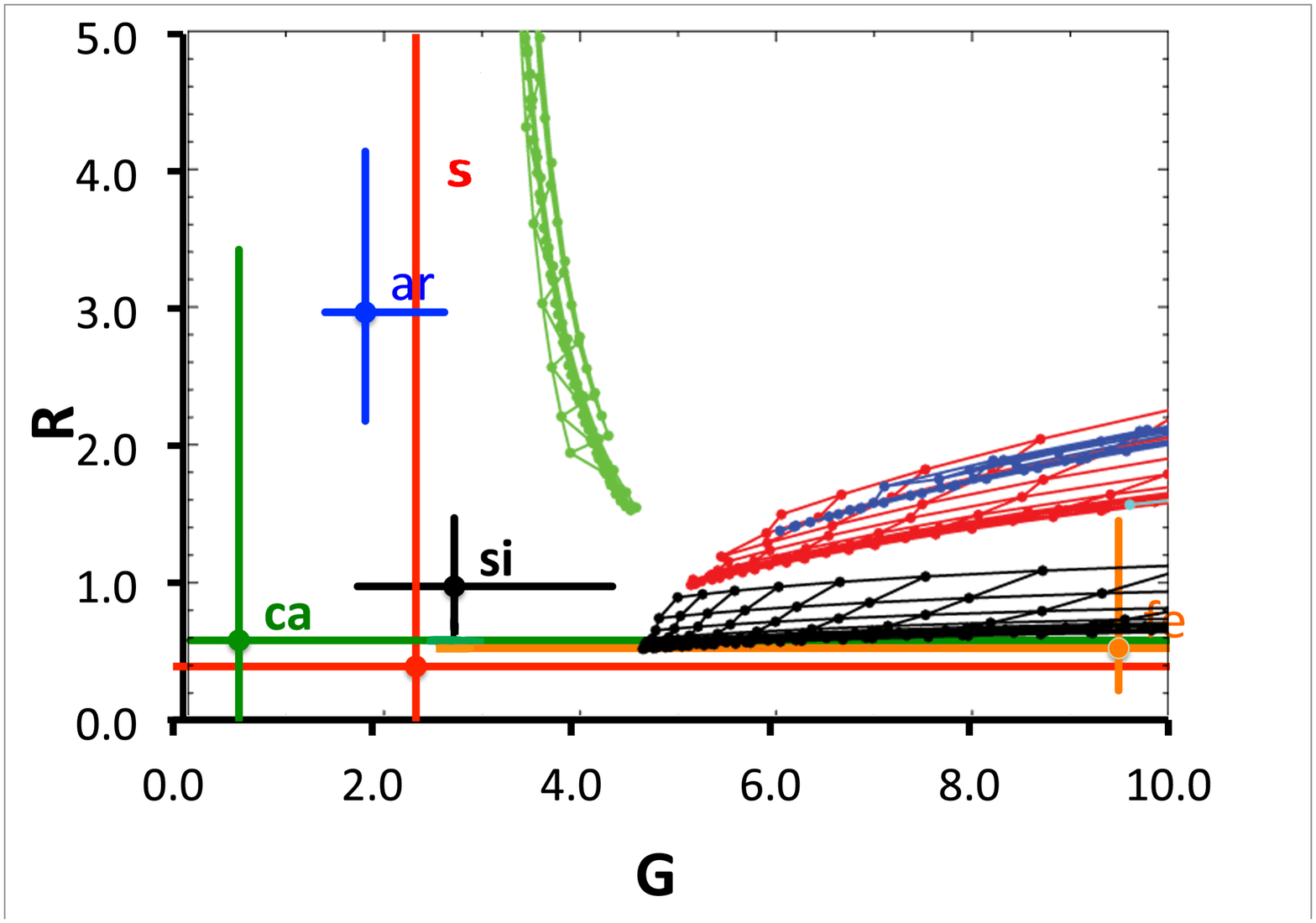}% Here is how to import EPS art 
\caption{\label{figheratio} Values of the two ratios R and G for the narrow components 
from the various He-like ions. Bars show the values allowed by the spectral fits.  The 
grids show the corresponding regions spanned by photoionization models. Colors are as 
in figure \ref{fighewidth}:  black=silicon, red=sulfur, blue=argon, green=calcium, orange=iron.  Model 
grid values for iron are outside the plotted range.}
\end{figure*}

Diagnostic information can be derived from He-like lines when they are 
described in terms of two ratios: R=$f$/$i$ and G=($f$+$i$)/$r$ \citep{gabr69}.  The 
R ratio depends most sensitively on density, with R $>$ 1 corresponding to density below 
a critical value which differs according to the element.  The G ratio depends on temperature
and on the nature of the excitation mechanism.  Recombination following photoionization 
produces G $> >$ 1, while electron impact excitation produces G $<$ 1.  In figure 
\ref{figheratio} we plot values for these ratios allowed by the spectral fits.  These ratios
utilize values for the $r$ line intensity which are solely from the nebular component of the fit; 
the wind component and its contribution to the $r$ line are not included. 

In a photoionized gas, the properties 
are conveniently described by the ionization parameter, which is proportional to the ratio of the ionizing flux 
$F$ to the gas density $n$, log($\xi$)=4$\pi F/n$.  
Figure \ref{figheratio} also shows the results of a grid of photoionization ({\sc xstar}) models 
spanning a range of conditions 1$\leq$log$\xi \leq$3, and 10$^5\leq$T$\leq 10^6$K.  
Colors are as 
follows:  black=silicon, red=sulfur, blue=argon, green=calcium, orange=iron.  Model 
grid values for iron are outside the plotted range.  Photoionized gases are 
typically assumed to have temperature determined by radiative equilibrium 
between photoionization heating and radiative cooling; here we allow the temperature to vary independently.
These models correspond to gas with temperature and ion fractions appropriate to photoionized conditions, 
and also excitation produced solely by recombination and collisions with thermal  electrons.
For conditions where the He-like ion is abundant the R ratio is  weakly dependent on temperature
or ionization parameter, while the G ratio is temperature dependent.  This corresponds to the behavior
of the grids for Si and S.  For Ar and Ca, the range of temperature and ionization parameter spanned by our grid 
is below the peak of the ion fractional abundance, and 
the $f$ and $i$ lines are excited primarily by electron impact collisional excitation.  As a result 
the R ratio also depends on the temperature, owing to the temperature dependence of the rate coefficients.
The measured values for Fe are also shown by the orange bars in figure \ref{figheratio}.  

The blending of the 
$i$ and $f$ lines with each other and with the wind lines results in large error bars in the R and G values for S, Ca, and Fe.
In spite of this, there are clear trends to the line ratio values.  The fact that R is $\leq$1 
for Fe implies a very high density for the formation of this line, i.e. $\geq 10^{17}$ cm$^{-3}$.  If so, the 
Fe line formation must occur very close to the continuum source in order to achieve the ionization parameter 
needed to produce Fe$^{24+}$, i.e. log($\xi)\geq$3.   More likely, as we will discuss below, the low energy part 
of the Fe$^{24+}$ line blend is likely to be affected by absorption, and the true R ratio may be greater than we observe.
If all the He-like lines are emitted from the same region, then a trend of the value of R with the atomic 
number of the parent ion might be expected, since for lower Z ions the density would be closer to the critical density, while 
for higher Z ions the density could  be below the critical density.   Figure \ref{figheratio} shows no such trend, possibly due 
to the large error bars on R for several ions.

The model G ratios are $\geq$4, as expected for recombination at temperatures expected for photoionization.  The fact that the 
measured G ratios are $\sim$2 for Si and Ar is an indication of additional contributions to the $r$ line intensity.  A likely 
explanation is that resonance scattering  contributes to the $r$ line intensity even in the nebular component, i.e. the 
component which does not show dynamical evidence for being in a wind.  
It is also notable that the measured R ratio is $\leq$1 for most elements.  This is in contrast to values $\geq$2 expected for 
low density recombination for these elements \citet{Baut00}.  This is an indication that the density is $\geq 10^{12}$ cm$^{-3}$.  
It is also possible that the R ratio we observe is the result of radiative excitation from the $1s2s(^3S)$ upper level 
of the $f$ line to one of the $1s2p(^3P)$ levels.  This possibility cannot be evaluated in the absence of any measurement of the 
flux at the appropriate energy, $\simeq$14.3 eV or $\sim$867 \AA\ .  With this caveat,  in what follows we will adopt the
collisional mechanism.
The models used here and in the remainder of this paper adopt a density of $10^{14}$ cm$^{-3}$.

\subsection{Dust Scattering Scenario for the 1.838 keV Line}

The feature at 1.838 keV, which we attribute to the $f$ line of Si$^{12+}$ resembles the feature associated 
with dust scattering from silicate grains.  Depending on the grain size, dust scattering can produce 
a notch-like feature in the total extinction which may appear as an apparent narrow emission feature at energies below the threshold for absorption.  This has been illustrated by \cite{corr16} and by 
\cite{schu16} in $Chandra$ data from GX 3+1, and using laboratory data by \cite{zeeg17}.  This apparent emission 
feature appears at an energy which is not significantly different than the energy of the feature that we attribute to 
the Si XIII $f$ line; the lab energy of this line is 1.8398 
\footnote{https://physics.nist.gov/PhysRefData/ASD/lines\_form.html}.  

One test of the nebular scenario is whether the nebular component is present in other ions besides Si XIII; 
if the 1.838 keV feature 
is due to dust scattering, it should be strongest near the Si K edge and weaker or absent for other elements 
which are not depleted on dust to the same extent.  
The depletion of S onto dust is much less than that of Si, approximately 0.4 \citet{drai03}, so any apparent emission 
associated with dust scattering on this element would be expected to be much weaker than for Si.
The region of the spectrum 
containing the lines of He-like and H-like S shows emission near the the $f$ line of S XV  at 2.43 keV, and also emission 
near the $i$ lines near 2.44 keV.  The lab energies of these lines are 2.430 and 2.448 keV, respectively.
However, the shape of the spectrum near these features does not match well with the {\sc photemis} model.  This is illustrated 
in figure \ref{figsnarrow}.   

Further evidence in favor of nebular emission as the origin of the 1.838 keV feature 
is the fact that the broad feature near 1.845 keV, if it is associated 
with the P-Cygni emission from the $r$ line, is not likely to be broader than the corresponding feature associated with 
the Si XIV L$\alpha$ line, near 2 keV.  If so, the additional emission from the Si XIII $i$ lines is needed in order to 
explain the observed width of the 1.845 keV feature.  A similar argument applies to the P-Cygni emission from other 
He-like $r$ lines, notably the S XV line in figure \ref{figsnarrow}.

In addition, the strength of the nebular emission component varies with orbital phase, as 
shown in section \ref{orbitalphasevariability}.  This is not expected to occur if the emission near the Si edge were 
due to dust scattering.  

Fits in which the strength of the nebular component is set to zero and a Gaussian is included at 1.84 keV 
do not provide as good C-statistic values as 
fits in which the nebular component is included. 
For example, we find C-statistic/dof=225720/32768 for the Obsid 7268 spectrum summed over orbital phase without the nebular 
component, as compared with 214748/32768 for the same spectrum fit with the nebular component.  
In what follows, we will adopt the nebular emission scenario for the line at 1.838 keV
but we cannot conclusively rule out the possibility that this line is associated with dust scattering.

\subsection{Si K Lines}

As shown in the previous subsections, the most abundant ion stage of most elements is the He-like stage.  It 
is of interest to set limits on the strengths of features arising from ions of lower ion stage.  This can be done 
by searching for K lines from ion stages with 3 or more electrons.  The existence of these lines and their use 
as tracers of lower ion stages has been validated using the spectra from other objects, such as Vela X-1 
\citep{sako02}.  Table \ref{table3} contains a list of these lines for the obsids 6601 and 7268 spectra.

\begin{deluxetable}{crrrrrrrrrr}
\tabletypesize{\scriptsize}
\tablecaption{Si K Line Fluxes (cm$^{-2}$ s$^{-1}$) \label{table3}}
\tablewidth{0pt}
\tablehead{
\colhead{ion}&\colhead{wavelength (A)}&\colhead{energy (keV)}&\colhead{Obsid\_6601}&\colhead{Obsid\_7268}}
\startdata
V&7.188&1.725&1.98$^{+0.129}_{-1.80}$&1.48$^{+0.193}_{-0.180}$\\
VI&7.125&1.740&0.687$^{+0.108}_{-0.150}$&0.830$^{+0.140}_{-0.134}$\\
VII&7.059&1.756&$\leq$0.002&$\leq$1.389\\
VIII&6.990&1.774&$\leq$0.013&$\leq$0.187\\
IX&6.926&1.790&$\leq$0.013&$\leq$0.042\\
X&6.856&1.808&$\leq$0.009&$\leq$0.012\\
XI&6.782&1.828&$\leq$0.037&$\leq$0.074\\
XII&6.906&1.795&$\leq$0.007&$\leq$0.017\\
\enddata
\end{deluxetable}

This shows that most of these lines are not detected in these spectra.  This a manifestation of the 
fact that the average ionization state of the wind in Cyg X-3 is higher than it is in other HMXBs such as 
Vela X-1.
 
\subsection{Ionization Balance \label{ionizationbalance}}

In addition to decomposing the line emission into nebular and wind components, it is useful to explore
the ionization balance in the line emitting gas. 
Motivated by the decomposition of the He-like lines into narrow and wind components, we fit the 
spectrum to a physical model based on realistic ionization balance and emission physics. 

The challenge of fitting a single photoionization model to the Cyg X-3 spectrum can be inferred from examination of 
table \ref{table1}:  the spectrum is dominated by lines from H- and He-like ions of Si, S, Ar, Ca and Fe, and the intensities 
of the H- and He-like lines from each element are comparable.  In contrast, equilibrium ionization balance calculations 
are affected by the fact that high Z elements have  greater rate coefficients for 
recombination than low Z elements.  Therefore high Z elements require a greater ionization rate to achieve a given ionization state.  
This is compounded by the fact that the ionization cross sections for high Z elements are smaller, so a greater ionizing radiation 
flux is needed to achieve a given ionization rate.  This is true for either thermal electron impact ionization or photoionization.
Furthermore, most plausible ionizing spectral energy distributions decrease with energy in the 1-10 keV band.  Therefore, the
ionization parameter needed to achieve a given ionization state is much greater for high Z elements than it is for low Z elements.
The most apparent manifestation of this is the contrast between the lowest Z element with good line detections, Si, compared with the 
highest Z element, which is Fe.  Both the H-like and He-like lines from both elements are detected and have comparable strength ratios.
Calculations of photoionized ion balance \citep{kall01} show that this implies an ionization parameter 
log($\xi)\simeq$2 for Si, and log($\xi)\geq$4 for Fe.   This is illustrated in figure \ref{figtxi}.
    In what follows we refer to this as the ionization balance discrepancy.

\begin{figure*}[p] 
\includegraphics*[angle=90, scale=0.6]{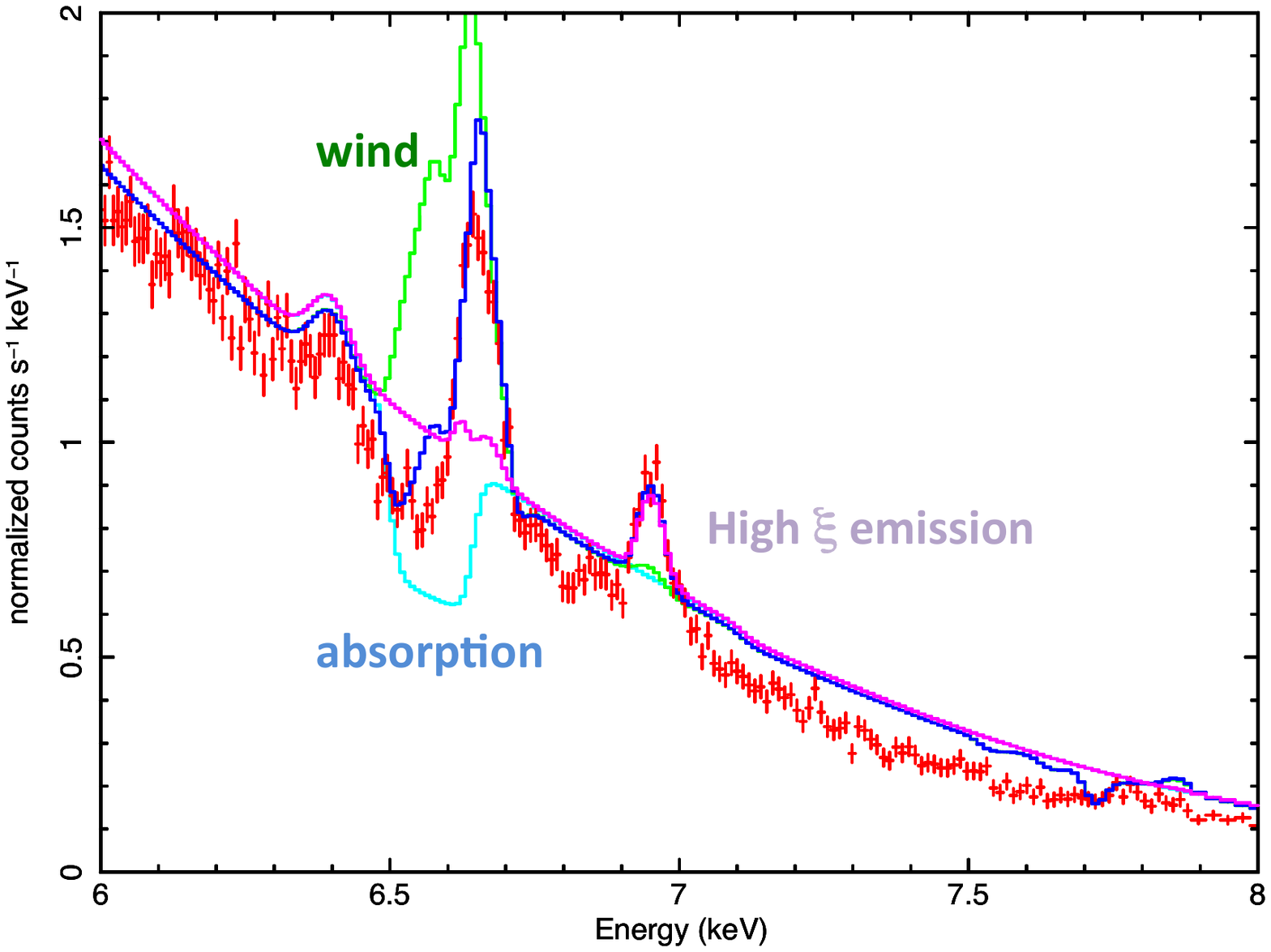}% Here is how to import EPS art 
\caption{\label{figfe} Iron line region for Obsid 7268 showing components of the model fit:
High ionization nebular emission + neutral iron emission 
(purple), wind emission (green), notch absorption (light blue),
total model (dark blue), and observed data (red).}
\end{figure*}

A further problem is the fact that gas at log($\xi)\simeq$2 radiates strong iron lines at energies below the energy of the 
He-like line at 6.7 keV.  Such line emission is not consistent with the observed spectrum and can be excluded at a very high 
level of confidence.  This incongruity is difficult to resolve.  One possible explanation, which we adopt here, is that there is 
absorption which removes the iron K emission associated with the log($\xi$)=2 gas.  We parameterize this absorption using the 
ad hoc {\sc xspec} 'notch' model component.   The notch model is supplied by {\sc xspec}; it is a multiplicative model with 
value of unity every where except  over a region of specified width around a specified centroid energy, in which 
the model is reduced by a constant factor $\leq$1.   Further evidence for this comes from the weak dip in the spectrum 
visible near 6.6 keV in the spectra in figure \ref{fig1}.
We allow the values of the notch energy, width and depth to vary during fitting.
For the total spectrum fit to Obsid 7268 we obtain a value for the energy parameter $\simeq$ 6.56 keV.  We fix this value for the phase fits.
All our model fits obtain similar values for the other parameters:  width $\sim$0.1 keV, and depth $\sim$0.3.  
Other evidence in favor of this explanation are discussed in the following 
subsection.  The discussion section examines alternative scenarios for the apparent disparity between the conditions 
needed to explain the iron lines compared with lower energy lines.
The high ionization gas responsible for the highly ionized iron emission in this scenario radiates negligible emission from lighter 
elements and so does not significantly affect the spectrum at lower energies.
As discussed below, we also find evidence for a 6.4 keV iron line which we parameterize by a single Gaussian emission model.
The contributions of the various components to the iron line are illustrated in figure \ref{figfe}.

\begin{deluxetable}{crrrrrrrrrr}
\tabletypesize{\scriptsize}
\tablecaption{Model Components \label{modelcomponents}}
\tablewidth{0pt}
\tablehead{
\colhead{number}&\colhead{name}}
\startdata
1&cold absorption\\
2& notch absorbing iron\\
3& photoionized nebular emission with log($\xi)\sim$2 - 3 \\
4& 6.4 keV Gaussian emission\\
5& photoionized nebular emission with log($\xi)\sim$4 - 5), \\
6& wind emission\\
7& Diskbb \\
8& Blackbody\\
\enddata
\end{deluxetable}

The model which we use to fit the global spectrum thus consists of 8 components.  These are listed in 
table \ref{modelcomponents}.  In what follows we refer to them by number.
This corresponds to the {\sc xspec} model 
command 'model  wabs*notch(photemis + gaussian + photemis + windabs*(diskbb + bbody))'.
We have also experimented with other choices for the neutral absorption, including 
the 'tbabs' model.  None of the conclusions of this paper are affected by this choice, with
the exception of the quantitative value of the absorbing column, which depends on 
the abundances used; the 'wabs' model uses fixed abundances.
We force the element abundances to be the same for all the model components which depend on them, i.e. components
3,5,6.  
Parameters of the fits are tabulated in table \ref{globalfittable} for the orbital phase summed data and 
table \ref{globalfittable2} for the phase resolved data 
from obsids 6601 and 7268.  Also given are the values of the statistical quantities
describing the the goodness of fit, the C-statistic and $\chi^2$; model fits minimize the value of the C-statistic.
Owing to limited statistics, values of parameters for model components 2, 4 and 5, i.e. the notch, Gaussian iron line and 
high $\xi$ components, contain degeneracies and cannot all be minimized uniquely.  Therefore we have frozen the notch width 
and the $\xi$ value for the high $\xi$ component for most of the spectral fits.  Similarly, the element abundances cannot all 
be uniquely fit for all spectra, so we have used the abundances taken from the spectrum with the best statistics, Obsid 7268, and 
used them for fitting to most of the other spectra.  There are indications of differing abundances, eg. the Fe abundance for Obsid 
16622 and the Mg abundance for many of the earlier spectra.  The latter may be due to contributions to the low energy part of the spectrum 
from the dust scattering halo.

The normalizations of the photoionized nebular components are related to the emission measure of the gas and the 
distance according to the relation $\kappa = \frac{{\rm EM}}{4 \pi D^2} \times 10^{-10}$.  
The normalization of component 3, the log($\xi)\simeq$2 - 3 photoionized component, is 429 in the total spectrum for Obsid 7268.
The normalization of component 5, the log($\xi)\simeq$4 - 5 photoionized component, is 1035 in this spectrum.  
These correspond to emission measures of EM=$2.7 \times 10^{58}$ cm$^{-3}$ 
and EM=$6.4 \times 10^{58}$ cm$^{-3}$ for these two components, respectively, assuming 
a source distance of 7.4 kpc.
 Abundances in tables \ref{globalfittable} and \ref{globalfittable2} are given in the 
traditional logarithmic H=12 format; values are generally within 0.5 dex of solar,  so we do not clearly identify any 
strong evidence for processed material.  However, we constrain abundances only for elements heavier than Mg, and we cannot 
constrain the abundance relative to H.  We choose Fe to be fixed at a value which produces the total line strength we 
observe. We have not allowed the value of the turbulent speed to vary during our fitting, due to the degeneracy between the 
width of the nebular line component and the wind component for many lines.  Thus we regard the chosen value, 1600 km s$^{-1}$, as 
an approximate upper limit.

\subsection{Iron K Line}
\label{felinesection}

As shown in the preceding discussion, the He-like iron K line does not show 
obvious evidence for a P-Cygni profile.  This may be due to the fact that the
splitting between the resonance and intercombination lines at iron is less than
for lower Z elements, compared with the instrument resolution.  

Also notable is the fact that the strong detectable components of the iron line 
consist of the H-like and He-like features centered at 6.97 and 6.6 keV respectively.
The HETG in first order cannot resolve the 
 energy differences of the line centroids of
$r$ and $i$,  and between $i$ and $f$ components of the He-like iron line, 
which are 0.033, and 0.031 keV respectively.  The width of the He-like 
feature is 0.015$^{+0.005}_{-0.0075}$ keV.  

The absence of lower ionization features constrains global spectral fits
which include self-consistent ionization balance of iron and lower Z elements, as
discussed in the previous subsection.  Figure \ref{figfe} shows the spectrum in the vicinity of the 
iron lines.  This reveals a trough centered near 6.55 kev in the total spectrum, and it is possible that this is associated 
with absorption by intermediate stages of iron.  If so, it would account for the existence 
but non-detection of of lower ionization stages of iron, which must coexist with the He-like 
Si and S observed at lower energies.  A similar feature is observed from the source 
Circinus X-1 \citep{schu08}.We emphasize that there is no evidence for such an absorbing 
component elsewhere in the spectrum.
As discussed in the previous section, the iron lines do not show evidence for a wind emission component at the
same ionization parameter as revealed in the lines from lighter elements, Si, S, Ar, Ca. 

The ionization balance of iron relative to lower Z elements depends on the shape 
of the ionizing radiation spectrum.  A power law with photon index $\geq$ 1 has less flux near the 
ionization thresholds for iron than it does for Si etc., and so produces a greater ionization balance 
discrepancy than if the ionizing spectrum is flat or inverted in the 1-10 keV band.  In fact, 
a spectrum which increases $\propto \varepsilon^3$ is needed to make the iron H/He transition occur at the 
same ionization parameter as for Si.  Such a spectrum is so different 
from anything we observe that we will not consider it further.  
An alternative is that the Si -- Ca elements are ionized by a spectrum which is soft, such that it does not ionize 
iron significantly.  Experiments show that a kT=1 keV thermal bremsstrahlung spectrum has this property.
A thermal bremsstrahlung component has been detected in the Cyg X-3 continuum spectrum \citep{kolj13} but it 
has a higher temperature (3-6 keV) than would be needed to account for the Si -- Ca ionization observed
 here.  A
second component ionized by a much harder spectrum would be needed to account for the H- and He-like iron emission.
There is evidence for such a harder component from the RXTE spectra.  However, the degree of fine-tuning required for 
the low energy component and the fact that there is no direct evidence for such a component (i.e. the observed continuum 
in the Chandra HETG band is more energetic) lead us to consider this scenario to be 
 implausible and we do not consider it further.  
It is also possible that the region responsible for most of the Fe K emission is closer to the continuum source than the 
region responsible for the lower Z element emission, and that the Fe K region absorbs the continuum illuminating the lower-Z emission region.  We have attempted to simulate this, using {\sc xstar}, and conclude that fairly fine tuning of the Fe K region is needed in order to preferentially absorb the photons above 6 keV.  However, this scenario is difficult to rule out 
conclusively.

%The fact that the orbital phase dependence of the iron lines is similar to that of the other lines suggests that they all 
%originate in the same region.  If so, our distinction between high ionization and medium ionization components is not 
%an indication of truly distinct gas components.  

Also apparent in figure \ref{figfe} is emission near 6.4 keV in the total spectrum for Obsid 7268.  This is consistent with 
fluorescence from near-neutral material.  This feature has equivalent width 3.0$\pm{1.5}$eV.  
Thus it is crudely consistent with fluorescence from a Thomson thick near-neutral reprocessor with log($\xi$) $\leq$ 0
with a covering fraction of $\sim$0.02 relative to the source of continuum.  The star itself 
is expected to have a covering fraction approximately 0.07.   
Accurate measurements of the 6.4 keV line strength are hampered by weakness of the feature, by the fact that the 
continuum is not tightly constrained at energies $\geq$ 7 keV due to the response of the instrument, and by the 
apparent absorption near $\sim$ 6.5 keV.  The errors in table \ref{globalfittable} are statistical and likely 
do not fully reflect the systematic uncertainty associated with the continuum placement.

\subsection{Orbital Phase Variability \label{orbitalphasevariability}}

\begin{figure*}[p] 
\includegraphics*[angle=0, scale=0.8]{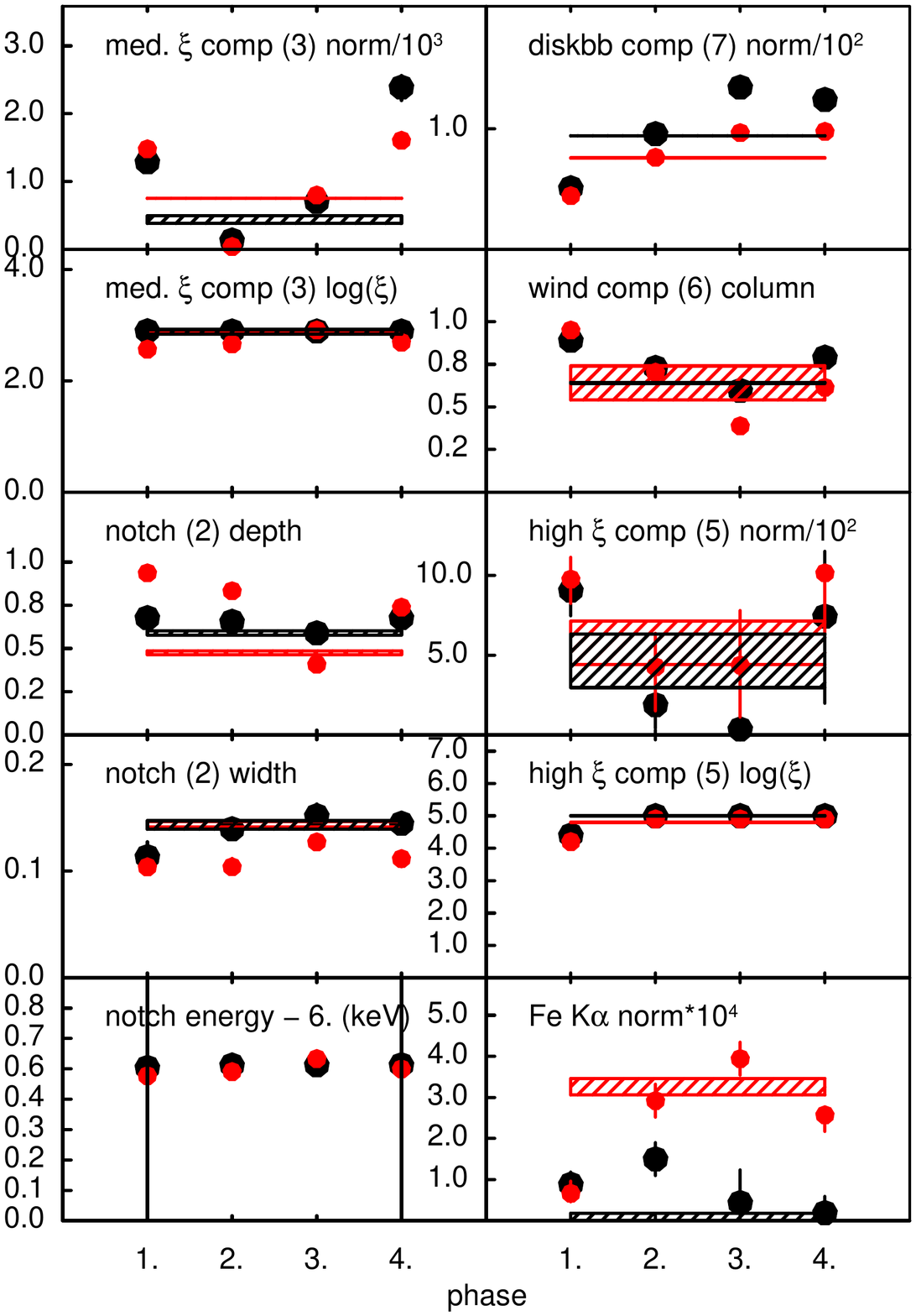}
\caption{\label{figphase} Variation in spectral fit quantities with orbital phase.  Hatched boxes  show regions for phase average; black=Obsid\_7268, red=Obsid\_6601.}
\end{figure*}

Changes in the line centroids with orbital phase have been previously reported by \citet{vilh09} and \citet{star03}, 
and have been used to derive mass functions for the Cyg X-3 system.  The spectra from our four phase bins 
are shown as panels in figures \ref{fig1} -- \ref{fig6} and the results of model fits are shown in 
table \ref{globalfittable}.   Changes in the model fit parameters with 
phase for Obsids 7268 (black) and 6601 (red) are shown in figure \ref{figphase}.   This shows the values of various quantities in phase bins as points, plus the value of 
the total spectrum as a dashed box.  Values which are not repeated across columns in table  \ref{globalfittable},
eg. the element abundances and the energy of the iron-absorbing notch component 2,
are held constant using values obtained from 
the total spectrum fits. 

Variability with phase is clearly apparent, and can be summarized as follows: 

(i) The strength of the diskbb continuum, 
component 7 normalization, peaks at phase bin number 3.  This identifies this bin as closest to inferior conjunction, 
typically defined as phase 0.5 in previous X-ray studies \citep{zdzi12}.  Superior conjunction 
occurs in phase bin 1, which is also the bin with the minimum continuum normalization in our fits. 
 The contrast between the maximum and minimum continuum normalizations is only a factor of $\simeq$2.5.  This is less than the 
maximum intensity contrast in the lightcurve shown in figure \ref{figlightcurve}, and likely results from the fact that the true intensity minimum is 
narrow compared with our phase bins.  
%The time delay of the dust scattered emission may also play a role in smoothing orbital variations.

(ii) The photoionized emission component responsible for the narrow lines from all elements lighter than iron, i.e. component 3,
is strongest during phase bin 4, and the upper limits during the other phases are $\leq$0.05 of the phase bin 4 value for 
obsid 7268.  This 
is a notable result owing to the fact that emission is not likely to be narrowly beamed in one direction relative to the 
binary system.  Obsid 6601 shows a lower normalization for this component across all 
phase bins but still peaks in bin 4.   Pileup 
may affect the detection of strong line emission at the high fluxes in Obsid 6601.

(iii) Component 5, the emission component responsible for the H- and He-like  iron lines, shows similar behavior in that it has a 
maximum in phase bin 4.  However, it is also strong in phase bin 1.  We point out that the He-like 
iron line is also produced by component 6, the wind component.  There is detectable He-like iron emission in phase bins 2 and 3, 
but in our model they can be produced by the wind emission plus notch absorption.  

(iv) The notch absorption is required at all orbital phases.  Its width has a maximum in phase bin 3.

(v) The wind component is detected at all phases, though its column varies by a factor $\sim$2 in the {\sc xspec} column parameter, 
which is a logarithmic quantity. The minimum occurs at phase 3.  Thus the wind is anti-correlated with the X-ray continuum.  

(vi)  The 6.4 keV line component shows a maximum in phase bin 3 for Obsid 6601, but not for Obsid 7268.

Many of these conclusions can be verified by examination of the spectra in figures \ref{fig1} -- \ref{fig6}.  
The strength of the high ionization component is reflected in the H-like iron line near 6.97 keV.  This is strongest in 
phase bins 1 and 4.  The notch appears weakest in phase bin 1, though this is not born out by the fits.  A likely explanation is the 
interaction with the wind emission, which is relatively strong during bin 1.  
%The results from obsids 7268 and 6601 are crudely consistent with each other, though many of the features are stronger 
%in Obsid 6601.  This does not conform to the expectation that the total flux would be lower in Obsid 6601 than in Obsid 7268
%owing to either pileup or to the loss of light in the dust-scattered halo using TE mode.  Thus we infer that the differences 
%between these two datasets are real, and possible stronger than we measure, owing to the above effects.  
We can think about these results in terms of two sets of model components:  those which are strongest in bins 1 and/or 4, and those
which are strongest in bin 3.  The former include the wind component 5, the high $\xi$ component 5, the medium $\xi$ component 3.  The 
latter include the diskbb component 7, the iron K$\alpha$ component 4, and the notch component 2.   

%Note that none of the other lines displayed in figure \ref{figfephase} have phase dependence similar to the iron lines.  This is 
%in contrast to the results of \citet{vilh09}, who claim that the Si$^{13+}$ L$\alpha$ line is similar.

\subsection{Obsid 16622 \label{16622}}

The data from obsids 425, 426 1456 and 16622, together with obsids 7268 and 6601, allow us to study the long term variability of 
Cyg X-3.  Results of fitting to all the spectra, summed over the binary orbit, are given in table \ref{globalfittable2}.  This 
shows that obsids 7268 and 6601 correspond to the highest intensity states.  Since they also have the longest observing times, 
they provide the most sensitive measurements of the various spectral components.  Obsids 425 and 426 have intensities which are 
comparable, but with shorter observation times.  Obside 16622 was obtained at the lowest flux state of all of these and differs
significantly in its spectral properties.    The flux during this observation was 2.15 $\times 10^{-8}$ erg cm$^{-2}$ s$^{-1}$ 
(89 mcrab) compared with $\sim$ 6 $\times 10^{-8}$  erg cm$^{-2}$ s$^{-1}$ ($\sim$240 mcrab) for the other obsids. 
Obsid 1456 was similar to Obsid 16622 in terms of state activity, and the observations have some features in common.  
However, the flux during Obsid 1456 is slightly less extreme than Obsid 16622, and also Obsid 1445 does not have as small 
statistical errors on measured spectral quantities, so we will base our subsequent discussion on Obsid 16622.

Principal differences between Obsid 16622 and the observations from higher flux states include a significantly harder 
continuum spectrum.  We fit the other obsids with a 1.7 keV disk blackbody, which peaks at $\simeq$ 3.2 keV.  Obsid 16622 
requires a hard power law or a blackbody with kT=2 keV.  The combination of this continuum and the neutral absorption 
causes the detected continuum to peak near 4.5 keV.  In this sense, the contrast between this low flux state and the 
other obsids crudely resembles the difference between the low hard and high soft states of black hole transients.

In addition, the absorbing column for Obsid 16622 is lower, 2.28 $\times 10^{22}$ cm$^{-2}$ 
compared with 3.2 -- 5.4 $\times 10^{22}$ cm$^{-2}$  for the other obsids.   If much of the low energy absorption 
comes from the wind from the companion then this would suggest the companion as the source of this variability.  
%Alternatively, this could be an artifact of the  variability of the X-ray source together with the effects of the 
%dust scattered halo, as described in subsection \ref{dustsection}.  

The line emission in Obsid 16622 crudely resembles that seen in other obsids in that the four line emitting components 
are all significantly detected, and with parameters similar to those found in the other fits.  On the other hand, the 
the 6.4 keV iron K line has a similar flux compared with the other spectra, and therefore it has an equivalent width of 101$\pm$5 eV
compared with 3 eV for Obsid 7268. 

In addition, our model cannot adequately fit to Obsid 16622 using the same elemental abundances as used for all the other 
obsids.  Those abundances included an enhancement of a factor 4 for the Si/Fe ratio, and a factor 2.6 for the S/Fe ratio compared 
with solar \citep{ande89}.  These abundances
greatly over-produce the Si and S lines when used in fitting to Obsid 16622, for which the best-fit Si/Fe ratio is $\sim$0.4 and the 
best-fit S/Fe ratio is $\sim$1 compared with solar.   The abundances of \citet{aspl09} have Fe which is smaller than that of \citet{ande89} by $\simeq$30$\%$, so all these ratios will be larger when interpreted in this context.
These differences between fit results for different Obsids 
are indicative of shortcomings in the physical assumptions 
of our spectral model, since it is unlikely that the elemental abundances actually vary in this way. 
It is worth noting that WR winds are expected to have abundances which show evidence for nuclear processing.  An example is the 
WN8 star analysis by \citet{hera01} which shows that Si and Ca are over-abundant and S, Ar, and Fe under-abundant compared to solar.
%It is possible 
%that the effects of dust scattering could lead to apparent abundance effects, since scattered emission could tend 
%to dilute the true line emission preferentially at low energies and thereby mimic lower elemental abundances. 

\subsection{Long Term Variability \label{otherspec}}

\begin{figure*}[p] 
\includegraphics*[angle=0, scale=0.8]{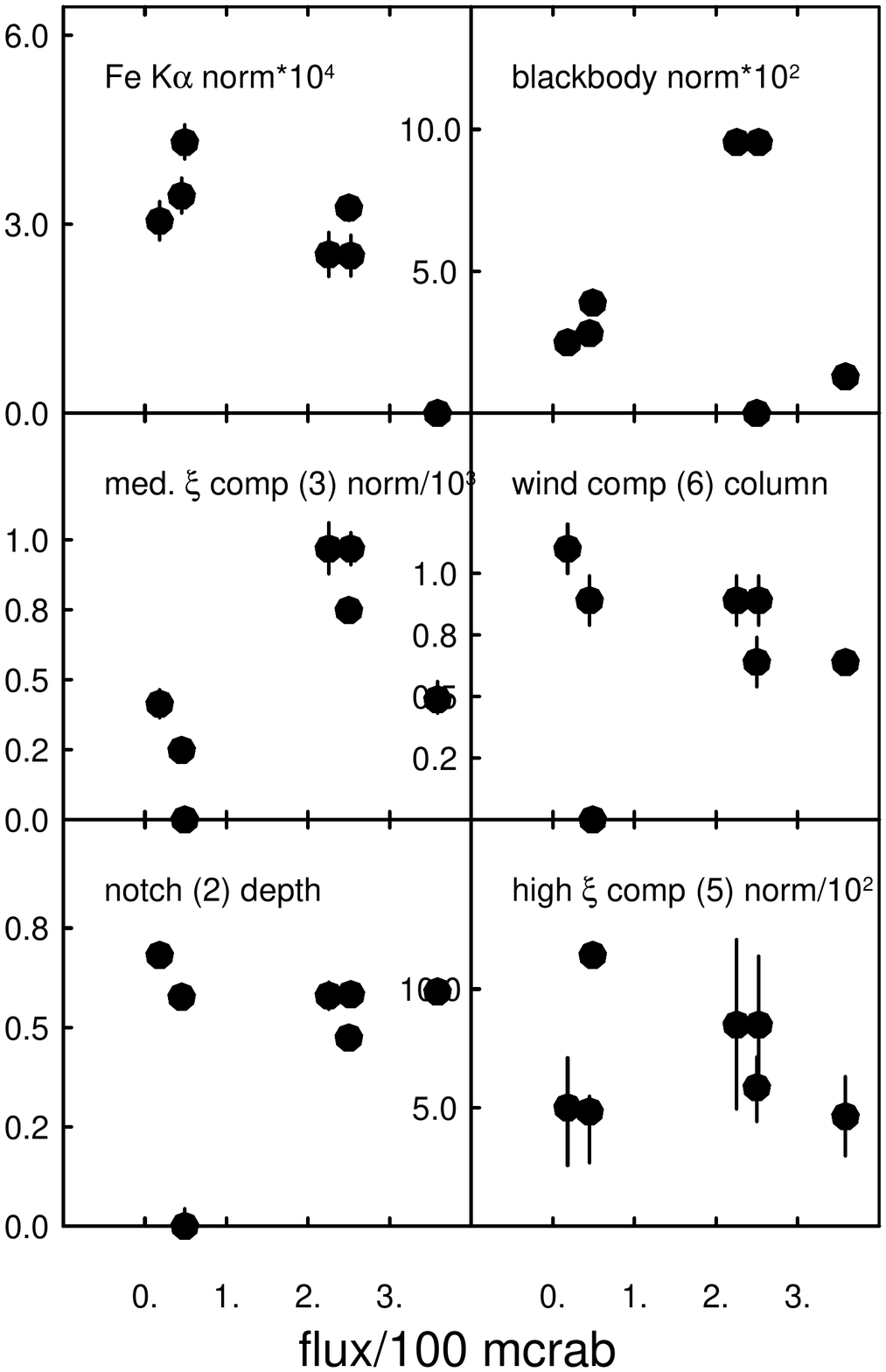}
\caption{\label{figphase2} Variation in strengths of the various components plotted vs. flux in component 7, the diskbb continuum.}
\end{figure*}

Figure \ref{figphase2} plots the parameters describing the various spectral components for the five obsids.  These are 
plotted vs. the total flux in 2 - 10 keV band.  This clearly shows that the medium $\xi$ nebular component 3
is strongest during the highest flux states, i.e. during obsids 6601 and 7268.   The iron K$\alpha$ line shows the 
opposite behavior, and is strongest during low flux states.  
These results, together with the results of the phase resolved fitting in figure \ref{figphase}, suggest that the blackbody and 
the iron K$\alpha$ line are associated with gas close to the compact object.  The nebular components are likely
associated with the size scale of the binary orbit, and also are strongest when the source luminosity is greatest.   
This suggests that the X-ray luminosity of the source is determined by the wind from the star, i.e. the highest luminosity
states coincide with the strongest wind.  It also confirms the identification of the disk or accretion structure close to the compact 
object as the origin for the fluorescent K$\alpha$ line, and suggests that this feature and the associated 
accretion flow structure (i.e. the accretion disk) are anti-correlated with the wind from the companion.

\section{Discussion}

The principal results obtained so far can be summarized as follows: (i) The 
He-like lines show that the centroid of the emission is not at the energy 
of the resonance or intercombination lines, but is rather between these two lines;  
(ii) The resonance line also shows blue-shifted 
absorption similar to a P-Cygni profile.  This is similar to what is observed
in the H-like lines \citep{vilh09}. (iii) The He-like triplet intercombination 
and forbidden components are also present and are significantly narrower than 
the resonance line component. (iv) These results obtain in differing degrees for
the He-like ions accessible to the Chandra HETG:  Si, S, Ar, Ca.  Iron 
differs in that the wind component is undetectable. 
(v) There is a trend that the
wind component becomes weaker and possibly slower as the element atomic number increases, while
the nebular component is consistent with having the same speed for all ions.
For iron, it is possible that the wind is present but has a 
velocity which makes it indistiguishable from the narrower 
triplet lines. (vi) The density in the gas emitting the triplet component can 
be inferred from the R ratio, and it is consistent with a density greater than 
$10^{12}$ cm$^{-3}$. The R ratio for Ar is significantly greater, however, and this 
does not have an obvious explanation.
 (vii) Orbital-phase resolved spectra show that the nebular components 
are strongest close to eclipse transitions, while the dominant continuum component and 
iron K$\alpha$ line are strongest near inferior conjunction. 
(viii) Fitting to global models shows that the ionization balance 
of iron and lower Z elements cannot be modeled self-consistently; this could 
be due to absorption of the iron K emission from intermediate ion stages, or due to 
some ionization process which has not yet been explored. (ix) Comparison of 
spectra taken at different flux states reveals that, when the flux is low, 
the continuum is harder than in the other state and resembles a hard power law 
or hot blackbody.

\subsection{Spherical Wind HMXB Models}

\begin{figure*}[p] 
\includegraphics*[angle=90, scale=0.6]{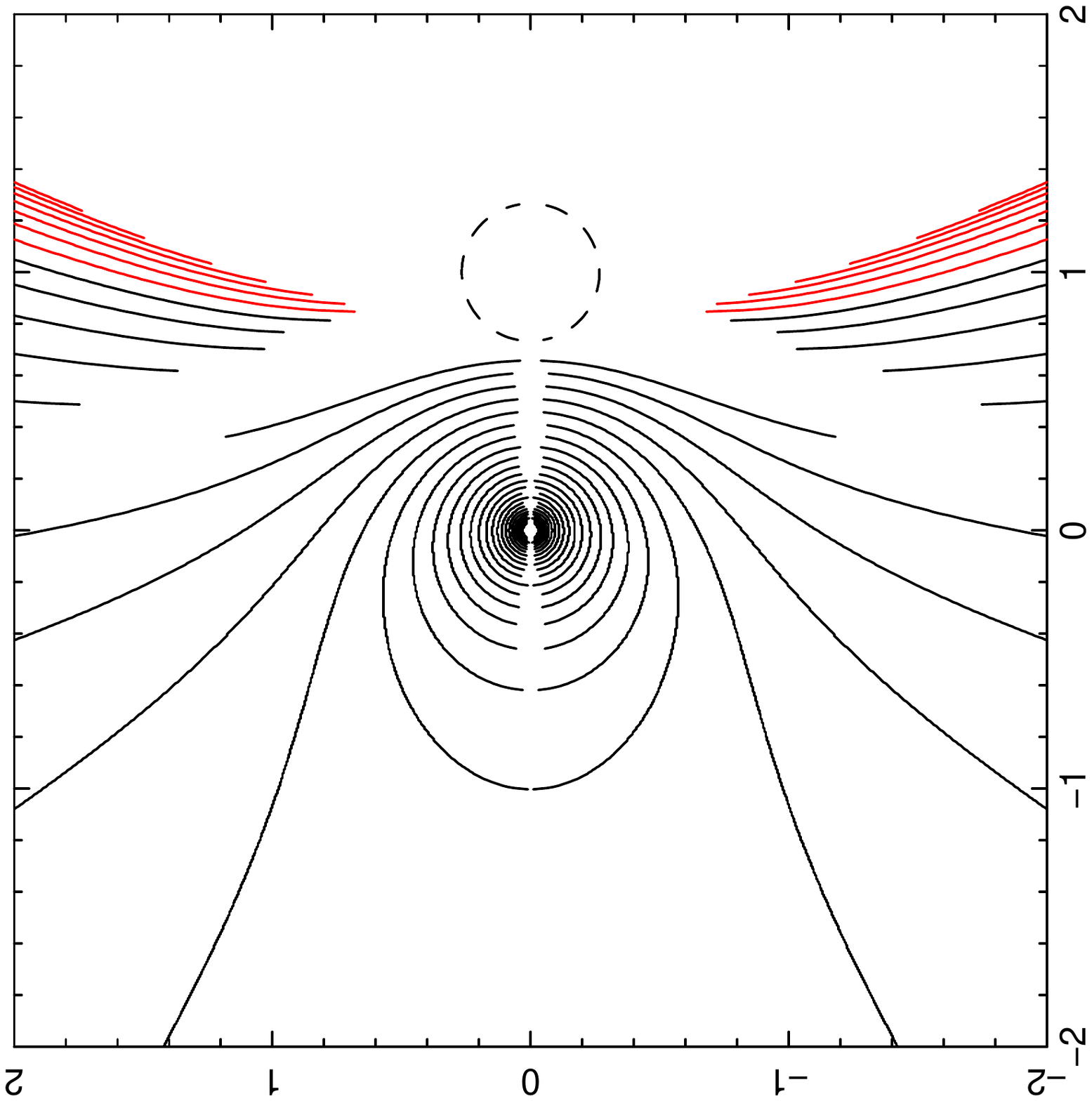}% Here is how to import EPS art 
\caption{\label{figqvolume} Surfaces of constant q in the orbital plane. Axes 
are position in units of 10$^{12}$ cm.  The companion star is shown as the dashed circle.  The solid curve 
show surfaces of constant q.  Black corresponds to q$\geq$1 and red corresponds to q$<$1.  The X-ray source is at the center 
of the concentric black surfaces.  This illustrates the fact that the X-ray source dominates the ionization throughout the half-space
containing it.}
\end{figure*}

It is of interest to compare the conditions implied by the observations with 
simple theoretical predictions for the HMXB winds.  Such predictions were 
first explored by \cite{hatc77}.  In the simplest case, it 
is expected that the wind originates from the surface of the primary star and 
flows outward in spherical symmetry.  The wind velocity starts with a value 
close to the sound speed at the photospheric temperature ($\sim 10^5$K), 
and accelerates to a terminal speed $\sim 1000$ km s$^{-1}$ within a few 
stellar radii.  X-rays originate from the compact source and illuminate the 
wind.  The ionization of the wind can be described by the ionization parameter
$\xi=L/nr^2$, where $L$ is the X-ray source luminosity in the 1 - 1000 Ry 
energy range, $n$ is the gas density, and $r$ is the distance from the X-ray 
source.  If so, the surfaces of constant $\xi$ are approximately spheres,
nested around the X-ray source, or else open surfaces nested around the 
primary star.  Which obtains depends on the quantity $q=\xi/\xi_x$, where
$\xi_x=L/n_xa^2$ and $n_x$ is the wind density at the orbit of the 
compact object, and $a$ is the orbital separation.  The $q=1$ surface is a 
plane dividing the orbital volume in half; closed surfaces surrounding 
the X-ray source have $q\geq$1.
This is illustrated in figure \ref{figqvolume}.

It is straightforward to calculate the quantity of gas vs. ionization state 
using this simple model.  The relevant quantity is the emission measure defined
as $EM=\int{dV n^2}$ where $n$ is the nucleon density.  The estimates depend 
on the properties of the wind and X-ray source.  Here we adopt the following 
values:  X-ray luminsity $L=2.5 \times 10^{38}$ erg s$^{-1}$ \citep{vilh09}; 
orbital separation $a=3.4 R_\odot=2.37 \times 10^{11}$ cm; primary radius 
$R_*=0.92 R_\odot=6.4 \times 10^{10}$ cm; wind mass loss rate 
$\dot M=10^{-5} M_\odot {\rm yr}^{-1}$; wind terminal speed 
$v_\infty=$1500 km s$^{-1}$.  If so, the value for the predicted 
gas density near the X-ray source is $n_x\simeq 5.2 \times 10^{12}$ 
cm$^{-3}$ and the fiducial ionization parameter is 
$\xi_x=8.6 \times 10^2$ erg cm s$^{-1}$.  This is the ionization parameter on the q=1 plane, i.e. the plane which bisects the binary volume. 
Thus, the X-ray source will produce 
an ionization parameter of this value or greater throughout its half-space, if
the effects of attenuation are neglected.
In the absence of X-ray ionization, the ionization balance in the wind is 
expected to resemble a $\sim 10^5$K coronal gas.  That is, most elements will be 
3 -- 4 times ionized.   Results of photoionization modeling (eg. \citet{kall04}) shows that 
photoionization will increase the local ionization balance above that expected for 
a $\sim 10^5$K gas when $\xi\geq$10.  Thus, the X-rays are expected to 
significantly change the wind ionization and temperature compared with 
that of a single WR star throughout more than half of the binary volume.

The total line emission from an effectively optically thin medium can be conveniently 
described by the luminosity $L_{line}=EM j/n^2$ where  $EM$ is the emission measure and 
$j/n^2$ is the emissivity per nucleus.   For a medium in which the temperature and ionization balance 
are constant throughout, the emission measure depends on the gas 
density distribution, and we can calculate it for various simple scenarios.  Of course, 
the true emissivity is a function of temperature and ionization parameter, and so an 
accurate discussion must consider the distribution of emission measure over these quantities.

In the absence of X-rays, the wind is likely driven by UV radiation pressure 
from the primary.  X-ray ionization will change the UV opacity of the wind, 
and therefore also affect the wind driving.  Qualitatively, X-ray 
ionization will likely decrease the UV opacity, and hence the wind speed 
throughout the region where X-ray ionization is dominant.  We can estimate 
where this occurs by assuming that the wind driving is cut off whenever the ionization 
parameter in the wind exceeds a critical value.
Ionization of the wind by the compact source will cause the wind to be slower near 
the compact object and increase the mass accretion rate relative to the un-ionized wind
\citep{vand17}.

%which we crudely take to log$\xi$=10.
%Then we can  calculate the location of log$\xi$=10 
%on the line of centers connecting the X-ray source and companion.  
For the parameter values chosen here, the $q=1$ surface occurs approximately 
0.95 $R_\odot$ from the surface of the primary star; the 
density there is $n_{q=1}\simeq 2.5 \times 10^{13}$ cm$^{-3}$.  The surface 
where $\xi \simeq 10$, i.e. where the X-rays might be expected to affect the 
wind driving, occurs very close to the surface of the primary, at 
a distance of 0.06 $R_\odot$.  The density there is 
7.9 $\times 10^{14}$ cm$^{-3}$.   

\begin{figure*}[p] 
\includegraphics*[angle=90, scale=0.6]{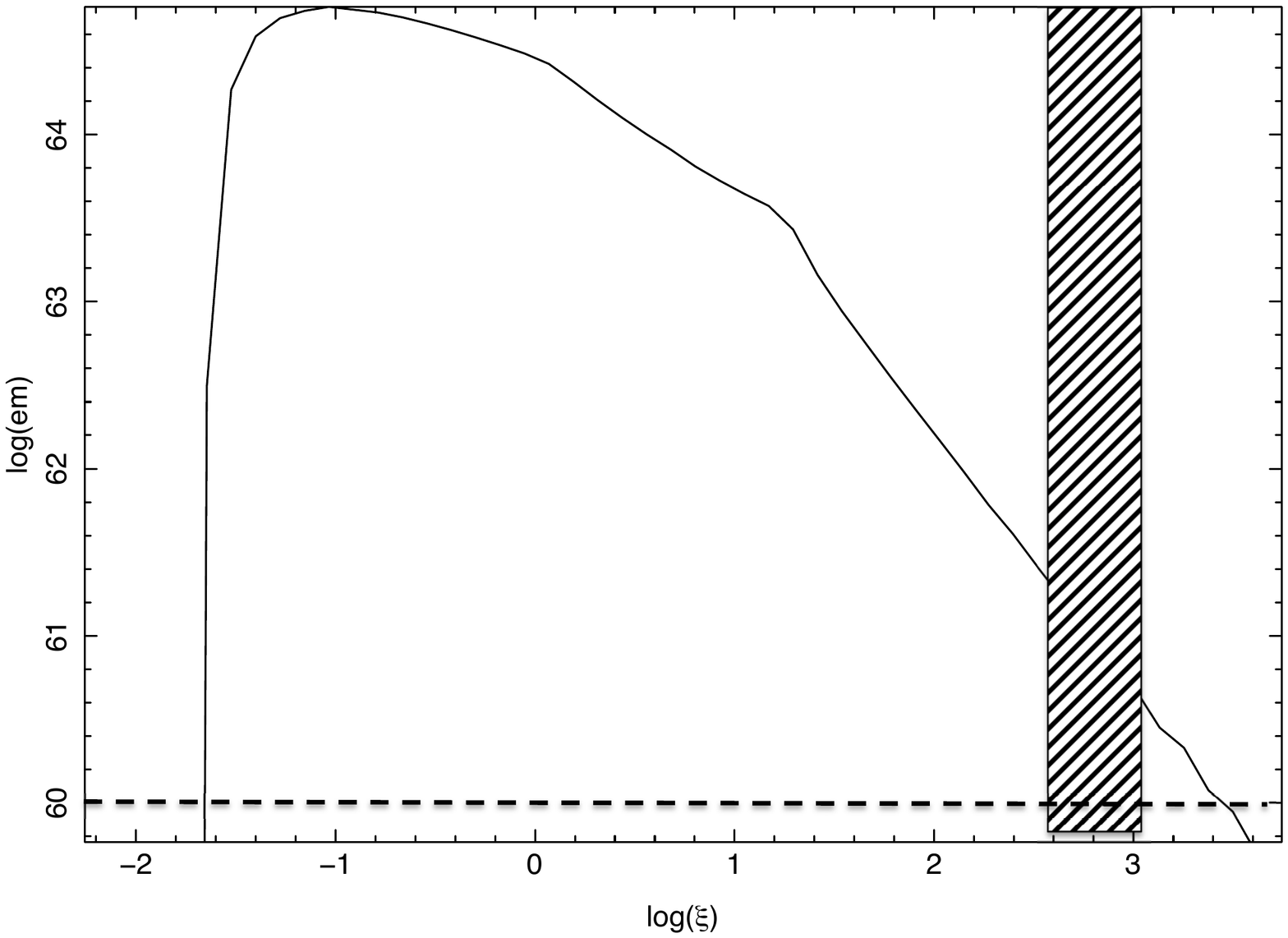}% Here is how to import EPS art 
\caption{\label{figemwind} Emission measure of an X-ray ionized wind with parameters 
similar to a single star Wolf-Rayet wind vs. ionization parameter. Values of EM and the 
range of $\xi$ for our nebular fits are shown as dashed and hatched regions, respectively.}
\end{figure*}

From the discussion in the previous section we can see that the
emission measure needed to account for the observed lines is crudely $\sim 10^{60}$ cm$^{-3}$.
The simple spherical wind model is expected to produce an emission measure which can be 
written $EM=\frac{(\dot M/\mu m_H)^2}{4 \pi a v_x} F(q,a)$ where $F(q,a)$ is a function with value of 
order unity. Crudely, $EM=10^{59} \left(\frac{\dot M}{10^{-6} M_\odot/yr}\right)^2 \left(\frac{a}{1.7 \times 10^{12} {\rm cm}}\right)^{-1} \left(\frac{v_x}{1000 {\rm km s}^{-1}}\right)^{-2} F(q,a) {\rm cm}^{-3}$.
Figure \ref{figemwind} shows the value of this quantity calculated numerically.  
The quantity plotted is the emission measure summed over all regions which have ionization parameters
within bins of width 0.1 in log$\xi$, with the log$\xi$ value corresponding to the values on 
the horizontal axis of the figure.
The simple spherical wind model produces emission measures which 
are greater than those implied by the observations, by a factor $\leq$10, at ionization parameters 
comparable to what we infer from the observed line strengths.  
This crude agreement suggests that the shadow wind + photoionized nebula explanation of the 
spectrum is not far from being correct.  Closer agreement, if found, would likely be 
fortuitous, since the X-ray source almost certainly modifies the wind dynamics, either through 
its gravity or through its effect on the ionization and wind driving.
%This must imply that the simple spherical wind model 
% over-predicts the average gas density, and thus the strong X-ray ionization 
%acts to suppress the wind throughout much of the binary volume.

\subsection{Orbital Phase Dependence}

The phase dependence of the continuum is crudely consistent with previous studies \citep{zdzi12}, which suggests that
absorption and scattering by the wind from the primary is the dominant process affecting the observed intensity. 
If so, the flux observed during the X-ray minimum is due to the partial transmission of the wind which occurs because the 
system inclination is not high enough to cause a complete eclipse of the source.  However, we detect 
no significant change in the absorbing column with orbital phase.  Such a change would be expected due to photoelectric absorption
in the partially ionized wind material.  \citet{zdzi12} claim to find stronger orbital modulation at low energies in their analysis 
of RXTE, Swift, and Integral data.   An alternative explanation for the broad band variability is that the disk is tilted 
or warped compared with the orbital plane. The disk would then be viewed at varying inclinations, and the foreshortening of the 
disk emission would account for the orbital variability.  

As stated in the previous section, we can think about the results of the phase dependent spectral fits 
 in terms of two sets of model components:  those which are strongest in bins 1 and/or 4, and those
which are strongest in bin 3.  The former include the wind component 6, the high $\xi$ component 5, the medium $\xi$ component 3.  The 
latter include the diskbb component 7, the iron K$\alpha$ component 4, and the notch component 2.   
During phase bins 1 and 4, the direct line of sight to the compact object is most obscured by the companion star wind.  Thus, 
we attribute the components which are strongest at these phases to gas which is extended relative to the compact source.
Such gas is likely associated with the companion wind, or the accretion flow on scales comparable to the binary separation.
Thus, the X-ray wind component which we fit, and both the nebular emission components are likely to represent companion wind 
material which is ionized and whose dynamics are affected by the compact X-ray source. 
 The wind component 6 is strongest as the system emerges from superior conjunction.
Thus, if this is gas which is dynamically associated with the companion star, then it trails the star in its orbit.
This is consistent with the expectation of a 'shadow wind', as discussed in the following section. The medium $\xi$ component 
shows opposite behavior, for Obsid 7268, and is strongest as the system approaches superior conjunction.  This suggests that
it either leads the companion star in its orbit if it is dynamically associated with the companion star, or else it trails the 
compact object if it is associated with the compact object.     In our phase-dependent fits we have held the {\sc windabs} covering fraction constant.  However, figures \ref{fig5} and 
\ref{fig6} show that the strength of the P-Cygni absorption in the Si$^{12+}$ $r$ line and the Si$^{13+}$ L$\alpha$ line appear to be stronger 
in phase bins 3 and 4.  This suggests that the wind preferentially trails the companion star in its orbit, as expected 
for a shadow wind.  This will be discussed in more detail below.

The components which are strongest in phase bin 3 we attribute to emission which is compact on the scale of the binary 
separation.  Thus, the Fe K$\alpha$ line, the notch absorption, and the diskbb emission all fall in this category.  These 
components all resemble components which are observed in other black hole candidate sources.  If so, this system 
may resemble those, on scales much less than the binary separation.  
The notch absorption required by our fits could be attributed to a disk wind.  If so, the fact that its depth 
is a maximum at phase bin 3 argues against orbital variability in the viewing angle of the disk, unless the disk 
is warped.

\subsection{Shadow Wind}

A spherical wind over-predicts the amount of material needed to account for the magnitude of the X-ray line emission in the 
Cyg X-3 system.  It also does not provide a consistent explanation for the dynamical information we have 
derived from the spectrum.  That is, we expect the X-ray source to ionize the wind above its mean 
ionization state throughout approximately half the wind volume.  The lines accessible to $Chandra$ HETG
observations correspond to an ionization state which is significantly higher than the mean ionization state 
the wind would have in the absence of X-ray ionization.   We observe H- and He-like ions of elements 
as light as Si; the un-X-ray-ionized wind is expected to have 3-5 times ionized ions of Si and other elements.
The dynamics of the wind depend on the ionization state, since the wind is thought to be driven by 
UV radiation pressure on lines in the ensemble of metal ions.  H- and He-like ions have UV opacities 
with are much less than those of the 3-5 times ionized ions.  Thus, we expect that the dynamics
in the X-ray-ionized zone will be significantly different than in the un-ionized wind.  In particular, 
there should be negligible wind driving, and the wind might be expected to 'coast' in the X-ray ionized region.

The presence of P-Cygni lines in both He- and H-like lines appears to conflict with this scenario, in particular
since the wind speed we measure is comparable to the full speed expected from a Wolf-Rayet wind.  This 
suggests that the X-ray ionized gas has been accelerated as if it were not ionized beyond the wind background
ionization level.  A possible source for such gas is the 'shadow wind' \citep{blon94}, which is the wind 
formed in the region which is shadowed from X-ray ionization by the primary star.  This wind is expected 
to have a velocity law which is unaffected by the X-rays.  However, the rapid orbital motion of the system
will bring this gas out of the shadow and expose it to X-ray ionization.  The X-ray ionization timescale is 
expected to be $t_{ion}\simeq 4 \pi \varepsilon/(n \xi \sigma_{PI})$ where $\sigma_{PI}$ is the photoionization 
cross section; this timescale is $\sim 10^{-3}$ s for plausible parameters.  This can be compared with the 
flow timescale $t_{dyn}\sim a/v $, and this quantity has a value $\sim 10^{+3}$ s.  Thus, the shadow wind 
will preserve its velocity structure as it becomes exposed to X-rays and shows X-ray ionized line features.
The shadow wind is not spherical but rather resembles a narrow conical region which wraps around 
the star due to orbital motion.  The shadow wind has been proposed as the origin for the variability in the 
strength of the IR lines by \citet{vank96}.

The {\sc windabs} model fit provides the quantities needed to estimate the wind mass flux.  The column 
density parameter values in tables \ref{globalfittable}  and \ref{globalfittable2} are 
log(column density/10$^{22}$ cm$^{-2}$); typical parameter values in the range 0.4 -- 0.9 
correspond to column densities 2.5 -- 8 $\times 10^{22}$ cm$^{-2}$.  This corresponds to a wind 
mass loss rate approximately $\dot M_{wind} = 0.8 -- 2.6 \times 10^{-7} M_{\odot}$ yr$^{-1}$ 
if the wind were 
spherical with the X-ray source at its center, and assuming an inner radius corresponding to 
the primary star, 6.4 $\times 10^{10}$ cm and a wind speed of 1600 km s$^{-1}$.  Of course, these 
assumptions are not appropriate in this case, and do not account for the fact that the wind 
is demonstrably non-spherical.  However, this value can be compared with the mass flux 
needed to fuel the X-ray source, which is 
$\dot M_{accretion}\simeq 4.1 \times 10^{-8} M_{\odot}$ yr$^{-1}$ assuming a luminosity 
2.46 $\times 10^{38}$ erg s$^{-1}$ and an efficiency of 0.1.   This demonstrates that there is 
an adequate supply of gas in the wind to fuel the accretion unless the spherical estimate is 
too high by a factor $\geq$5.

An additional question is the origin of the nebular line emission.  This gas has a velocity derived from line 
widths (sigma) which are a fraction of the value expected as the terminal velocity of the wind, approximately half.
This therefore resembles the expectation for the gas which is accelerated in the wind region close 
to the primary star, where the density is high and the ionization parameter is low, and then 
coasts in the region where the X-ay source dominates the ionization.  This scenario therefore provides a 
density measurement, via the He-like G ratio, which applies to the volume of the wind where the X-rays 
dominate.

%The fact that the 6.4 keV feature shows weak orbital modulation suggests that it originates in a region which 
%is large compared with the binary orbit.  This would appear to argue against an origin either close to the 
%compact object or else on the illuminated face of the companion star.

\subsection{Summary}

The Chandra HETG spectrum of Cyg X-3 reveals rich detail in its variability and variety of emission mechansisms.  Line emission 
originates from resonance scattering in a wind, whose dynamics are possibly associated with the UV/optical driven wind 
from the companion star.  In addition there is emission from 'nebular' gas, i.e. gas which is dominated by recombination and without
the obvious kinematic signatures of a wind.  There is potential confusion between one of the strongest features associated with the 
nebular component, i.e. the $f$ line of Si XIII at 1.84 keV, and the artifact of the scattering and absorption of X-rays by the K shell 
of Si in dust; simple models which attempt to simulate this process do not provide as good a fit as the models including nebular emission.
The iron lines reveal a distinct component of gas at much higher ionization, and 
also a component from fluorescence by neutral or near-neutral material.  Attempts to unify the iron emission with that 
of lower-Z elements implies a need for an additional absorption component, possibly associated with a disk wind.  Variability
with orbital phase reinforces the fact that these components are distinct. The wind and nebular components are anti-correlated with the 
peak of the continuum emission, suggesting an origin associated with the companion star.  The fluorescence and iron-absorbing components 
are strongest when the continuum source is strongest, suggesting an association with the compact object.  We have also
examined the long term variability by comparing all the archival observations with significant exposure times.  This shows that
during X-ray low states a distinct continuum component emerges.  We represent this by a 2 keV blackbody, though a hard power law can provide a comparable fit.  This component is crudely anti-correlated with the total luminosity and the diskbb continuum.  The iron 
K$\alpha$ line is also more prominent at low  continuum flux.  The wind and nebular components are likely
associated with the size scale of the binary orbit, and also are strongest when the source luminosity is greatest.   
This suggests that the X-ray luminosity of the source is determined by the wind from the star, i.e. the highest luminosity
states coincide with the strongest wind.  It also confirms the identification of the disk or accretion structure close to the compact 
object as the origin for the fluorescent K$\alpha$ line and the blackbody, and suggests that these features and the associated 
accretion flow structure (i.e. the accretion disk) are anti-correlated with the wind from the companion.

Questions which are not resolved by our fits to these data include the origin, or other explanation, for the notch component 
2.  That is, what is responsible for the apparent absorption suppressing the K lines from intermediate stages of iron 
(i.e. L shell ions).   In the previous sections we have discussed, and provisionally discarded, various possibilities associated 
with the ionization balance in the emitting gas.  We have adopted a partially ionized absorbing wind explanation, though 
this raises additional questions, which we have not addressed, about the energetics, mass flux, and ionization of this wind.   

\acknowledgements{ MM wished to acknowledge support from NASA under contract NAS8-03060. This research has made use of data obtained from the Chandra Data Archive and software provided by the Chandra X-ray Center (CXC).  
 TK was supported in part by
NASA grant 14-ATP14-0022 thru the Astrophysics Theory Program.}

%
%

%------------------------------------------------------------------------------------------------------- 
\bibliographystyle{apj}

\clearpage
\begin{turnpage}
\begin{deluxetable}{crrrrrrrrrr}
\tabletypesize{\scriptsize}
\tablecaption{Line IDs, wavelengths, widths and fluxes from Obsid 7268. \label{table1}}
\tablewidth{0pt}
\tablehead{
\colhead{index}&\colhead{$\varepsilon$}&\colhead{$\lambda$}&\colhead{ion}&\colhead{lower level} &\colhead{upper level} &\colhead{ID}&\colhead{$\sigma$} &\colhead{Flux}&\colhead{v$_{off}$}}
\startdata
&keV&$\AA$&&&&&km s$^{-1}$&cm$^{-2}$ s$^{-1}$&km s$^{-1}$\\
      1&$   6.971^{+ 0.001}_{- 0.001}$&$   1.779^{+ 0.000}_{- 0.000}$&fe\_xxvi&1s1.2S\_1/2      & 1s0.2p1.2P\_3/2 &L$\alpha$&$    200.^{+      8.}_{-      8.}$&$   2.2^{+   0.1}_{-   0.4} \times 10^{ -3}$&     90.\\
      2&$   6.948^{+ 0.003}_{- 0.003}$&$   1.784^{+ 0.004}_{- 0.001}$&fe\_xxvi&1s1.2S\_1/2      & 1s0.2p1.2P\_1/2 &L$\alpha$&$    192.^{+      4.}_{-     17.}$&$   4.7^{+   2.1}_{-   1.1} \times 10^{ -4}$&    170.\\
      3&$   6.697^{+ 0.000}_{- 0.000}$&$   1.851^{+ 0.000}_{- 0.000}$&fe\_xxv &1s2.1S\_0        & 1s1.2p1.1P\_1   &r        &$    198.^{+      4.}_{-     10.}$&$   1.5^{+   0.1}_{-   0.5} \times 10^{ -3}$&    140.\\
      4&$   6.669^{+ 0.001}_{- 0.001}$&$   1.859^{+ 0.000}_{- 0.000}$&fe\_xxv &1s2.1S\_0        & 1s1.2p1.3P\_1   &i        &$    200.^{+      2.}_{-     50.}$&$   2.5^{+   0.0}_{-   1.1} \times 10^{ -3}$&    -50.\\
      5&$   6.641^{+ 0.001}_{- 0.001}$&$   1.867^{+ 0.000}_{- 0.000}$&fe\_xxv &1s2.1S\_0        & 1s1.2s1.3S\_1   &f        &$    198.^{+      2.}_{-     62.}$&$   1.1^{+   0.3}_{-   0.1} \times 10^{ -3}$&   -190.\\
      6&$   5.133^{+ 0.014}_{- 0.014}$&$   2.416^{+ 0.001}_{- 0.006}$&ca\_xx  &1s1.2S\_1/2      & 1s0.4p1.2P\_3/2 &L$\gamma$&$    199.^{+     92.}_{-     75.}$&$   2.8^{+   1.0}_{-   1.0} \times 10^{ -4}$&   -190.\\
      7&$   4.867^{+ 0.002}_{- 0.002}$&$   2.547^{+ 0.001}_{- 0.001}$&ca\_xx  &1s1.2S\_1/2      & 1s0.3p1.2P\_3/2 &L$\beta$ &$    200.^{+     37.}_{-     23.}$&$   4.6^{+   0.8}_{-   1.3} \times 10^{ -4}$&   -190.\\
      8&$   4.859^{+ 0.020}_{- 0.019}$&$   2.552^{+ 0.007}_{- 0.010}$&ca\_xx  &1s1.2S\_1/2      & 1s0.3p1.2P\_1/2 &L$\beta$ &$    200.^{+     23.}_{-     31.}$&$   2.4^{+   1.0}_{-   0.0} \times 10^{ -4}$&    190.\\
      9&$   4.825^{+ 0.013}_{- 0.013}$&$   2.570^{+ 0.002}_{- 0.007}$&ca\_xix &1s2.1S\_0        & 1s1.4p1.1P\_1   &1 - 3    &$    200.^{+     25.}_{-     28.}$&$   3.0^{+   0.8}_{-   1.2} \times 10^{ -4}$&   -180.\\
     10&$   4.580^{+ 0.001}_{- 0.001}$&$   2.707^{+ 0.003}_{- 0.001}$&ca\_xix &1s2.1S\_0        & 1s1.3p1.1P\_1   &1 - 3    &$    199.^{+     25.}_{-     21.}$&$   5.7^{+   0.9}_{-   1.1} \times 10^{ -4}$&    190.\\
     11&$   4.148^{+ 0.001}_{- 0.001}$&$   2.989^{+ 0.003}_{- 0.001}$&ar\_xvii&1s1.2S\_1/2      & 1s0.4p1.2P\_3/2 &L$\gamma$&$    116.^{+     62.}_{-     29.}$&$   4.7^{+   0.5}_{-   0.7} \times 10^{ -4}$&    190.\\
     12&$   4.105^{+ 0.000}_{- 0.000}$&$   3.020^{+ 0.004}_{- 0.000}$&ca\_xx  &1s1.2S\_1/2      & 1s0.2p1.2P\_3/2 &L$\alpha$&$    200.^{+      8.}_{-     17.}$&$   1.7^{+   0.1}_{-   0.1} \times 10^{ -3}$&    150.\\
     13&$   4.097^{+ 0.000}_{- 0.000}$&$   3.026^{+ 0.008}_{- 0.000}$&ca\_xx  &1s1.2S\_1/2      & 1s0.2p1.2P\_1/2 &L$\alpha$&$    200.^{+      2.}_{-     57.}$&$   7.5^{+   0.9}_{-   1.4} \times 10^{ -4}$&    200.\\
     14&$   3.935^{+ 0.001}_{- 0.001}$&$   3.151^{+ 0.001}_{- 0.001}$&ar\_xvii&1s1.2S\_1/2      & 1s0.3p1.2P\_3/2 &L$\beta$ &$    200.^{+     42.}_{-     24.}$&$   5.2^{+   1.1}_{-   1.0} \times 10^{ -4}$&     30.\\
     15&$   3.933^{+ 0.011}_{- 0.011}$&$   3.153^{+ 0.013}_{- 0.008}$&ar\_xvii&1s1.2S\_1/2      & 1s0.3p1.2P\_1/2 &L$\beta$ &$    138.^{+     35.}_{-     28.}$&$   2.0^{+   0.7}_{-   0.0} \times 10^{ -4}$&    100.\\
     16&$   3.902^{+ 0.000}_{- 0.000}$&$   3.177^{+ 0.001}_{- 0.000}$&ca\_xix &1s2.1S\_0        & 1s1.2p1.1P\_1   &r        &$    194.^{+     30.}_{-      6.}$&$   1.5^{+   0.1}_{-   0.1} \times 10^{ -3}$&     10.\\
     17&$   3.890^{+ 0.000}_{- 0.000}$&$   3.187^{+ 0.007}_{- 0.000}$&ca\_xix &1s2.1S\_0        & 1s1.2p1.3P\_2   &i        &$    200.^{+     10.}_{-     13.}$&$   9.1^{+   1.0}_{-   1.3} \times 10^{ -4}$&   -190.\\
     18&$   3.883^{+ 0.000}_{- 0.000}$&$   3.193^{+ 0.009}_{- 0.000}$&ca\_xix &1s2.1S\_0        & 1s1.2p1.3P\_1   &i        &$    199.^{+      4.}_{-     31.}$&$   2.0^{+   0.6}_{-   1.9} \times 10^{ -4}$&     60.\\
     19&$   3.872^{+ 0.000}_{- 0.000}$&$   3.202^{+ 0.007}_{- 0.000}$&ar\_xvii&1s2.1S           & 1s1.4p1.1P      &1 - 4    &$    199.^{+      2.}_{-     44.}$&$  10.0^{+   1.2}_{-   1.0} \times 10^{ -4}$&    190.\\
     20&$   3.861^{+ 0.002}_{- 0.002}$&$   3.211^{+ 0.002}_{- 0.002}$&ca\_xix &1s2.1S\_0        & 1s1.2s1.3S\_1   &f        &$    199.^{+      2.}_{-     49.}$&$   3.9^{+   1.3}_{-   0.8} \times 10^{ -4}$&     20.\\
     21&$   3.682^{+ 0.002}_{- 0.002}$&$   3.367^{+ 0.002}_{- 0.002}$&ar\_xvii&1s2.1S           & 1s1.3p1.1P      &1 - 3    &$    199.^{+      2.}_{-     49.}$&$   3.4^{+   1.0}_{-   1.1} \times 10^{ -4}$&    140.\\
     22&$   3.321^{+ 0.000}_{- 0.000}$&$   3.733^{+ 0.003}_{- 0.000}$&ar\_xvii&1s1.2S\_1/2      & 1s0.2p1.2P\_3/2 &L$\alpha$&$    195.^{+     28.}_{-      6.}$&$   2.3^{+   0.1}_{-   0.2} \times 10^{ -3}$&    160.\\
     23&$   3.316^{+ 0.000}_{- 0.000}$&$   3.739^{+ 0.009}_{- 0.000}$&ar\_xvii&1s1.2S\_1/2      & 1s0.2p1.2P\_1/2 &L$\alpha$&$    193.^{+      2.}_{-     46.}$&$   9.2^{+   1.2}_{-   1.3} \times 10^{ -4}$&    190.\\
     24&$   3.275^{+ 0.001}_{- 0.001}$&$   3.786^{+ 0.001}_{- 0.001}$&s\_xvi  &1s1.2S\_1/2      & 1s0.4p1.2P\_3/2 &L$\gamma$&$    198.^{+      2.}_{-     42.}$&$   6.4^{+   1.2}_{-   1.2} \times 10^{ -4}$&    100.\\
     25&$   3.137^{+ 0.000}_{- 0.000}$&$   3.952^{+ 0.005}_{- 0.000}$&ar\_xvii&1s2.1S           & 1s1.2p1.1P      &r        &$    200.^{+      2.}_{-     34.}$&$   1.1^{+   0.1}_{-   0.1} \times 10^{ -3}$&    190.\\
     26&$   3.127^{+ 0.003}_{- 0.003}$&$   3.965^{+ 0.000}_{- 0.004}$&ar\_xvii&1s2.1S           & 1s1.2p1.3P      &i        &$    200.^{+      2.}_{-     36.}$&$   7.5^{+   1.7}_{-   0.9} \times 10^{ -4}$&   -190.\\
     27&$   3.107^{+ 0.000}_{- 0.000}$&$   3.990^{+ 0.001}_{- 0.001}$&s\_xvi  &1s1.2S\_1/2      & 1s0.3p1.2P\_3/2 &L$\beta$ &$    199.^{+      2.}_{-     28.}$&$   1.6^{+   0.1}_{-   0.1} \times 10^{ -3}$&    -20.\\
     28&$   3.104^{+ 0.000}_{- 0.000}$&$   3.995^{+ 0.011}_{- 0.000}$&ar\_xvii&1s2.1S           & 1s1.2s1.3S      &f        &$    199.^{+      2.}_{-     34.}$&$   1.1^{+  12.6}_{-  12.7} \times 10^{ -5}$&     50.\\
     29&$   3.099^{+ 0.000}_{- 0.000}$&$   4.001^{+ 0.007}_{- 0.000}$&s\_xv   &1s2.1S\_0        & 1s1.5p1.1P\_1   &1 - 5    &$    200.^{+      2.}_{-     39.}$&$   6.5^{+   1.3}_{-   1.2} \times 10^{ -4}$&    190.\\
     30&$   2.621^{+ 0.000}_{- 0.000}$&$   4.731^{+ 0.009}_{- 0.000}$&s\_xvi  &1s1.2S\_1/2      & 1s0.2p1.2P\_3/2 &L$\alpha$&$    200.^{+     70.}_{-      6.}$&$   4.2^{+   0.4}_{-   0.4} \times 10^{ -3}$&    190.\\
     31&$   2.618^{+ 0.000}_{- 0.000}$&$   4.736^{+ 0.010}_{- 0.000}$&s\_xvi  &1s1.2S\_1/2      & 1s0.2p1.2P\_1/2 &L$\alpha$&$    200.^{+      2.}_{-     62.}$&$   2.1^{+   0.4}_{-   0.3} \times 10^{ -3}$&    190.\\
     32&$   2.505^{+ 0.001}_{- 0.001}$&$   4.950^{+ 0.013}_{- 0.003}$&si\_xiv &1s1.2S\_1/2      & 1s0.4p1.2P\_3/2 &L$\gamma$&$    200.^{+      2.}_{-     62.}$&$   1.5^{+   2.0}_{-   0.0} \times 10^{ -4}$&    190.\\
     33&$   2.459^{+ 0.000}_{- 0.000}$&$   5.042^{+ 0.014}_{- 0.000}$&s\_xv   &1s2.1S\_0        & 1s1.2p1.1P\_1   &r        &$    200.^{+     83.}_{-     13.}$&$   1.4^{+   0.2}_{-   0.2} \times 10^{ -3}$&    200.\\
     34&$   2.450^{+ 0.004}_{- 0.004}$&$   5.060^{+ 0.000}_{- 0.009}$&s\_xv   &1s2.1S\_0        & 1s1.2p1.3P\_2   &i        &$    200.^{+     51.}_{-     10.}$&$   1.5^{+   0.2}_{-   0.2} \times 10^{ -3}$&   -190.\\
     35&$   2.445^{+ 0.000}_{- 0.000}$&$   5.070^{+ 0.014}_{- 0.001}$&s\_xv   &1s2.1S\_0        & 1s1.2p1.3P\_1   &i        &$    200.^{+     23.}_{-     13.}$&$   4.2^{+   2.0}_{-   1.8} \times 10^{ -4}$&    190.\\
     36&$   2.432^{+ 0.001}_{- 0.001}$&$   5.098^{+ 0.003}_{- 0.002}$&s\_xv   &1s2.1S\_0        & 1s1.2s1.3S\_1   &f        &$    200.^{+     14.}_{-     19.}$&$   4.0^{+   2.1}_{-   2.1} \times 10^{ -4}$&   -190.\\
     37&$   2.375^{+ 0.000}_{- 0.000}$&$   5.220^{+ 0.009}_{- 0.000}$&si\_xiv &1s1.2S\_1/2      & 1s0.3p1.2P\_3/2 &L$\beta$ &$    200.^{+     14.}_{-     17.}$&$   8.8^{+   2.4}_{-   2.2} \times 10^{ -4}$&    200.\\
     38&$   2.005^{+ 0.000}_{- 0.000}$&$   6.185^{+ 0.008}_{- 0.000}$&si\_xiv &1s1.2S\_1/2      & 1s0.2p1.2P\_3/2 &L$\alpha$&$    200.^{+     81.}_{-      2.}$&$   6.4^{+   0.3}_{-   0.3} \times 10^{ -3}$&    200.\\
     39&$   2.003^{+ 0.000}_{- 0.000}$&$   6.190^{+ 0.017}_{- 0.000}$&si\_xiv &1s1.2S\_1/2      & 1s0.2p1.2P\_1/2 &L$\alpha$&$    200.^{+      2.}_{-     34.}$&$   2.2^{+   0.2}_{-   0.5} \times 10^{ -3}$&    200.\\
     40&$   1.883^{+ 0.001}_{- 0.001}$&$   6.583^{+ 0.003}_{- 0.003}$&mg\_xii &1s1.2S\_1/2      & 1s0.5p1.2P\_3/2 &L$\delta$&$    198.^{+      6.}_{-     39.}$&$   2.6^{+   2.1}_{-   2.0} \times 10^{ -4}$&    130.\\
     41&$   1.864^{+ 0.000}_{- 0.000}$&$   6.652^{+ 0.018}_{- 0.000}$&si\_xiii&1s2.1S\_0        & 1s1.2p1.1P\_1   &r        &$    199.^{+     10.}_{-     24.}$&$   1.4^{+   0.3}_{-   0.3} \times 10^{ -3}$&    190.\\
     42&$   1.856^{+ 0.005}_{- 0.005}$&$   6.681^{+ 0.001}_{- 0.018}$&si\_xiii&1s2.1S\_0        & 1s1.2p1.3P\_2   &i        &$    198.^{+     76.}_{-      3.}$&$   3.2^{+   6.9}_{-   1.5} \times 10^{ -4}$&   -190.\\
     43&$   1.853^{+ 0.000}_{- 0.000}$&$   6.693^{+ 0.018}_{- 0.001}$&si\_xiii&1s2.1S\_0        & 1s1.2p1.3P\_1   &i        &$    199.^{+     76.}_{-      3.}$&$   2.2^{+   4.7}_{-   2.3} \times 10^{ -4}$&    190.\\
     44&$   1.840^{+ 0.000}_{- 0.000}$&$   6.737^{+ 0.001}_{- 0.001}$&mg\_xii &1s1.2S\_1/2      & 1s0.4p1.2P\_3/2 &L$\gamma$&$    200.^{+     14.}_{-     12.}$&$   2.3^{+   0.2}_{-   0.4} \times 10^{ -3}$&    -30.\\
     45&$   1.839^{+ 0.000}_{- 0.000}$&$   6.743^{+ 0.011}_{- 0.001}$&mg\_xii &1s1.2S\_1/2      & 1s0.4p1.2P\_1/2 &L$\gamma$&$    200.^{+      6.}_{-     23.}$&$   2.6^{+   2.0}_{-   2.0} \times 10^{ -4}$&    190.\\
     46&$   1.745^{+ 0.000}_{- 0.000}$&$   7.105^{+ 0.002}_{- 0.002}$&mg\_xii &1s1.2S\_1/2      & 1s0.3p1.2P\_3/2 &L$\beta$ &$    200.^{+      8.}_{-     19.}$&$   7.4^{+   6.1}_{-   1.0} \times 10^{ -4}$&    -50.\\
     47&$   1.743^{+ 0.002}_{- 0.002}$&$   7.112^{+ 0.019}_{- 0.009}$&mg\_xii &1s1.2S\_1/2      & 1s0.3p1.2P\_1/2 &L$\beta$ &$    200.^{+      8.}_{-     19.}$&$   2.3^{+   2.6}_{-   2.2} \times 10^{ -4}$&    190.\\
     48&$   1.728^{+ 0.000}_{- 0.000}$&$   7.176^{+ 0.008}_{- 0.001}$&al\_xiii&1s1.2S\_1/2      & 1s0.2p1.2P\_3/2 &L$\alpha$&$    200.^{+     10.}_{-     13.}$&$   1.4^{+   0.2}_{-   0.3} \times 10^{ -3}$&    200.\\
     49&$   1.599^{+ 0.001}_{- 0.001}$&$   7.752^{+ 0.029}_{- 0.004}$&al\_xii &1s2.1S\_0        & 1s1.2p1.1P\_1   &r        &$    200.^{+     14.}_{-     10.}$&$   1.7^{+   0.3}_{-   0.6} \times 10^{ -3}$&   -190.\\
     50&$   1.472^{+ 0.000}_{- 0.000}$&$   8.420^{+ 0.002}_{- 0.002}$&mg\_xii &1s1.2S\_1/2      & 1s0.2p1.2P\_3/2 &L$\alpha$&$    200.^{+    672.}_{-      4.}$&$   8.0^{+   1.4}_{-   0.7} \times 10^{ -3}$&     30.\\
     51&$   1.471^{+ 0.000}_{- 0.000}$&$   8.430^{+ 0.023}_{- 0.000}$&mg\_xii &1s1.2S\_1/2      & 1s0.2p1.2P\_1/2 &L$\alpha$&$    200.^{+    146.}_{-      8.}$&$   2.3^{+   0.8}_{-   1.0} \times 10^{ -3}$&    190.\\
     52&$   1.351^{+ 0.000}_{- 0.000}$&$   9.175^{+ 0.025}_{- 0.002}$&mg\_xi  &1s2.1S\_0        & 1s1.2p1.1P\_1   &r        &$    200.^{+    235.}_{-      4.}$&$   8.9^{+   1.9}_{-   2.1} \times 10^{ -3}$&    190.\\
     53&$   1.344^{+ 0.001}_{- 0.001}$&$   9.222^{+ 0.001}_{- 0.009}$&mg\_xi  &1s2.1S\_0        & 1s1.2p1.3P\_2   &i        &$    200.^{+    257.}_{-      2.}$&$   1.2^{+   0.3}_{-   0.2} \times 10^{ -2}$&   -190.\\
     54&$   1.342^{+ 0.000}_{- 0.000}$&$   9.237^{+ 0.025}_{- 0.001}$&mg\_xi  &1s2.1S\_0        & 1s1.2p1.3P\_1   &i        &$    200.^{+    226.}_{-      2.}$&$   2.0^{+   4.6}_{-   0.7} \times 10^{ -3}$&    200.\\
     55&$   1.331^{+ 0.001}_{- 0.001}$&$   9.315^{+ 0.005}_{- 0.007}$&mg\_xi  &1s2.1S\_0        & 1s1.2s1.3S\_1   &f        &$    200.^{+    305.}_{-      2.}$&$   1.0^{+   1.0}_{-   0.1} \times 10^{ -2}$&    -30.\\
     56&$   1.168^{+ 0.003}_{- 0.003}$&$  10.612^{+ 0.019}_{- 0.028}$&fe\_xxiv&2s1.2S\_1/2      & 2s0.3p1.2P\_3/2 &         &$    200.^{+    330.}_{-      2.}$&$   9.2^{+  99.7}_{-   3.0} \times 10^{ -3}$&   -190.\\
     57&$   1.163^{+ 0.001}_{- 0.001}$&$  10.661^{+ 0.036}_{- 0.012}$&fe\_xxiv&2s1.2S\_1/2      & 2s0.3p1.2P\_1/2 &         &$    200.^{+    341.}_{-      2.}$&$   9.2^{+ 114.5}_{-   3.2} \times 10^{ -3}$&    -50.\\
\enddata
\tablenotetext{a}{Where uncertainties are not given they are less than the numerical precision given.}
\end{deluxetable}
\end{turnpage}
\clearpage

\clearpage
\begin{turnpage}
\begin{deluxetable}{crrrrrrrrrr}
\tabletypesize{\scriptsize}
\tablecaption{RRC wavelengths, IDs,  widths and fluxes. \label{table2}}
\tablewidth{0pt}
\tablehead{
\colhead{index}&\colhead{$\varepsilon$\tablenotemark{a}}&\colhead{$\lambda$\tablenotemark{a}}&\colhead{ion}&\colhead{lower level} &\colhead{upper level} &\colhead{ID}&\colhead{kT} &\colhead{Flux}&\colhead{v$_{off}$}}
\startdata
&keV&$\AA$&&&&&km s$^{-1}$&cm$^{-2}$ s$^{-1}$&km s$^{-1}$\\
      1&$   5.126^{+ 0.021}_{- 0.020}$&$   2.419^{+ 0.006}_{- 0.010}$&ca\_xix &1s2.1S\_0        & continuum       &RRC      &$    199.^{+     62.}_{-     84.}$&$   6.8^{+  18.8}_{-   0.0} \times 10^{ -5}$&    190.\\
      2&$   4.424^{+ 0.001}_{- 0.001}$&$   2.802^{+ 0.000}_{- 0.000}$&ar\_xvii&1s1.2S\_1/2      & continuum       &RRC      &$    199.^{+    213.}_{-    136.}$&$   4.0^{+   1.1}_{-   1.0} \times 10^{ -4}$&    100.\\
      3&$   4.119^{+ 0.000}_{- 0.000}$&$   3.010^{+ 0.000}_{- 0.000}$&ar\_xvii&1s2.1S           & continuum       &RRC      &$    103.^{+    918.}_{-      4.}$&$   4.5^{+   0.3}_{-   1.0} \times 10^{ -4}$&    120.\\
      4&$   3.226^{+ 0.010}_{- 0.010}$&$   3.844^{+ 0.001}_{- 0.012}$&s\_xv   &1s2.1S\_0        & continuum       &RRC      &$    197.^{+      2.}_{-     44.}$&$   2.2^{+   1.6}_{-   1.5} \times 10^{ -4}$&   -170.\\
      5&$   2.674^{+ 0.007}_{- 0.007}$&$   4.636^{+ 0.001}_{- 0.013}$&si\_xiv &1s1.2S\_1/2      & continuum       &RRC      &$    200.^{+     49.}_{-     84.}$&$   1.1^{+   0.5}_{-   0.6} \times 10^{ -3}$&   -140.\\
      6&$   2.438^{+ 0.000}_{- 0.000}$&$   5.085^{+ 0.000}_{- 0.001}$&si\_xiii&1s2.1S\_0        & continuum       &RRC      &$    200.^{+     25.}_{-     12.}$&$   1.5^{+   0.3}_{-   0.3} \times 10^{ -3}$&    -30.\\
      7&$   1.998^{+ 0.000}_{- 0.000}$&$   6.205^{+ 0.000}_{- 0.000}$&fe\_xxiv&2s0.2p1.2P\_1/2  & continuum       &RRC      &$    155.^{+      6.}_{-     39.}$&$   3.4^{+   0.2}_{-   0.5} \times 10^{ -3}$&   -110.\\
      8&$   1.855^{+ 0.000}_{- 0.000}$&$   6.682^{+ 0.000}_{- 0.001}$&fe\_xxii&2s1.2p1.1P\_1    & continuum       &RRC      &$    200.^{+     12.}_{-     19.}$&$   2.4^{+   0.4}_{-   0.4} \times 10^{ -3}$&    180.\\
      9&$   1.674^{+ 0.000}_{- 0.000}$&$   7.408^{+ 0.002}_{- 0.000}$&fe\_xxi &2p2.3P\_2        & continuum       &RRC      &$    200.^{+     10.}_{-     12.}$&$   1.8^{+   0.8}_{-   0.8} \times 10^{ -3}$&    120.\\
     10&$   1.642^{+ 0.000}_{- 0.000}$&$   7.552^{+ 0.008}_{- 0.002}$&fe\_xxi &2p2.1S\_0        & continuum       &RRC      &$    200.^{+     12.}_{-     12.}$&$   1.0^{+   0.6}_{-   0.6} \times 10^{ -3}$&    200.\\
     11&$   1.583^{+ 0.000}_{- 0.000}$&$   7.832^{+ 0.001}_{- 0.000}$&fe\_xx  &2p3.4S\_3/2      & continuum       &RRC      &$    200.^{+     14.}_{-     10.}$&$   3.5^{+   1.0}_{-   0.9} \times 10^{ -3}$&   -180.\\
     12&$   1.566^{+ 0.000}_{- 0.000}$&$   7.919^{+ 0.001}_{- 0.000}$&fe\_xx  &2p3.2D\_3/2      & continuum       &RRC      &$    200.^{+     19.}_{-      6.}$&$   5.0^{+   1.1}_{-   1.0} \times 10^{ -3}$&   -160.\\
     13&$   1.543^{+ 0.000}_{- 0.000}$&$   8.036^{+ 0.001}_{- 0.001}$&fe\_xx  &2p3.2P\_3/2      & continuum       &RRC      &$    200.^{+     21.}_{-      6.}$&$   3.4^{+   1.1}_{-   0.7} \times 10^{ -3}$&   -190.\\
     14&$   1.457^{+ 0.000}_{- 0.000}$&$   8.512^{+ 0.001}_{- 0.000}$&fe\_xix &2p4.3P\_2        & continuum       &RRC      &$    199.^{+    205.}_{-      4.}$&$   1.1^{+   0.2}_{-   0.2} \times 10^{ -2}$&   -120.\\
     15&$   1.446^{+ 0.000}_{- 0.000}$&$   8.576^{+ 0.001}_{- 0.000}$&fe\_xix &2p4.3P\_0        & continuum       &RRC      &$    199.^{+    218.}_{-      4.}$&$   1.1^{+   0.2}_{-   0.2} \times 10^{ -2}$&    190.\\
     16&$   1.444^{+ 0.000}_{- 0.000}$&$   8.587^{+ 0.023}_{- 0.001}$&fe\_xix &2p4.3P\_1        & continuum       &RRC      &$    198.^{+    190.}_{-      4.}$&$   1.3^{+   6.8}_{-   0.3} \times 10^{ -3}$&    190.\\
     17&$   1.246^{+ 0.000}_{- 0.000}$&$   9.947^{+ 0.000}_{- 0.002}$&ca\_xix &1s1.2p1.3P\_0    & continuum       &RRC      &$    200.^{+    315.}_{-      2.}$&$   9.8^{+  40.2}_{-   2.4} \times 10^{ -3}$&     70.\\
     18&$   1.245^{+ 0.004}_{- 0.004}$&$   9.957^{+ 0.036}_{- 0.030}$&ca\_xix &1s1.2s1.1S\_0    & continuum       &RRC      &$    200.^{+    320.}_{-      2.}$&$   9.8^{+  16.9}_{-   5.2} \times 10^{ -3}$&    -90.\\
\enddata
\end{deluxetable}
\end{turnpage}
\clearpage

\clearpage
\begin{turnpage}

\begin{deluxetable}{crrrrrrrrrrrrrr}
\tabletypesize{\scriptsize}
\tablecaption{Parameters for Global Fit, Nebula plus Wind, Phase-Summed Spectra \label{globalfittable}}
\tablewidth{0pt}
\tablehead{
\colhead{component}&\colhead{parameter}&\colhead{Obsid\_6601}&\colhead{Obsid\_1456$_{000}$}&\colhead{Obsid\_1456$_{002}$}&\colhead{Obsid\_425}&\colhead{Obsid\_16622}&\colhead{Obsid\_426}&\colhead{Obsid\_7268}}
\startdata
    1&NH&    4.25$_{    -0.01  }^{+    0.01  }$ &    3.75$_{    -0.05 }^{+    0.04 }$ &    4.02$_{    -0.10 }^{+    0.09 }$ &    5.68$_{    -0.02 }^{+    0.03 }$ &    2.28$_{    -0.07 }^{+    0.06 }$ &    5.34$_{    -0.03 }^{+    0.03 }$ &    4.16$_{    -0.01  }^{+    0.01  }$\\                                                                                                                          
    2&notch energy - 6. (keV)&    0.56$_{    -6.00}^{+   -6.00}$ &    0.55$_{    -6.00}^{+   -6.00}$ &    0.55$_{    -6.00}^{+   -6.00}$ &    0.55$_{    -6.00}^{+   -6.00}$ &    0.50$_{    -6.00}^{+   -6.00}$ &    0.55$_{    -6.00}^{+   -6.00}$ &    0.57$_{    -6.00}^{+   -6.00}$\\                                                                                                                       
    3&notch (2) width&    0.16 &    0.10  &    0.10  &    0.10  &&    0.10  &    0.16\\                                                                                                                                                                                                                                                                                                                          
    4&notch (2) depth&    0.47$_{    -0.01  }^{+    0.01 }$ &    0.58$_{    -0.03 }^{+    0.03 }$ &    0.68$_{    -0.03 }^{+    0.03 }$ &    0.58$_{    -0.03 }^{+    0.03 }$ &$\leq$    0.04 &    0.58$_{    -0.03 }^{+    0.03 }$ &    0.59$_{    -0.01 }^{+    0.01 }$\\                                                                                                                                      
    5&mg$^{a,b}$&    7.58$_{    -0.07 }^{+    0.06 }$ &&&&&&\\                                                                                                                                                                                                                                                                                                                                                            
    6&al&    6.16$_{    -0.23}^{+    0.15}$ &&&&&&\\                                                                                                                                                                                                                                                                                                                                                              
    7&si&    8.06$_{    -0.01 }^{+    0.01 }$ &&&&&&\\                                                                                                                                                                                                                                                                                                                                                            
    8&s&    7.67$_{    -0.03 }^{+    0.04 }$ &&&&&&\\                                                                                                                                                                                                                                                                                                                                                             
    9&ar&    7.14$_{    -0.06 }^{+    0.05 }$ &&&&&&\\                                                                                                                                                                                                                                                                                                                                                            
   10&ca&    7.19$_{    -0.15}^{+    0.06 }$ &&&&&&\\                                                                                                                                                                                                                                                                                                                                                             
   11&fe&    8.27 &&&&&&\\                                                                                                                                                                                                                                                                                                                                                                                        
   12&z&0.001&&&&&&\\                                                                                                                                                                                                                                                                                                                                                                                                  
   13&medium $\xi$ comp (3) log($\xi$)$^c$&    2.90$_{    -0.01  }^{+    0.01  }$ &    2.90 &    2.90 &    2.90 &    2.20 &    2.90 &    2.89$_{    -0.05 }^{+    0.03 }$\\                                                                                                                                                                                                                                            
   14&medium $\xi$ comp (3) norm$^d$&    0.75$_{     0.00}^{+    0.00  }$ &    0.25$_{     0.00}^{+    0.03 }$ &    0.41$_{    -0.05 }^{+    0.05 }$ &    0.97$_{    -0.06 }^{+    0.06 }$ &$\leq$    0.01 &    0.97$_{    -0.09 }^{+    0.09 }$ &    0.43$_{    -0.05 }^{+    0.06 }$\\                                                                                                                         
   15&Fe K$\alpha$ norm*10$^4$$^e$&    3.26$_{    -0.20}^{+    0.20}$ &    3.45$_{    -0.27}^{+    0.28}$ &    3.05$_{    -0.30}^{+    0.31}$ &    2.50$_{    -0.32}^{+    0.33}$ &    4.30$_{    -0.26}^{+    0.28}$ &    2.52$_{    -0.35}^{+    0.35}$ &$\leq$    0.18\\                                                                                                                                          
   16&high $\xi$ comp (5) log($\xi$)$^c$&    4.80 &    4.50 &    5.00 &    5.00 &    4.20 &    5.00 &    5.00\\                                                                                                                                                                                                                                                                                                      
   17&high $\xi$ comp (5) norm/10$^2$$^d$&    5.85$_{    -1.44}^{+    1.28}$ &    4.83$_{    -2.16}^{+    0.65}$ &    5.00$_{    -2.44}^{+    2.09}$ &    8.49$_{    -2.85}^{+    2.89}$ &   11.43 $_{    -0.19}^{+    0.20}$ &    8.49$_{    -3.55}^{+    3.59}$ &    4.63$_{    -1.66}^{+    1.68}$\\                                                                                                              
   18&wind comp (6) column$^f$&    0.64$_{    -0.10 }^{+    0.10 }$ &    0.89$_{    -0.10 }^{+    0.10 }$ &    1.10$_{    -0.10 }^{+    0.10 }$ &    0.89$_{    -0.10 }^{+    0.10 }$ &   -0.23$_{    -0.10 }^{+    0.10 }$ &    0.89$_{    -0.10 }^{+    0.10 }$ &    0.64\\                                                                                                                                        
   19&vturb (km s$^{-1}$)&         1600.00    &&&&&&\\                                                                                                                                                                                                                                                                                                                                                                          
   20&cfrac&    1.80 &&&&&&\\                                                                                                                                                                                                                                                                                                                                                                                     
   21&z$_{wind}$&0.001&&&&&&&\\                                                                                                                                                                                                                                                                                                                                                                                            
   22&tin&    1.72 &&&&&&\\                                                                                                                                                                                                                                                                                                                                                                                       
   23&diskbb comp (7) norm$^g$&    0.76 &    0.07  &    0.05  &    0.13$_{     0.00}^{+    0.01  }$ &&    0.13 &    0.94\\                                                                                                                                                                                                                                                                                    
   24&kT (8)&    2.00 &    1.97 &    2.39 &    1.34 &    2.00 &    1.34 &    2.00\\                                                                                                                                                                                                                                                                                                                              
   25&blackbody norm*10$^2$$^g$&&    2.81$_{    -0.05 }^{+    0.07 }$ &    2.49$_{    -0.10 }^{+    0.11}$ &    9.55$_{    -0.10}^{+    0.03 }$ &    3.89$_{    -0.04 }^{+    0.04 }$ &    9.55$_{    -0.21}^{+    0.16}$ &    1.29$_{    -0.06 }^{+    0.06 }$\\                                                                                                                                                    
   26&cstat&        61824.94     &        15565.53     &        13101.69     &        21325.52     &        13387.40     &        21325.50     &      214748.36     \\                                                                                                                                                                                                                                          
   27&dof&        32761.00     &        32761.00     &        32761.00     &        32761.00     &        32761.00     &        32761.00     &        32761.00    \\                                                                                                                                                                                                                                             
   28&chisq&        68105.80     &         7344.08    &         4871.88    &        14424.98     &         5068.53    &        14425.15     &        65343.20    \\                                                                                                                                                                                                                                              
   29&flux/100 mCrab&    2.53 &    0.83 &    0.60 &    2.55 &    0.86 &    2.33 &    3.44\\                                                                                                                                                                                                                                                                                                                      
   30&tobs&        49200.00     &        12100.00     &         8419.00    &        18959.77     &        28625.00     &        15700.00     &        69900.00    \\                                                                                                                        
\enddata
\tablenotetext{a}{Abundances are given in log value relative to hydrogen, with H=12}
\tablenotetext{b}{Where values are not repeated across columns, they are held constant for all fits}
\tablenotetext{c}{$\xi$ has units erg cm s$^{-1}$}
\tablenotetext{d}{See discussion in text for normalization units}
\tablenotetext{e}{iron line flux is in units of cm$^{-2}$ s$^{-1}$}
\tablenotetext{f}{log(wind column) where wind column is in units of 10$^{22}$ cm$^{-2}$}
\tablenotetext{g}{Continuum norm for 1.72 keV diskbb spectrum:  norm is $(R_{in}/D_{10})^2 \cos\theta$, where where $R_{in}$ is apparent 
inner disk radius in km, $D_{10}$ the distance to the source in units of 10 kpc, and $\theta$ the angle of the disk 
($\theta= 0$ is face-on)}

\end{deluxetable}
\end{turnpage}
\clearpage

\clearpage
\begin{turnpage}
\begin{deluxetable}{crrrrrrrrrrrrrr}
\tabletypesize{\scriptsize}
\tablecaption{Parameters for Global Fit, Nebula plus Wind, Phase-binned Spectra \label{globalfittable2}}
\tablewidth{0pt}
\tablehead{
 \colhead{component}&\colhead{parameter}&\colhead{7268 phase 1}&\colhead{7268 phase 2}&\colhead{7268 phase 3}&\colhead{7268 phase 4}&\colhead{6601 phase 1}&\colhead{6601 phase 2}&\colhead{6601 phase 3}&\colhead{6601 phase 4}}
\startdata
    1&NH&    4.09$_{    -0.02 }^{+    0.02 }$ &    4.40$_{    -0.01 }^{+    0.01 }$ &    4.20$_{    -0.01 }^{+    0.01 }$ &    4.30$_{    -0.01 }^{+    0.01 }$ &    4.17$_{    -0.02 }^{+    0.02 }$ &    4.36$_{    -0.01 }^{+    0.01 }$ &    4.01$_{    -0.01  }^{+    0.01  }$ &    3.90\\                                                                                                                  
    2&notch energy - 6. (keV)&    0.56$_{    -6.01}^{+   -6.00}$ &    0.57$_{    -6.00}^{+   -6.00}$ &    0.57$_{    -6.01}^{+   -6.00}$ &    0.57$_{    -6.00}^{+   -6.00}$ &    0.53$_{    -6.00}^{+   -6.00}$ &    0.55$_{    -6.00}^{+   -6.00}$ &    0.59$_{    -6.00}^{+   -6.00}$ &    0.55$_{    -6.00}^{+   -6.00}$\\                                                                                   
    3&notch (2) width&    0.13$_{    -0.01  }^{+    0.02 }$ &    0.16$_{    -0.01  }^{+    0.01  }$ &    0.17$_{    -0.01 }^{+    0.01 }$ &    0.16$_{    -0.01 }^{+    0.01  }$ &    0.12$_{    -0.01  }^{+    0.01  }$ &    0.12$_{    -0.01  }^{+    0.01  }$ &    0.14$_{    -0.01  }^{+    0.01  }$ &    0.12\\                                                                                             
    4&notch (2) depth&    0.68$_{    -0.04 }^{+    0.03 }$ &    0.66$_{    -0.02 }^{+    0.02 }$ &    0.59$_{     0.02 }^{+    0.02 }$ &    0.68$_{    -0.02 }^{+    0.01 }$ &    0.94$_{    -0.03 }^{+    0.02 }$ &    0.83$_{    -0.03 }^{+    0.03 }$ &    0.41$_{    -0.02 }^{+    0.02 }$ &    0.74\\                                                                                                       
    5&mg$^{a,b}$&    7.58$_{    -0.07 }^{+    0.06 }$ &&&&&&\\                                                                                                                                                                                                                                                                                                                                                            
    6&al&    6.16$_{    -0.23}^{+    0.15}$ &&&&&&\\                                                                                                                                                                                                                                                                                                                                                              
    7&si&    8.06$_{    -0.01 }^{+    0.01 }$ &&&&&&\\                                                                                                                                                                                                                                                                                                                                                            
    8&s&    7.67$_{    -0.03 }^{+    0.04 }$ &&&&&&\\                                                                                                                                                                                                                                                                                                                                                             
    9&ar&    7.14$_{    -0.06 }^{+    0.05 }$ &&&&&&\\                                                                                                                                                                                                                                                                                                                                                            
   10&ca&    7.19$_{    -0.15}^{+    0.06 }$ &&&&&&\\                                                                                                                                                                                                                                                                                                                                                             
   11&fe&    8.27 &&&&&&\\                                                                                                                                                                                                                                                                                                                                                                                        
   12&z&0.001&&&&&&\\                                                                                                                                                                                                                                                                                                                                                                                                  
   13&medium $\xi$ comp (3) log($\xi$)$^c$&    2.89 &    2.89$_{    -0.07 }^{+    0.01  }$ &    2.89$_{    -0.07 }^{+    0.01  }$ &    2.89$_{    -0.05 }^{+    0.03 }$ &    2.57 &    2.66 &    2.91 &    2.69\\                                                                                                                                                                                                      
   14&medium $\xi$ comp (3) norm$^d$&    1.29$_{     0.00}^{+    0.01  }$ &    0.13$_{     0.00}^{+    0.02 }$ &    0.70$_{     0.00}^{+    0.04 }$ &    2.39$_{    -0.19}^{+    0.17}$ &    1.48$_{     0.00}^{+    0.08 }$ &    0.04 $_{     0.00 }^{+    0.01  }$ &    0.80$_{    -0.09 }^{+    0.09 }$ &    1.60\\                                                                                           
   15&Fe K$\alpha$ norm*10$^4$$^e$&    0.88$_{    -0.30}^{+    0.30}$ &    1.50$_{    -0.40}^{+    0.40}$ &    0.44$_{     0.00}^{+    0.80}$ &    0.19$_{     0.40}^{+    0.40}$ &    0.66$_{    -0.20}^{+    0.30}$ &    2.92$_{    -0.40}^{+    0.40}$ &    3.94$_{    -0.40}^{+    0.40}$ &    2.57\\                                                                                                            
   16&high $\xi$ comp (5) log($\xi$)$^c$&    4.40 &    5.00 &    5.00 &    5.00 &    4.20 &    4.90 &    4.90 &    4.90\\                                                                                                                                                                                                                                                                                            
   17&high $\xi$ comp (5) norm/10$^2$$^d$&    9.05$_{    -1.59}^{+    1.04}$ &    1.88$_{    -1.46}^{+    2.22}$ &    0.36$_{    -0.76}^{+    2.83}$ &    7.43$_{    -5.44}^{+    4.07}$ &    9.75$_{    -1.51}^{+    1.37}$ &    4.25$_{    -2.74}^{+    2.10}$ &    4.36$_{    -3.25}^{+    3.43}$ &   10.14 \\                                                                                                    
   18&wind comp (6) column$^f$&    0.89$_{    -0.00  }^{+    0.01 }$ &    0.73$_{    -0.02 }^{+    0.03 }$ &    0.60$_{    -0.02 }^{+    0.03 }$ &    0.79$_{    -0.02 }^{+    0.02 }$ &    0.95 &    0.70 &    0.39 &    0.61\\                                                                                                                                                                                     
   19&vturb&1600.0&&&&&&&\\                                                                                                                                                                                                                                                                                                                                                                                             
   20&cfrac&1.80&&&&&&&\\                                                                                                                                                                                                                                                                                                                                                                                             
   21&z$_{wind}$&0.001&&&&&&&\\                                                                                                                                                                                                                                                                                                                                                                                            
   22&tin&1.72&&&&&&&\\                                                                                                                                                                                                                                                                                                                                                                                               
   23&diskbb comp (7) norm$^g$&    0.51 &    0.96 &    1.35 &    1.24 &    0.44 &    0.76 &    0.97 &    0.98\\                                                                                                                                                                                                                                                                                               
   24&kT (8)&2.00&&&&&&&\\                                                                                                                                                                                                                                                                                                                                                                                            
   25&blackbody norm*10$^2$$^g$&&&&&&&&\\                                                                                                                                                                                                                                                                                                                                                                             
   26&cstat&        59227.92     &        66186.24     &        65288.02     &        65198.33     &        21812.77     &         2257.27    &        29146.67     &        27815.88    \\                                                                                                                                                                                                                      
   27&dof&        32761.00     &        32761.00     &        32761.00     &        32761.00     &        32761.00     &        32761.00     &        32761.00     &        32761.00    \\                                                                                                                                                                                                                       
   28&chisq&        16371.30     &        19699.67     &        21359.17     &        21462.95     &        12296.02     &        15580.80     &        22375.15     &       214748.36     \\                                                                                                                                                                                                                    
   29&flux/100 mCrab&    1.68 &    3.06 &    4.44 &    3.96 &    1.28 &    2.36 &    3.31 &    3.20\\                                                                                                                                                                                                                                                                                                            
\enddata 
\tablenotetext{a}{Abundances are given in log value relative to hydrogen, with H=12}
\tablenotetext{b}{Where values are not repeated across columns, they are held constant for all fits}
\tablenotetext{c}{$\xi$ has units erg cm s$^{-1}$}
\tablenotetext{d}{See discussion in text for normalization units}
\tablenotetext{e}{iron line flux is in units of cm$^{-2}$ s$^{-1}$}
\tablenotetext{f}{log(wind column) where wind column is in units of 10$^{22}$ cm$^{-2}$}
\tablenotetext{g}{Continuum norm for 1.72 keV diskbb spectrum:  norm is $(R_{in}/D_{10})^2 \cos\theta$, where where $R_{in}$ is apparent 
inner disk radius in km, $D_{10}$ the distance to the source in units of 10 kpc, and $\theta$ the angle of the disk 
($\theta= 0$ is face-on)}
\end{deluxetable}
\end{turnpage}
\clearpage

\end{document}